\documentclass[11pt]{article}
\pdfoutput=1
\usepackage[centertags]{amsmath}
\usepackage[square, comma, sort&compress,numbers]{natbib}
\usepackage{array,multirow}
\numberwithin{equation}{section}
\usepackage{amssymb,amsfonts}
\usepackage{graphicx}
\usepackage{color}
\usepackage{mathtools,bm}
\usepackage{accents}
\usepackage{physics}
\usepackage{authblk}
\usepackage[title,titletoc]{appendix}
\usepackage{cancel}

\usepackage{epsfig}
\usepackage{bbold}

\usepackage{wrapfig}
\usepackage{float}
\usepackage{soul}
\usepackage{braket}

\newtheorem{prop}{Proposition}[section]
\newtheorem{lemma}[prop]{Lemma}

\newcommand{\sgn}{\mathop{\mathrm{sgn}}}

\def\cA{\mathcal{A}}
\def\cB{\mathcal{B}}
\def\cC{\mathcal{C}}
\def\cD{\mathcal{D}}

\def\cJ{\mathcal{J}}

\def\cM{\mathcal{M}}

\def\cS{\mathcal{S}}

\numberwithin{equation}{section} \makeatletter
\@addtoreset{equation}{section}

\newcommand{\ga}{\alpha}
\newcommand{\gad}{{\dot{\alpha}}}
\newcommand{\gb}{\beta}
\newcommand{\gbd}{{\dot{\beta}}}
\renewcommand\gg{\gamma}
\newcommand{\ggd}{{\dot{\gamma}}}

\newcommand{\yd}{\bar{y}}
\newcommand{\zd}{\bar{z}}

\newcommand{\pd}{{\bar{p}}}

\newcommand{\Hd}{{\bar{H}}}

\newcommand{\hmt}{\vartriangle}

\newcommand{\lp}{\left(}
\newcommand{\rp}{\right)}
\newcommand{\lb}{\left[}
\newcommand{\rb}{\right]}

\newcommand{\rd}{\right.}
\newcommand{\ld}{\left.}

\newcommand{\Um}{{\underline{m}}}
\newcommand{\Un}{{\underline{n}}}
\newcommand{\Uk}{{\underline{k}}}
\newcommand{\Ul}{{\underline{l}}}
\newcommand{\Up}{{\underline{p}}}
\newcommand{\Uq}{{\underline{q}}}

\makeatletter
\newcommand{\pushright}[1]{\ifmeasuring@#1\else\omit\hfill$\displaystyle#1$\fi\ignorespaces}
\newcommand{\pushleft}[1]{\ifmeasuring@#1\else\omit$\displaystyle#1$\hfill\fi\ignorespaces}
\makeatother
\usepackage{jheppub}
\makeatletter
\def\@fpheader{\vspace{-.1cm}}
\makeatother

\title{\centering{Bilinear Fronsdal currents in the AdS\textsubscript{4} higher-spin theory}}

\author[1,2]{Yu. A. Tatarenko}
\author[1,2]{M. A. Vasiliev}

\affiliation[1]{
I.E. Tamm Department of Theoretical Physics, \\
P.N. Lebedev Physical Institute, Leninsky ave. 53, 119991 Moscow, Russia}
\affiliation[2]{
Moscow Institute of Physics and Technology, \\
Institutsky lane 9, 141700, Dolgoprudny, Moscow region, Russia}

\emailAdd{tatarenko.iua@phystech.edu}
\emailAdd{vasiliev@lpi.ru}

\abstract{We analyse higher-spin theory with general coupling constant $\eta$ at the second order, focusing on the gauge non-invariant vertices $\Upsilon(\omega,\omega)$, $\Upsilon(\Omega,\omega,C)$ and $\Upsilon(\omega,C)$, that are shown to generate nontrivial currents in the Fronsdal equations. Explicit expressions for the currents are found in the frame-like formalism counterpart of the TT gauge worked out in the paper. The nonlinear higher-spin theory is shown to generate all types of Metsaev's currents with the coupling constants manifestly expressed via the complex coupling constant $\eta$ of
the higher-spin theory. It is shown that all currents in the higher-spin theory are conformal in the TT gauge
except for those bilinear in the higher-spin gauge fields $\omega$.}
 
\dedicated{To the memory of Christian Fronsdal}

\preprint{FIAN/TD/8-24}

\begin{document}

\maketitle
\flushbottom
\newpage
%%%%%%%%%%%%%%%%%%%%%%%%%%%%%%%%
\section{Introduction}\label{Introduction}
%%%%%%%%%%%%%%%%%%%%%%%%%%%%%%%%

The consistent system of nonlinear equations for the 4d higher-spin (HS) fields was originally found in \cite{Vasiliev:1990en} while its modern formulation used in this paper was proposed in \cite{Vasiliev:1992av}. This construction is formulated in terms of the star-product realization of the HS algebra  (see review \cite{Vasiliev:1999ba} and references therein). Due to nonlocality of the star product the question of space-time locality is highly nontrivial. During the recent years,  this problem has been   extensively  studied \cite{Gelfond:2018vmi, Didenko:2018fgx, Didenko:2019xzz, Gelfond:2019tac, Didenko:2020bxd, Didenko:2022qga, Didenko:2023vna, Sleight:2017pcz, Lysov:2022nsv, Neiman:2023orj}. Usually it is analysed within the  perturbative approach to nonlinear HS equations starting with $AdS$ background and studying the perturbations of the HS gauge potentials $\omega$ (one-forms) and HS Weyl tensors
$C$ (zero-forms). In these terms, the HS equations have  the structure (in the sequel the wedge symbols are implicit)
\begin{align*}
    \dd \omega &=
        - \omega \star \omega + \Upsilon(\omega, \omega, C) + \Upsilon(\omega, \omega, C, C) + ...\,,\\
    \dd C &= -[\omega, C]_\star + \Upsilon(\omega, C, C) + ...\,
\end{align*}
with $\omega$ and $C$  treated as generating functions for the fields within the unfolding procedure \cite{Vasiliev:1988sa} which expresses their components via space-time derivatives of physical (Fronsdal) fields contained in the de Rham differential $\dd = \dd x^{\underline{n}}\,\pdv{}{x^{\underline{n}}}$. In this setup, for a fixed spin, $\omega$ contains a finite number of derivatives while $C$ contains an infinite number of them \cite{Vasiliev:1988sa}. Along with the star-product properties, the
latter fact may lead to the nonlocality starting from the terms  quadratic in $C$. HS equations have a freedom in expressing fields of the second perturbation order via the lower-order ones. In \cite{Boulanger:2015ova} it was shown that the seemingly simplest 
solution leads to the nonlocal $C^2$ vertices. Later on, in the series of papers \cite{Gelfond:2018vmi, Didenko:2018fgx, Didenko:2019xzz, Gelfond:2019tac, Didenko:2020bxd} it was shown how the local 
$C^2$ vertices can be obtained using the so-called shifted homotopy approach.

As a result, the $C^2$ vertices are well-studied. For instance, in \cite{Misuna:2017bjb} it was shown that the $C^2$ vertices resulting from the unfolded HS equations match those obtained by other approaches \cite{Sleight:2016dba,Metsaev:2005ar}. On the other hand, the manifestly spin-local gauge non-invariant vertices bilinear in $\omega$ and $C$  are less explored. In this paper we fill in this gap showing that the gauge non-invariant vertices generate nontrivial currents in the Fronsdal equations \cite{Fronsdal:1978rb} and finding their explicit form as  was done for the $C^2$ currents in \cite{Misuna:2017bjb}. The relation of these currents to the cubic vertices obtained in the light-cone formalism is established (for 4d Minkowski fields light-cone cubic vertices were obtained in \cite{Bengtsson:1983pd,Bengtsson:1986kh}, the full classification of cubic $P$-even HS vertices in Minkowski space of arbitrary dimension was obtained in \cite{Metsaev:2005ar} and those in $AdS_4$ were found in \cite{Metsaev:2018xip}). See also papers \cite{Berends:1984wp,Berends:1984rq,Berends:1985xx,Metsaev:1991mt,Metsaev:1993ap,Metsaev:1999ui,Bekaert:2005jf,Bengtsson:2006pw,Metsaev:2007rn,Manvelyan:2010jr,Sagnotti:2010at,Fotopoulos:2010ay,Vasiliev:2011knf,Joung:2012rv,Bengtsson:2012jm,Buchbinder:2012xa,Francia:2016weg,Buchbinder:2017nuc,Kuzenko:2022hdv,Buchbinder:2023xlb,Buchbinder:2024pjm}
where cubic HS vertices were studied within various formalisms in the lowest order that does not determine the coupling constants.

An interesting output of this analysis is the interpretation of the two types of independent $P$-even vertices $V_{\text{min}}$ and $V_{\text{max}}$ identified by Metsaev, that carry either $s_1+s_2+s_3-2\min\{s_1,s_2,s_3\}$ for $V_{\text{min}}$ or  $s_1+s_2+s_3$ for $V_{\text{max}}$ derivatives in the vertices involving integer spins $s_1$, $s_2$ and $s_3$. Usually, independent vertices in a theory are associated with the independent coupling constants. The question is whether this is the case in the HS theory. The answer obtained in this paper is positive. Namely, the HS theory contains a complex coupling constant $\eta =|\eta| \exp i\vartheta$ equivalent to two real ones, $|\eta|$ and $\vartheta$.  We show that the coupling constant associated with $V_{\text{min}}$ is phase independent while that associated with $V_{\text{max}}$ does depend on the phase. Apart from $P$-even vertices, HS theory may contain $P$-odd ones, which  are also found here along with the corresponding coupling constant. The coupling constants of the higher-derivative currents (both $P$-even and $P$-odd) depend on $\vartheta$ in a way analogous to \cite{Giombi:2012ms}, where the case of $s_3=0$ was considered. Having the structure analogous 
to that of the boundary correlators of \cite{Maldacena:2012sf} this result is anticipated to be important in the context of the 
HS holographic duality  \cite{Klebanov:2002ja, Sezgin:2002rt, Sezgin:2003pt} in presence of Chern-Simons interactions at the boundary \cite{Aharony:2011jz, Giombi:2011kc}.

Another useful output of our analysis is a frame-like  HS analogue of the  TT-gauge of the metric-like formalism \cite{Joung:2011ww}.

The rest of the paper is organized as follows. In section \ref{Higher-spin_equations} we recall the basics of 4d nonlinear HS theory. Details of the perturbative analysis of the HS equations are explained in section \ref{Perturbative_analysis}, where also  the formulae for all bilinear vertex functions in the  theory are collected. Obtained currents are analysed in section \ref{Analysis_of_currents} where we discuss restrictions on spins in the vertices, independence and nontriviality of the currents, and the gauge fixing. Also in this section we comment on the derivation of  conserved currents and charges  from the cubic HS action. Section \ref{Projection} contains details of the transition from unfolded to the Fronsdal currents along with the explicit expressions for the resulting Fronsdal currents. Technical details are presented in appendices A -- D.

%%%%%%%%%%%%%%%%%%%%%%%%%%%%%%%%
\section{Higher-spin equations}\label{Higher-spin_equations}
%%%%%%%%%%%%%%%%%%%%%%%%%%%%%%%%

We analyse the system of nonlinear HS equations of \cite{Vasiliev:1992av}, that has the form\footnote{For reviews see \cite{Vasiliev:1999ba} or \cite{Didenko:2014dwa} which is more detailed but contains neither fermions nor topological fields.}
\begin{subequations}\label{Vasiliev}
\begin{align}
    &\dd W + W\star W=0\,,\label{Vas1}\\ 
    &\dd B + [W, B]_\star=0\,,\label{Vas2}\\ 
    &\dd S + [W, S]_\star =0\,,\label{Vas3}\\ 
    &S \star S =i(
            \theta^A \theta_A +
            \eta B \star \gamma +
            \bar{\eta} B \star \bar{\gamma}
        )\,,\label{Vas4}\\  
    &[S, B]_{\star}=0\,. \label{Vas5}
\end{align}
\end{subequations} 
The \emph{master fields} $W$, $B$ and $S$ are valued in the HS algebra generated by commuting spinor variables 
$Y^A = (y^\ga, \yd^\gad)$, $Z^A = (z^\ga, \zd^\gad)$ (indices $\ga$ and $\gad$ take 2 values), 
anticommuting $Z$-differentials $\theta^A = (\theta^\ga, \Bar{\theta}^\gad)$ such that
\begin{equation}\label{Z-diff}
    \{\dd x^{\underline{n}}, \theta^A\} = \{\theta^A, \theta^B\}=0\,,
\end{equation}
and \emph{exterior Klein operators} $K=(k,\bar{k})$ obeying the following  relations:
\begin{subequations}\label{ExtKlein}
\begin{align}
    k f(z,\zd,y,\yd;\theta,\Bar{\theta}) = f(-z,\zd,-y,\yd;-\theta,\Bar{\theta}) k\,,\\
    \bar{k} f(z,\zd,y,\yd;\theta,\Bar{\theta}) = f(z,-\zd,y,-\yd;\theta,-\Bar{\theta})\bar{k}\,,\\
    [k, \bar{k}] = 0\,,\qquad k k = \bar{k} \bar k  = 1\,.
\end{align}
\end{subequations}

Associative \emph{star product}  of functions $f(Z, Y; \theta; K) $ is defined  via the integral formula\footnote{Note that HS algebra combines the two types of product: the one with commutation relations \eqref{Z-diff}, \eqref{ExtKlein} and star product 
\eqref{star-product}  on the functions  of $Y$ and $Z$.}
\begin{equation}\label{star-product}
    f(Z, Y; \theta; K) \star g(Z, Y; \theta; K)
        := \int dU dV f(Z + U, Y + U; \theta; K)e^{iU_A V^A }g(Z - V, Y + V; \theta; K)\,.
\end{equation}
Here $U_A V^A := u_\ga v^\ga + \bar{u}_\gad \bar{v}^\gad$ where indices are raised and lowered by the $\mathfrak{sp}(2)$-symplectic form $u_\ga = u^\gb \epsilon_{\gb\ga},\,u^\ga = \epsilon^{\ga\gb}u_\gb,\,\epsilon_{\gb}{}^\ga = \delta^\ga_\gb$.
The integration measure is normalised so that $1\star 1=1$.

So defined star product admits \emph{inner Klein operators} $\varkappa := e^{iz_\ga y^\ga}$, $\bar{\varkappa} := e^{i\zd_\gad \yd^\gad}$ obeying commutation relations analogous to \eqref{ExtKlein} except that they commute with $\theta^A$:
\begin{subequations}\label{InnKlein}
\begin{align}
    \varkappa \star f(z,\zd,y,\yd;\theta,\Bar{\theta}) =
        f(-z,\zd,-y,\yd;\theta,\Bar{\theta}) \star \varkappa\,,\\
    \bar{\varkappa} \star f(z,\zd,y,\yd;\theta,\Bar{\theta}) =
        f(z,-\zd,y,-\yd;\theta,\Bar{\theta}) \star \bar{\varkappa}\,,\\
    [\varkappa, \bar{\varkappa}]_\star = 0\,,\qquad
    \varkappa\star\varkappa = \bar{\varkappa}\star\bar{\varkappa} = 1\,.
\end{align}
\end{subequations}
As a result, by virtue of the Schouten identity implying that $\theta^\alpha \theta_\alpha \theta_\beta=0$,
$$\gamma := k\varkappa \theta^\ga\theta_\ga\,,\qquad  \bar{\gamma} := \bar{k}\bar{\varkappa} \bar{\theta}^\gad\bar{\theta}_\gad$$ 
are central elements of the HS algebra.

The field $B$ is a zero-form, while $W$ and $S$ are one-forms in $x$- and $Z$-differentials, respectively. These master fields contain both physical and topological fields. The truncation to the  physical sector is
\begin{subequations}\label{DynCond}
\begin{align}
W(Z, Y; K | x) &= W(Z, Y; -K | x)\,,\label{DynW}\\
S(Z, Y; K | x) &= S(Z, Y; -K | x)\,,\label{DynS}\\
B(Z, Y; K | x) &= -B(Z, Y; -K | x)\,.\label{DynB}
\end{align}
\end{subequations}

The theory has a free complex coupling constant $\eta$. Note that its absolute value (if  non-zero)  can be rescaled away by  a field redefinition  analogously to pure gravity 
and Yang-Mills theory at the classical level, while its phase is an essential parameter \cite{Vasiliev:1992av}.

%%%%%%%%%%%%%%%%%%%%%%%%%%%%%%%%
\section{Perturbative analysis}\label{Perturbative_analysis}
%%%%%%%%%%%%%%%%%%%%%%%%%%%%%%%%

In this section the standard perturbative analysis of the HS equations in $AdS_4$ is recalled.

\subsection{Vacuum}\label{Vacuum}%%%%%%%%%%%%%%%%%%%%%%%%%%%%%%%%%%%%%%%%%%%%%%%%%%%%%%%%%%%%%%%%%%%%%%%

The vacuum solution reads
\begin{subequations}\label{VacSol}
\begin{align}
    & B_0 =\label{B_0}
        0\,,\\
    & S_0 =\label{S_0}
        \theta^A Z_A\,,\\
    & W_0(Y; K | x) \equiv\Omega(Y |x) \equiv\label{W_0}
        -\frac i4 (
            \varpi_{\ga \gb} y^\ga y^\gb + 
            2h_{\ga \gad} y^\ga \yd^\gad + 
            \bar{\varpi}_{\gad \gbd} \yd^\gad \yd^\gbd
        )\,,
\end{align}
\end{subequations}
where $\varpi_{\ga \gb}$, $\bar{\varpi}_{\gad \gbd}$ and $h_{\ga \gad}$ are, respectively, the Lorentz connection and vierbein one-forms of the $AdS_4$ background.
 From the first HS equation \eqref{Vas1} it follows that $\Omega$ satisfies the zero-curvature condition
\begin{equation}\label{AdS_eqn}
    \dd \Omega + \Omega \star \Omega = 0\,.
\end{equation}
As a result, the $AdS$-covariant \emph{vacuum derivative} is flat,
\begin{equation}
D_\Omega := \dd + [\Omega, \bullet]_\star \,,\qquad D_\Omega^2 = 0\,.
\end{equation}

\subsection{Perturbative expansion}\label{Perturbative_expansion}%%%%%%%%%%%%%%%%%%%%%%%%%%%%%%%%%%%

Following \cite{Sezgin:2000hr}
we consider master fields as formal power series in $|\eta|$
\begin{align*}
    W = \sum_{n=0}^\infty W_n, \quad W_n \propto |\eta|^n\,;\qquad
    S = \sum_{n=0}^\infty S_n, \quad S_n \propto |\eta|^n\,;\qquad
    B = \sum_{n=1}^\infty B_n, \quad B_n \propto |\eta|^{n-1}\,.
\end{align*}
The subscript refers to the perturbation order (the degree shift in $B_n$ is since $\eta$ contributes
explicitly to \eqref{Vas4}).

From the star-product definition \eqref{star-product} it follows that 
\begin{equation}\label{d_z_def}
    [S_0, \bullet]_\star = -2i \theta^A \pdv{}{Z^A} \equiv -2i \dd_Z\,.
\end{equation}
Hence,  \eqref{Vas5} implies that $B_1$ is $Z$-independent. Analogously, at higher orders one  faces
equations of the form
\begin{equation}
\label{fJ}
    \dd_Z f(Z, Y; \theta; K) = J(Z, Y; \theta; K)\,,
\end{equation}
where $J(Z, Y; \theta; K)$  composed from nonlinear combinations of the lower-order fields
is $\dd_Z$-closed as a consequence of the consistency of the HS equations \eqref{Vas1}-\eqref{Vas5}.

The general solution to \eqref{fJ} can be obtained using the \emph{homotopy trick} (see e.g. \cite{Didenko:2018fgx}):
\begin{equation}\label{hmt_sol}
    f(Z, Y; \theta; K) = \hmt_Q J(Z, Y; \theta; K) + h(Y; K) + \dd_Z \epsilon(Z, Y; \theta; K)\,,
\end{equation}
where $\epsilon(Z, Y; \theta; K)$ is an arbitrary function regular in $Z$ and $Y$, $h(Y; K)$ belongs to $\dd_Z$
cohomology  and $\hmt_Q$ is a \emph{shifted homotopy operator} defined as
\begin{equation}\label{hmt_def}
    \hmt_Q J(Z, Y; \theta; K) :=
        (Z^A + Q^A)\pdv{}{\theta^A} \int_0^1 \frac{dt}{t} J(tZ - (1-t)Q, Y; t\theta; K)\,
\end{equation}
with any $Z$-independent shift parameter $Q^A$.

The general structure of the perturbative analysis of the HS equations is as follows. Let all the fields of the order smaller than $n$ be known. Using \eqref{d_z_def} and equation \eqref{Vas5} one gets
% \begin{equation*}
%     \dd_Z B_n = -\frac i2 \sum_{k=1}^{n-1} [S_k, B_{n-k}]_\star\,.
% \end{equation*}
% Its solution
\begin{equation}\label{B_n}
    B_n = -\frac i2 \hmt_{Q_B} \sum_{k=1}^{n-1} [S_k, B_{n-k}]_\star + C_n(Y;K)\,,
\end{equation}
where $C_n$ is a $\dd_Z$-cohomological part (the $\dd_Z$-exact part is absent since $B$ is a zero-form).

Equations \eqref{Vas4} and \eqref{Vas3} yield
% \begin{equation*}
%     \dd_Z S_n = -\frac i2 \sum_{k=1}^{n-1} S_k \star S_{n-k}
%         - \frac 12 \eta B_n \star \gamma - \frac 12 \bar{\eta} B_n \star \Bar{\gamma}\,,
% \end{equation*}
\begin{equation}\label{S_n}
    S_n = -\frac i2 \hmt_{Q_S} \lp
            \sum_{k=1}^{n-1} S_k \star S_{n-k}
            - i\eta B_n \star \gamma
            - i\bar{\eta} B_n \star \Bar{\gamma}
        \rp + \dd_Z \epsilon_n(Z, Y; K)\,,
\end{equation}
% \begin{equation*}
%     \dd_Z W_n = -\frac i2 D_\Omega S_n - \frac i2 \sum_{k=1}^{n-1}[W_k, S_{n-k}]\,,
% \end{equation*}
\begin{equation}\label{W_n}
    W_n = -\frac i2 \hmt_{Q_W} \lp D_\Omega S_n + \sum_{k=1}^{n-1}[W_k, S_{n-k}]_\star \rp + \omega_n(Y;K)\,.
\end{equation}
Here $\omega_n$ is in $\dd_Z$-cohomology like $C_n$. These two fields encode the physical content of the theory. 
Conventionally, HS fields are described by the full $\dd_Z$-cohomological part of $W$ and $B$,
$$\omega := \sum \omega_n\,,\qquad C := \sum C_n.$$ In this paper we deal with the second perturbative order  so that $n \in \{1, 2\}$.

Note that, formally, $\omega_n$ and $C_n$ with different $n$ can be treated as independent
fields obeying the
differential equations resulting from \eqref{Vas1} and \eqref{Vas2}:
\begin{align}
    D_\Omega \omega_n &=\label{omega_n_eq}
        - \sum_{k=1}^{n-1} W_k \star W_{n-k}
        + \frac i2 D_\Omega \hmt_{Q_W} \lp D_\Omega S_n + \sum_{k=1}^{n-1}[W_k, S_{n-k}]_\star \rp\,,\\
    D_\Omega C_n &=\label{C_n_eq}
        - \sum_{k=1}^{n-1} [W_k, B_{n-k}]_\star
        + \frac i2 D_\Omega \hmt_{Q_B} \sum_{k=1}^{n-1} [S_k, B_{n-k}]_\star\,.
\end{align}
As a consequence of  consistency of the system in $Z$-directions 
the RHSs of these equations are $Z$-independent, i.e., the dependence on $Z$ on the RHSs must cancel out.
Also, by construction, as is not hard to check straightforwardly, these equations are consistent as an exterior system 
in $x$ directions, i.e.,
the infinite system of equations \eqref{omega_n_eq}, \eqref{C_n_eq} on the infinite set of fields $\omega_n$ and $C_n$ forms
a consistent  unfolded system in  $x$-space. As such its form is somewhat reminiscent of integrable hierarchies.
Though we hope to explore this interesting similarity in more detail elsewhere let us briefly comment on this 
interesting option here. 

\subsection{Speculation on the relation to integrable hierarchies}

On the one hand, consistency of the system \eqref{omega_n_eq},
\eqref{C_n_eq} follows from the expansion of the generating system in  powers of the coupling constant $|\eta|$
and, as such, may look trivial since an expansion of a nonlinear system into power series in the coupling constant
does not produce an integrable hierarchy. However, the unfolded formulation of the system in question is very 
special being \emph{universal}
in terminology of \cite{Bekaert:2004qos}. In simple words this implies that it remains consistent with the differential
${\rm d}$ extended to a larger space with any number of new coordinates in addition to $x$. This makes it possible to
introduce an infinite number of higher times $t_k$ associated with higher hierarchy equations. To make the system integrable
it remains to restrict  the connection to an Abelian subalgebra of the HS algebra which may be not too difficult. 
For the examples of integrable hierarchies resulting from HS equations see e.g. \cite{Gutperle:2014aja, Beccaria:2015iwa}.

\subsection{Simplest vertices}

For the analysis of $\omega C$ vertices we will use conventional homotopy with vanishing shift parameters since it properly
reproduces \cite{Vasiliev:1992av}
the form of free HS unfolded equations  \cite{Vasiliev:1986td,Vasiliev:1988sa} known as \emph{First on-mass shell theorem}. 
Though conventional homotopy violates locality of some $C^2$ vertices \cite{Boulanger:2015ova}, as shown in \cite{Vasiliev:2016xui} in the second order of perturbation theory, after an appropriate change of variables they take manifestly local form, that can also be recovered directly by virtue of an appropriate shifted homotopy \cite{Didenko:2018fgx}\footnote{One can see from \cite{Didenko:2018fgx} that $\omega C$ and $C^2$ vertices result from independent  $\dd_Z$-closed structures. Hence the shift 
parameters for them can be chosen  independently.}. 

 In these terms equations \eqref{omega_n_eq} and \eqref{C_n_eq} take the form
\begin{align}
    D_\Omega \omega &=\label{omega_eq}
        \Upsilon(\Omega, \Omega, C)
        - \omega \star \omega + \Upsilon(\Omega,\omega,C)
        + \Upsilon(\Omega, \Omega, C, C)+\ldots\,,\\
    D_\Omega C &=\label{C_eq}
        -[\omega, C]_\star + \Upsilon(\Omega, C, C)+\ldots\,,
\end{align}
where ellipses denotes higher-order terms.

The term $\Upsilon(\Omega, \Omega, C)$ reproduces the  First on-mass shell theorem:
\begin{align}
    \Upsilon(\Omega, \Omega, C) =
        \frac i2 \eta \Hd(\bar{\partial},\bar{\partial}) C(0, \yd; K) k + c.c.\,,\label{WWC_res}
\end{align}
where $\bar{\partial}_\ga := \pdv{}{\yd^\ga}$ and the two-forms $\Hd(\bar{a},\bar{b})$ and $H(a,b)$ are defined as
\begin{align}
\begin{split}
    \Hd(\bar{a},\bar{b}) := \frac 12 \Hd_{\gad\gbd}\, \bar{a}^\gad \bar{b}^\gbd\,,\qquad
    \Hd_{\gad\gbd} := h_{\gg \gad} h^\gg{}_\gbd\,;\\
    H(a,b) := \frac 12 H_{\ga\gb}\, a^\ga b^\gb\,,\qquad
    H_{\ga\gb} := h_{\ga \ggd} h_\gb{}^\ggd\,.
\end{split}
\end{align}
In this paper, the symbol $c.c.$ swaps all dotted and undotted spinor indices, $k$ and $\bar{k}$, $\eta$ and $\bar{\eta}$, not affecting however the imaginary unit $i$ like that in \eqref{Vas4}. For brevity, we will omit arguments $Y$ and $K$ on the LHSs of the expressions like \eqref{WWC_res}. Hope this will not confuse the reader.

It is not difficult to obtain the expression for $\Upsilon(\Omega,\omega,C)$, found originally in \cite{Boulanger:2015ova,Gelfond:2017wrh} in a slightly different form 
(for detail see appendix \ref{VertDer}):
\begin{align}
\begin{split}
    \Upsilon(\Omega,\omega,C) =
        -\frac {i\eta}{2} h^{\ga \gad} \pdv{}{u^\gb}\pdv{}{v_\gb}
            &\left[ 
                u_\gg\pdv{\omega}{y_\gg} ,
                v_\ga\int\limits_0^1dt \pdv{}{\yd^\gad}
                    C(-t z, \yd; K) e^{it (zy)}k
            \right]_{\star\ z=0} -\\-
        \frac {i\eta}{2} h^{\ga \gad}
            &\left[ 
                \pdv[2]{\omega}{y^\ga}{\yd^\gad} ,
                \int\limits_0^1dt t C(-t z, \yd; K) e^{it (zy)}k
            \right]_{\star\ z=0} + c.c\,.
\end{split}\label{wWC_res}
\end{align}
Here $u^\ga$ and $v^\ga$ are auxiliary spinor variables that anticommute with $k$, introduced to account for  sign factors due to exterior Klein operators which are implicit. Also, we use shorthand notation for spinor contractions:
\begin{equation}
    (ab) := a_\ga b^\ga\,,\qquad (\bar{a}\bar{b}) := \bar{a}_\gad \bar{b}^\gad\,.
\end{equation}

The vertices $\Upsilon(\Omega, C, C)$ and $\Upsilon(\Omega, \Omega, C, C)$ were found in  \cite{Vasiliev:2016xui, Gelfond:2017wrh}. In \cite{Misuna:2017bjb} these vertices were projected to Fronsdal equations yielding explicit formulae for the respective HS currents. 
Expressions for $\Upsilon(\Omega, C, C)$ and $\Upsilon(\Omega, \Omega, C, C)$ (formulae \eqref{WWCC_exp} and \eqref{WCC_exp} below)
are taken from \cite{Gelfond:2017wrh}.

We introduce operators of differentiation with respect to auxiliary spinor variables $y^\ga$, $y^\ga_i$:
\begin{equation}\label{p_def}
    p_\ga := -i\pdv{}{y^\ga}\,,\qquad
    (p_i)_\ga := -i\pdv{}{y_i^\ga}\,,
\end{equation}
that are assumed to act on the entire expressions on the right.

Using that
\begin{align}\label{expform}
    &f(y_1 + a) = \exp{i(p_1a)}f(y_1)\,,\\
    &f(y; k) \star g(y; k) = \exp{ - i(p_1p_2) + i(p_1y) + i(p_2y)} f(y_1; k) g(y_2; k)\Big|_{y_i=0}\,,
\end{align}
all relevant vertices can be rewritten in the following form:
\begingroup\allowdisplaybreaks
\begin{align}
\begin{split}\label{ww_exp}
    \Upsilon(\omega,\omega) =&
        -\exp{ - i(p_1p_2) + i(p_1y) + i(p_2y) - i(\pd_1\pd_2) + i(\pd_1\yd) + i(\pd_2\yd)}
            \cdot\\\cdot& \omega(Y_1; K) \omega(Y_2; K)\bigg|_{Y_i=0}\,;
\end{split}\\
\begin{split}\label{wWC_exp}
    \Upsilon(\Omega,\omega,C) =&
        -\frac {i\eta}{2} \int\limits_0^1dt h(p_1,t\pd_1 + \pd_2)\cdot\\\cdot&
            \exp{ - it(p_1p_2) + i(1-t)(p_1y) - i(\pd_1\pd_2) + i(\pd_1\yd) + i(\pd_2\yd)}\cdot\\\cdot&
                \big[\omega(Y_1; K)C(Y_2;K) - e^{-2i(\pd_1\yd)}C(Y_2;K)\omega(-Y_1;K)\big]k\bigg|_{Y_i=0}
        + c.c.\,;
\end{split}\\
\begin{split}\label{WWCC_exp}
    \Upsilon(\Omega, \Omega, C, C) =&
        -\frac {i\eta\bar{\eta}}{4}
            \int\limits_{[0;1]^4}\frac{d^4t}{t_4^2} \delta(1-t_3-t_4)\delta'(1-t_1-t_2) \Hd(\pd,\pd)\cdot\\\cdot& 
                \exp{ it_1(p_1y) - it_2(p_2y) - it_3(\pd_1\pd_2) + it_2t_4(\pd_1\yd) - it_1t_4(\pd_2\yd)}
                    \cdot\\\cdot& C(Y_1; K)C(Y_2;K)k\Bar{k}\bigg|_{Y_i=0}
        + c.c.\,;
\end{split}\\
\begin{split}\label{wC_exp}
    \Upsilon(\omega, C) =&
        -\exp{ - i(p_1p_2) + i(p_1y) + i(p_2y) - i(\pd_1\pd_2) + i(\pd_1\yd) + i(\pd_2\yd)}
            \cdot\\\cdot& \big[\omega(Y_1; K)C(Y_2;K) - C(Y_1;K)\omega(Y_2; K)\big]\bigg|_{Y_i=0}\,;
\end{split}\\
\begin{split}\label{WCC_exp}
    \Upsilon(\Omega, C, C) =&
        \frac {i\eta}{2}
            \int\limits_0^1dt h(y,(1-t)\pd_1 - t\pd_2)\cdot\\\cdot& 
                \exp{ it(p_1y) - i(1-t)(p_2y) - i(\pd_1\pd_2) + i(\pd_1\yd) + i(\pd_2\yd)}
                    \cdot\\\cdot& C(Y_1; K)C(Y_2;K)k\bigg|_{Y_i=0}
        + c.c.
\end{split}
\end{align}
\endgroup
In the sequel we use notations
\begin{equation}
    h(a,\Bar{a}) := h_{\ga\gad} a^\ga \Bar{a}^\gad\,,\qquad
    \varpi(a,b) := \frac12 \varpi_{\ga\gb} a^\ga b^\gb\,.
\end{equation}

\subsection{Field pattern}\label{Field_pattern}%%%%%%%%%%%%%%%%%%%%%%%%%%%%%%%%%%%%%%%%%%%%%

In this paper we will use the following notation: for any formal power series in $Y$
\begin{equation}\label{f_m,n}
    f(Y) \equiv \sum\limits_{m,n} f_{m,n}(Y)
\end{equation}
with $f_{m,n}(Y)$ being homogeneous polynomials in $y^\alpha$ and $\yd^{\dot \alpha}$ of degrees $m$ and $n$, respectively.

Let us recall the pattern of $\omega (Y;K|x)$ and $C(Y;K|x)$ for a fixed spin. The free part of the dynamical equations \eqref{omega_eq}, \eqref{C_eq} reads:
\begin{align}
    D_\Omega \omega &=\label{omega_eq_free}
        \Upsilon(\Omega, \Omega, C)\,,\\
    D_\Omega C &=\label{C_eq_free}0\,.
\end{align}
Due to  \eqref{DynCond} the vacuum derivative acts differently on $\omega$ and $C$:
\begin{align}
    D_\Omega \omega(Y;K) &= \big( D_L + ih(y,\pd) + ih(p,\yd) \big)\omega(Y;K)\,,\label{D^adj}\\
    D_\Omega C(Y;K) &= \big( D_L - ih(y,\yd) - ih(p,\pd) \big) C(Y;K)\,,\label{D^tw}
\end{align}
where $$D_L := \dd + 2i\varpi(y,p) + 2i\bar{\varpi}(\yd,\pd)$$ is the \emph{Lorentz covariant derivative}. Hence, to determine the pattern of $\omega$ and $C$ one has to analyse  equations \eqref{omega_eq_free}, \eqref{C_eq_free} taking into account the relation between the frame-like and Fronsdal HS fields. The latter are represented by the two Lorentz-irreducible components denoted as $\phi_{s,s}(Y)$ and $\phi'_{s-2,s-2}(Y)$. As shown in \cite{Vasiliev:1980as, Vasiliev:1986td}, there is a gauge in which the frame-like field $\omega_{s-1,s-1}(Y;K)$ is expressed via the Fronsdal fields as follows (for simplicity we focus on the bosonic case and neglect the Lorentz-like purely gauge components; for more detail see section \ref{Gauge_fixing}):
\begin{equation}
    \omega_{s-1,s-1}(Y;K) = -h(p,\pd)\phi_{s,s}(Y;K) + h(y,\yd)\phi'_{s-2,s-2}(Y;K)\,.\label{omega-phi}
\end{equation}

The unfolded equations \eqref{omega_eq} and \eqref{C_eq} allow one to express step-by-step other components of $\omega$ via derivatives of the frame-like fields. $C$ is expressed via derivatives of $\omega$ by virtue of the First on-mass shell theorem \eqref{WWC_res}. This procedure yields the following structure for the spin-$s$ fields \cite{Vasiliev:1986td} (note that for fermions the summation index $m$ is half-integer)
\begin{align}
\begin{split}
    \omega(Y) &= \sum_{m=-s+1}^{s-1} \omega_{s-1+m, s-1-m}(Y) =\\&=
        \sum_{m=-s+1}^{s-1}\frac{1}{(s-1-m)!(s-1+m)!}
            \omega_{\ga(s-1+m)\ \gad(s-1-m)}y^{\ga(s-1+m)}\yd^{\gad(s-1-m)}\,;
\end{split}\label{omega}\\
\begin{split}
    C(Y) &= \sum_{n=0}^\infty\lp C_{n, 2s+n}(Y) + C_{2s+n, n}(Y) \rp =\\&=
        \sum_{n=0}^\infty \frac{1}{n!(2s+n)!}\lp
            C_{\ga(n)\ \gad(2s+n)}y^{\ga(n)}\yd^{\gad(2s+n)} +
            C_{\ga(2s+n)\ \gad(n)}y^{\ga(2s+n)}\yd^{\gad(n)} \rp\,.
\end{split}\label{C}
\end{align}
In these formulae the dependence of the fields on the exterior Klein operators, which is in accordance with \eqref{DynCond}, is implicit.  Also, we use the convention that $\alpha(n)$ stands for $n$ symmetrized indices ($\alpha(n) := \ga_1\ga_2\dots\ga_n$ and $y^{\ga(n)} := y^{\ga_1}\dots y^{\ga_n}$).

It should be noticed that, for the free system, the numbers $m$ and $n$ in \eqref{omega} and \eqref{C} determine the order of space-time derivatives acting on the Fronsdal field. Namely, $\omega_{s-1+m, s-1-m}(Y|x)$ contains up to $[|m|]$ derivatives of the Fronsdal field ($[\dots]$ denotes integer part of the number) while $C_{\ga(n)\ \gad(2s+n)}$ and $C_{\ga(2s+n)\ \gad(n)}$ contain  $([s]+n)$ derivatives. (For  explicit formulae see \eqref{comp-to-der}.)

%%%%%%%%%%%%%%%%%%%%%%%%%%%%%%%%
\section{Analysis of currents}\label{Analysis_of_currents}
%%%%%%%%%%%%%%%%%%%%%%%%%%%%%%%%

\subsection{Vertex content for different spins}\label{Vertex_content}%%%%%%%%%%%%%%%%%%%%%%%%%%%%%%

Let us start with a terminological remark. We call currents $J$ on the RHS of the Fronsdal equations, 
\begin{equation*}
    \Box \phi + \ldots = J\,,
\end{equation*}
\emph{Fronsdal currents} or just \emph{currents}. As a result of unfolding,
Fronsdal equations lead to the system of the type \eqref{omega_eq}, \eqref{C_eq}. The terms on the RHSs of these equations with the different dependence on the dynamical and background fields are called \emph{vertices} $\Upsilon$. The terms on the RHS of unfolded equations resulting from different Fronsdal currents are called \emph{unfolded currents}. In principle, unfolded currents do not coincide with vertices. For instance, an  unfolded current can be constituted by a sum of different vertices. Note that, within the $\sigma_-$  cohomology formalism \cite{Shaynkman:2000ts}, Fronsdal currents can be identified with the primary components of the unfolded current modules \cite{Gelfond:2003vh}.

Let us discuss the restrictions on spins, that contribute via different types of $\Upsilon$. The spin of $\Upsilon$ will be denoted by $s_1$ while those of fields inside $\Upsilon$ by $s_2$ and $s_3$. We consider in detail the simplest case of $\Upsilon(\omega,\omega)$ with all three spins assumed to be integer. Formulae for other types of vertices are obtained analogously.

Using \eqref{omega}, the contribution to \eqref{omega_eq} of $\Upsilon(\omega,\omega)$ \eqref{ww_exp} with fixed ordering of $s_2,\,s_3$ can be rewritten as
\begin{multline*}
    \sum_{m_1} (D_\Omega \omega(Y))_{s_1-1+m_1,s_1-1-m_1} =\\=
    \sum_{m_2,m_3,l,r} \frac{(-i)^{l+r} (p_1p_2)^l (\pd_1\pd_2)^r} {l!r!} 
        \omega_{s_2-1+m_2,s_2-1-m_2}(Y_1)\omega_{s_3-1+m_3,s_3-1-m_3}(Y_2)\bigg|_{Y_i=Y}
    + \dots\,,
\end{multline*}
where ellipsis denotes the other vertices. Equating the degrees in $y$ and $\yd$ on the both sides of these equations one finds that
\begin{subequations}\label{Y_numb}\begin{align}
    \deg y &=s_1 - 1 + m_1 = s_2 - 1 + m_2 + s_3 - 1 + m_3 - 2l\,,\label{y_numb}\\
    \deg \yd &= s_1 - 1 - m_1 = s_2 - 1 - m_2 + s_3 - 1 - m_3 - 2r\,.\label{yd_numb}
\end{align}\end{subequations}
Here $m_i$ are analogues of $m$ in \eqref{omega}  for the fields of spins $s_i$, so that 
$- s_i + 1 \leqslant m_i \leqslant s_i - 1$.   The numbers of contracted indices $l$ and $r$ obey
\begin{align}
\begin{split}\label{lr_ineq}
    &0\leqslant l\leqslant s_2 - 1 + m_2\,,
        \hspace{3cm} 0\leqslant l\leqslant s_3 - 1 + m_3\,,\\
    &0\leqslant r\leqslant s_2 - 1 - m_2\,,
        \hspace{3cm} 0\leqslant r\leqslant s_3 - 1 - m_3\,.
\end{split}
\end{align}

By virtue of \eqref{lr_ineq}, summation of  \eqref{y_numb} and \eqref{yd_numb}  leads to the \emph{triangle inequalities }
\begin{align}\label{tri_ineq}
    s_1 + s_2 \geqslant s_3 + 1\,,\qquad
    s_3 + s_1 \geqslant s_2 + 1\,,\qquad
    s_2 + s_3 \geqslant s_1 + 1\,.
\end{align}

Now it is easy to evaluate the number of derivatives $\#_{der}$ in the components of $\Upsilon(\omega,\omega)$ contributing to the deformation of the Fronsdal equations. As discussed in section \ref{Field_pattern}, $\#_{der} = |m_2|+|m_3|$. If $m_2,\ m_3>0$, then, according to \eqref{y_numb}, $\#_{der} = m_2+m_3 = s_1-s_2-s_3+2l+1+m_1$. From \eqref{Y_numb} we find that $l+r = s_2+s_3-s_1-1$. Consequently, $\#_{der} \leqslant s_2+s_3-s_1-1+m_1$. Since Fronsdal equations are contained  in the unfolded ones with $|m_1|\leqslant 1$, corresponding Fronsdal currents carry $\#_{der} \leqslant s_2+s_3-s_1$ derivatives. The same inequality can be proven for the
opposite signs of $m_2$ and $m_3$. Note that in the flat limit all lower-derivative terms in the currents disappear so that only the
term with the maximal number of derivatives 
$\#_{der} = s_2+s_3-s_1$ at $s_1 = \min\{s_1,s_2,s_3\}$ survives, that precisely corresponds to the $4d$ vertex with the minimal number of derivatives in the Metsaev classification \cite{Metsaev:2005ar}.

Other types of vertices can be analysed analogously. The final results valid both for bosons and for fermions are collected in the table ($\cS := [s_1]+[s_2]+[s_3]$):
\begin{table}[H]
\begin{tabular}{|l|l|l|}
\hline
\multirow{3}{7em}{$\begin{array}{l}
    s_1 + s_2 \geqslant s_3 + 1\\
    s_1 + s_3 \geqslant s_2 + 1\\
    s_2 + s_3 \geqslant s_1 + 1
\end{array}$}&
$\Upsilon(\omega,\omega)$ &
    $\#_{der} \leqslant \cS - 2s_1$ and
        $\#_{der} \leqslant \cS - 2\max\{s_2,s_3\}$\\ \cline{2-3}
&$\Upsilon(\Omega,\omega,C)$ & $\#_{der} \leqslant \cS - 2\min\{s_1,s_2,s_3\}$ \\ \cline{2-3}
&$\Upsilon(\Omega,\Omega,C,C)$ & $\#_{der} = \cS$ \\ \hline
%%%
\multirow{2}{5em}{$s_2 = s_1+s_3$}&
$\Upsilon(\Omega,\Omega,C,C)$ & $\#_{der} = \cS$ \\ \cline{2-3}
&$\Upsilon(\Omega,\omega,C)$ & $\#_{der} \leqslant \cS - 2\min\{s_1,s_3\}$ \\ \hline
%%%
\multirow{2}{5em}{$s_3 = s_1+s_2$}&
$\Upsilon(\Omega,\Omega,C,C)$ & $\#_{der} = \cS$ \\ \cline{2-3}
&$\Upsilon(\Omega,\omega,C)$ & $\#_{der} \leqslant \cS - 2\min\{s_1,s_2\}$ \\ \hline
%%%
$s_2 > s_1+s_3$ & $\Upsilon(\Omega,\omega,C)$ &
    $\#_{der} \leqslant \cS - 2\min\{s_1,s_3\}$
         and $\#_{der} \leqslant \cS$\\ \hline
%%%
$s_3 > s_1+s_2$ & $\Upsilon(\Omega,\omega,C)$ &
    $\#_{der} \leqslant \cS - 2\min\{s_1,s_2\}$
         and $\#_{der} \leqslant \cS$\\ \hline
%%%
$s_1 \geqslant s_2+s_3$ & $\Upsilon(\Omega,\Omega,C,C)$ &
    $\#_{der} = \cS$ and $\#_{der} = \cS - 2\min\{s_2,s_3\}$ \\ \hline
\end{tabular}\caption{Parameters of the vertices in the one-form sector}\label{spins_1}
\end{table}

It is worth to emphasize that the doubled restrictions on $\#_{der}$ in some cells of the Table \ref{spins_1} (e.g. $\#_{der} = \cS$ and $\#_{der} = \cS - 2\min\{s_2,s_3\}$ in the last line), correspond to vertices with different helicity signs of the dynamical fields. When the helicities of both fields have the same sign, $\#_{der}$ is represented by the first expression ($\#_{der} = \cS$ for $\Upsilon(\Omega,\Omega,C,C)$), while when the signs are opposite $\#_{der}$ is represented by the second one ($\#_{der} = \cS - 2\min\{s_2,s_3\}$ for $\Upsilon(\Omega,\Omega,C,C)$).

One should not be confused that for $\Upsilon(\omega,\omega)$ the inequality restricting $\#_{der}$ is not symmetric in all three spins. This is because this vertex forms a Fronsdal current along with $\Upsilon(\Omega,\omega,C)$ as explained in more detail in section \ref{Consistency_conditions}.

For lower-spins $s_1 \leqslant 1$, Fronsdal equations belong to the zero-form sector of the unfolded equations.
The structure of the respective vertices  $\Upsilon(\omega,C)$ and $\Upsilon(\Omega,C,C)$ is as follows:
\begin{table}[H]
\begin{tabular}{|l|l|l|}
\hline
\multirow{2}{5em}{$\Upsilon(\omega,C)$}&
$\#_{der} = \cS - 2\min\{s_1,s_2,s_3\}$
    & $s_2 + s_3 \geqslant s_1 + 1,\ s_1 = 1/2,\,1$\\ \cline{2-3}
&$\#_{der} = \cS$
    & $|s_2 - s_3| \geqslant s_1 + 1,\ s_1 = 0,\,1/2,\,1$\\ \hline
%%%
\multirow{3}{5em}{$\Upsilon(\Omega,C,C)$}&
$\#_{der} = \cS - 2\min\{s_1,s_2,s_3\}$
    & $s_2 + s_3 \leqslant s_1,\ s_1 = 1/2,\,1$\\ \cline{2-3}
&$\#_{der} = \cS$
    & $|s_2 - s_3| \leqslant s_1,\ s_1 = 1/2,\,1$\\ \cline{2-3}
&$\#_{der} = [s_2]+[s_3]$
    & $|s_2 - s_3| \leqslant 1,\ s_2 + s_3 \geqslant 1,\ s_1 = 0$\\ \hline
\end{tabular}\caption{Parameters of the vertices in the zero-form sector}\label{spins_0}
\end{table}
Note that in the case of $s_1 = s_2 = s_3 = 0$ there is no interaction vertex in the Table \ref{spins_0} in agreement
with the known fact that scalar vertex $\varphi^3$ is absent in the HS theory \cite{Sezgin:2003pt}.

\subsection{Consistency conditions and independent currents}\label{Consistency_conditions}%%%%%

As shown in \cite{Gelfond:2003vh} for $3d$ $C^2$ currents, a Fronsdal current  $J$  
is conserved iff the unfolded current 
$\Upsilon$ obeys its rank-two unfolded equations, that can be treated as 
the consistency condition (\emph{Bianchi identity})
\begin{equation}\label{Bianchi_id}
    D_\Omega \Upsilon = 0\,.
\end{equation}
This phenomenon has general applicability.

Generally, equation \eqref{Bianchi_id}  relates vertices with different numbers of the fields $C$ and $\omega$. Firstly let us show that equation \eqref{Bianchi_id} on $\Upsilon(\Omega,\Omega,C,C)$ is independent of the other vertices in the sense that at the second order in
$$D_\Omega\big (\Upsilon(\Omega,\Omega,C) + \Upsilon(\omega,\omega)+ \Upsilon(\Omega,\omega,C) + \Upsilon(\Omega,\Omega,C,C)\big)\,,$$ $\Upsilon(\Omega,\Omega,C,C)$ talks only to $\Upsilon(\Omega,\Omega,C)$ via the $C^2$ part of the equation on $C$.\footnote{Since   Bianchi identities are analysed at the second order, calculating $D_\Omega \Upsilon(\omega,\omega)$, $D_\Omega \Upsilon(\Omega,\omega,C)$ and $D_\Omega \Upsilon(\Omega,\Omega,C,C)$ one has to use the free equations \eqref{omega_eq_free}, \eqref{C_eq_free} while in $D_\Omega \Upsilon(\Omega,\Omega,C)$ one has to account for the second-order equation on $C$ \eqref{C_eq}.}
Indeed, $D_\Omega \Upsilon(\Omega,\Omega,C,C)$ is quadratic in $C$ by virtue of equation \eqref{C_eq_free} on $C$ at the first order. Equations of motion for $\omega$ \eqref{omega_eq_free} yield $D_\Omega \Upsilon(\omega,\omega)$ being either quadratic in $\omega$ or bilinear in $C$ and $\omega$. Thus, the question is whether $D_\Omega \Upsilon(\Omega,\omega,C)$ produces the $C^2$-terms.

To see that this does not happen we observe using \eqref{wWC_exp} that $\Upsilon(\Omega,\omega,C)$ has the following structure:
\begin{align}
\begin{split}
    \Upsilon(\Omega,\omega,C) =
        \big\{
            &\mathbf{K}_1(Y,P_1,P_2) h(p_1, \pd_1) \omega(Y_1)C(Y_2) +\\+
            &\mathbf{K}_2(Y,P_1,P_2) h(p_1, \pd_2) \omega(Y_1)C(Y_2)
        \big\}_{Y_{1,2}=0} + c.c.
\end{split}\label{wWC_structure}
\end{align}
The $C^2$-terms can appear in $D_\Omega \Upsilon(\Omega,\omega,C)$ only via the First on-mass shell theorem \eqref{WWC_res}. However, this mechanism does not contribute to the $\mathbf{K}_1$-terms since the latter contain both holomorphic and antiholomorphic derivatives of the arguments of $\omega(Y_1)$, hence not containing $\omega(y_1,0)$ or $\omega(0,\yd_1)$. Neither it contributes to
the $\mathbf{K}_2$-term since the $C^2$ part of its background derivative yields:
\begin{align}
\begin{split}
    D_\Omega (\mathbf{K}_2(Y,P_1,P_2) &h(p_1, \pd_2) \omega(Y_1)C(Y_2))_{Y_{1,2}=0}\big|_{C^2} 
        \propto\\\propto \mathbf{K}_2(Y,P_1,P_2) &h(p_1, \pd_2) H(p_1,p_1) C_{2s_2,0}(Y_1)C(Y_2) \,,
\end{split}
\end{align}
which is zero since the three-form $h_{\ga\gad} H_{\ga\ga}$ vanishes due to the Schouten identity. Here it is  also  accounted that $H(\pd_1,\pd_1)C(0,\yd_1)$ does not contribute analogously to the $\mathbf{K}_1$ case.

Thus,  $\Upsilon(\omega,\omega)$ and $\Upsilon(\Omega,\omega,C)$ do not contribute to the $C^2$ part of \eqref{Bianchi_id} while $\Upsilon(\Omega,\Omega,C,C)$ does not contribute to the other parts.

Now let us consider the Bianchi identity for $\Upsilon(\omega,\omega)$ and $\Upsilon(\Omega,\omega,C)$. As shown in more detail in Appendix \ref{ConsCond}, straightforward calculation yields (considering for brevity only the $\omega C$-ordered terms):
\begin{align*}
    (D_\Omega \Upsilon(&\omega,\omega))_{\omega C} =\\= -&\frac{i\eta}{4}
        (\pd_3\pd_2)^2
        e^{ - i(p_1y) - i (\pd_1\pd_2) + i (\pd_1\yd) + i (\pd_2\yd)}
    k H(Y_3)\omega(Y_1)C(Y_2)\bigg|_{Y_i = 0}
    + c.c.\,;\\
%%%
    (D_\Omega \Upsilon(&\Omega,\omega,C))_{\omega C} =\\= -&\frac{i\eta}{4}
        \big(
            (\pd_3\pd_1) + (\pd_3\pd_2)
        \big)^2
        e^{ -i (p_1p_2) - i (\pd_1\pd_2) + i (\pd_1\yd) + i (\pd_2\yd)} k
            H(Y_3)\omega(Y_1)C(Y_2)\bigg|_{Y_i = 0}
                +\\+
    &\frac{i\eta}{4}
        (\pd_3\pd_2)^2
        e^{ - i(p_1y) - i (\pd_1\pd_2) + i (\pd_1\yd) + i (\pd_2\yd)} k
            H(Y_3)\omega(Y_1)C(Y_2)\bigg|_{Y_i = 0}
    + c.c.\,;\\
%%%
    (D_\Omega \Upsilon(&\Omega,\Omega,C))_{\omega C} =\\= &\frac{i\eta}{4}
        \big(
            (\pd_3\pd_1) + (\pd_3\pd_2)
        \big)^2
        e^{ -i (p_1p_2) - i (\pd_1\pd_2) + i (\pd_1\yd) + i (\pd_2\yd)}
    k H(Y_3)\omega(Y_1)C(Y_2)\bigg|_{Y_i = 0}
    + c.c.
\end{align*}
As a result, at spins obeying triangle inequality \eqref{tri_ineq} when
 both $\Upsilon(\omega,\omega)$ and $\Upsilon(\Omega,\omega,C)$ are non-zero, 
 \eqref{Bianchi_id} involves both of them.

That \eqref{Bianchi_id} decomposes  into two independent parts signals that there are two types of conserved currents in the Fronsdal theory. Upon unfolding, in the sector respecting triangle inequalities \eqref{tri_ineq}, one of them is represented by $\Upsilon(\Omega,\Omega,C,C)$ with $\#_{der}=[s_1]+[s_2]+[s_3]$ (see Table \ref{spins_1}) while the other one by  $\Upsilon(\omega,\omega)+\Upsilon(\Omega,\omega,C)$ with $\#_{der} \leqslant [s_1]+[s_2]+[s_3] - 2\min\{s_1,s_2,s_3\}$. This agrees with the results of the light-cone analysis, that, modulo improvements, there are two types of $P$-even cubic vertices in $4d$ 
bosonic HS theory in Minkowski space: with $s_1+s_2+s_3$ or $s_1+s_2+s_3 - 2\min\{s_1,s_2,s_3\}$ derivatives\footnote{HS vertices are known since \cite{Bengtsson:1983pd, Bengtsson:1986kh}. Their  properties were analysed by Metsaev in \cite{Metsaev:2005ar}. In \cite{Metsaev:2018xip} the light-cone cubic vertices in $AdS_4$ were shown to match those on Minkowski background.}. In the flat limit from $AdS$ to Minkowski space the subleading derivative terms proportional to the powers of the cosmological constant vanish, so that only the highest derivative terms in the currents survive. Hence, $\Upsilon(\Omega,\Omega,C,C)$ corresponds to the light-cone vertex with $s_1+s_2+s_3$ derivatives while $\Upsilon(\omega,\omega)+\Upsilon(\Omega,\omega,C)$  to the other one. The  situation beyond the triangle inequalities
is analogous. For instance, if $s_1 \geqslant s_2 + s_3$, two branches of $\Upsilon(\Omega,\Omega,C,C)$ correspond to different light-cone vertices. (For more detailed discussion of such  currents  see \cite{Misuna:2017bjb}.) Notice that since the HS coupling constant $\eta$   is complex, both $P$-even and  $P$-odd currents are present. It turns out that the highest derivative currents with $\#_{der}=[s_1]+[s_2]+[s_3]$ contain both $P$-even and $P$-odd parts as  demonstrated  in section  \ref{Calculation_results} for the general case  and  \ref{Yang-Mills} for the Yang-Mills example.

\subsection{Conserved charges from higher-spin action}\label{Conserved_charges}%%%%%%%%%%%%%%%%%

The results of the previous section agree with those of \cite{Smirnov:2015waz}.\footnote{We warn the reader that the objects called `currents' in this paper and in \cite{Smirnov:2015waz} are  related in a somewhat tricky way. Hence, one should not expect the present paper expressions to coincide precisely with those of \cite{Smirnov:2015waz}.} In this section, we clarify some points of \cite{Smirnov:2015waz} reconsidering the analysis of currents with the emphasis on the general aspects of the formalism and simplifications of the final formulae. Also we interpret the resulting currents in the context of classification of \cite{Metsaev:2005ar,Metsaev:2018xip} and of this paper. Since the content of this section is not directly related to the rest of the paper,
the uninterested reader can skip this part going directly to section \ref{Gauge_fixing}.

Consider a gauge theory on a space-time manifold $\cM^d$ described by an action $S$  formulated in terms of the wedge products of differential forms $W^I(x)$ valued in some associative algebra\footnote{Note that here is an essential difference between the setup of this paper and that of \cite{Smirnov:2015waz} where the fields were  carrying color indices unrelated to any Lie or associative algebraic structure, that is possible at the lowest interaction order but not beyond. In this paper, we assume that all fields are valued in an arbitrary associative algebra as was originally suggested in \cite{Vasiliev:1988sa}. Such setup works at all orders being mathematically related to $A_\infty$ strong homotopy algebra \cite{Stasheff1963-ia, Stasheff1963-dp}.} $A$.  Analogously to  section \ref{Perturbative_analysis}, we split $W^I$ into background and dynamical part: $W^I = \Omega^I + \omega^I$. (In the sequel the dependence on $x$ will be mostly implicit.)

Let $S$ be invariant under the gauge transformations
\begin{equation}\label{gauge_transform}
    \delta_\epsilon \omega^I = \cD^I{}_J \epsilon^J + (\text{higher order terms})\,,
\end{equation}
where $\cD^I{}_J$ is the vacuum covariant derivative built from $\Omega^I$ obeying the flatness condition $\cD^2=0$. So, its quadratic part $S_2$ is invariant under the free field Abelian  gauge transformations $\delta_\epsilon \omega^I = \cD^I{}_J \epsilon^J$.

Let two expressions $F$ and $G$ be called on-shell equivalent if they coincide on-shell, i.e., at 
$\frac{\delta S_2}{\delta \omega^I}=0$,
\begin{equation}
    F\sim G :\qquad F-G \propto \frac{\delta S_2}{\delta \omega^I}\,.
\end{equation}
This notation will only be used in this section, while in the sequel all statements are on-shell.

From the Noether identity  
\begin{equation*}
    (\cD + (\text{higher order terms}))^I{}_J\fdv{S}{\omega^I} = 0\,,
\end{equation*}
it follows that  
\begin{equation}
    \cD^I{}_J \cJ_I \sim 0\,,\qquad
    \cJ_I:=\frac{\delta S_3}{\delta \omega^I}\,.
\end{equation}
For a $p$-form $\omega^I$, let $\xi^I = \xi^I(x)$ be a $(p-1)$-form obeying the covariant constancy condition,
\begin{equation}\label{xi_eq}
    \cD^I{}_J \xi^J =0\,,
\end{equation}
which admits nontrivial solutions since $\cD^I{}_J$ is flat. Then a $(d-1)$-form 
\begin{equation}
    \cJ_\xi := \xi^I \cJ_I
\end{equation}
is weakly closed,
\begin{equation}
    \dd \cJ_\xi \sim 0\,,
\end{equation}
and
\begin{equation}
    Q := \int_{\Sigma^{d-1}} \cJ_\xi
\end{equation} is an on-shell  conserved charge independent of the variations of $\Sigma^{d-1}$.

An auxiliary $(p-1)$-form $\xi^I$ corresponds to the global symmetry parameters 
associated with the Noether conserved charge $Q$. Indeed, these are the symmetry parameters that do not contribute to the gauge transformations \eqref{gauge_transform} at $\epsilon^I = \xi^I$. In the topologically trivial case of $\cM^d = \mathbb{R}^d$ only the sector of zero-forms $\xi^I$ matters with \eqref{xi_eq} admitting as many linearly independent solutions as the dimension of the gauge group in question. The free parameters in $\xi^I$ identify with the parameters of the vacuum global symmetry (for instance, translations and Lorentz rotations in the Poincar\'e case). In fact, $\xi^I$ are the unfolded dynamics analogues of the Killing vectors (tensors in the HS case). 

To apply this construction to the  HS theory we consider  a nonlinear HS action on $AdS_4$  proposed in \cite{Fradkin:1986qy}. Though this action does not reproduce the full system \eqref{Vasiliev}, it is consistent at the cubic order. Following  \cite{Vasiliev:1986qx} (see also \cite{Vasiliev:1999ba}) we define a supertrace in the HS algebra,
\begin{equation}\label{tr_def}
    \tr f(Y;K) := \tr_A f(0;0)\,,
\end{equation}
where $\tr_A$ is a trace over $A$. In these terms, the action of \cite{Fradkin:1986qy} reads as
\begin{align}
    S = \frac12 \int \tr{R(Y;K) \star \Tilde{R}(Y;K)}\,,
        \label{S_def}
 \end{align}
where the wedge symbols are implicit and, using notation $R_{m,n}$ \eqref{f_m,n},
\begin{align}
    R := D_\Omega \omega + \omega \star \omega\,,\qquad
    \Tilde{R} := \sum_{m,n} \sgn(m-n) R_{m,n}\,.\label{R_def}
\end{align}

As shown in \cite{Vasiliev:1980as,Vasiliev:1986td}, at the quadratic level this action reproduces the free HS equations of the Fronsdal theory \cite{Fronsdal:1978rb}. The gauge transformation law for $\omega$ reads as $\delta_\epsilon \omega = D_\Omega \epsilon + (\text{higher order terms})$. This yields the conserved charge 
\begin{equation}\label{Q}
    Q = \int\limits_{\Sigma^3} \tr{
        \xi(Y;K) \star [\omega(Y;K), \Tilde{R}^{(1)}(Y;K)]_{\star} }
    \equiv \int\limits_{\Sigma^3} \cJ_\xi\,,
\end{equation}
\begin{equation}\label{J}
     \cJ_\xi\ = \tr{
        \xi(Y;K) \star [\omega(Y;K), \Tilde{R}^{(1)}(Y;K)]_{\star} }
    \,,
\end{equation}
where $R^{(1)}(Y;K)$ is the linearized  part of $R(Y;K)$. Indeed,
\begin{equation*}
    \delta S =
        \int \tr{D_\Omega (\delta\omega) \star \Tilde{R} + [\delta\omega,\omega]_\star \star \Tilde{R}}\,,
\end{equation*}
where the first term does not contribute to $Q$ by virtue of \eqref{xi_eq} while the second one leads to \eqref{Q} by the cyclic property of trace. The expression for the charge $Q$ \eqref{Q} is far simpler than that of \cite{Smirnov:2015waz} being different by an exact form and on-shell trivial terms.

It is important that, by virtue of the First on-shell theorem \eqref{WWC_res},
\begin{equation}\label{RC}
    R^{(1)}(y,\yd;K) \sim \Upsilon(\Omega,\Omega,C)\,,\qquad
    R^{(1)}(y,\yd;K) \sim R^{(1)}(y,0;K) + R^{(1)}(0,\yd;K)\,.
\end{equation}
Note that 
\begin{equation*}
    \Tilde{R}^{(1)}(y,\yd;K) \sim R^{(1)}(y,0;K) - R^{(1)}(0,\yd;K)\,.
\end{equation*}

To prove nontriviality of currents, in \cite{Smirnov:2015waz}  $\cJ_\xi$  was argued to be expressible in the $\omega$-dependent form free from the zero-forms $C$ resulting from \eqref{RC} by adding an appropriate improvement (exact form). This was shown for the case of $s_1 < s_2 + s_3 - 1$ (while the currents are in fact nontrivial if $s_1 \leqslant s_2 + s_3 - 1$). We show that the result of \cite{Smirnov:2015waz} is true for general spins, if $\xi$ belongs to the center of $A$ (for instance, is proportional to unit matrix for $A=Mat_N$).

Indeed, from \eqref{omega_eq_free} and \eqref{xi_eq} it follows that
\begin{align*}
   \cJ_\xi
        &+ \dd \tr{\xi(Y) \star [\omega(Y), \omega(y,0) - \omega(0,\yd)]_\star }
        + \dd \tr{\xi(Y) \star [\omega(0,\yd), \omega(y,0)]}
        -\\&- \dd \tr{\xi(Y) \star
            \big( \omega(y,0) \star \omega(y,0) - \omega(0,\yd) \star \omega(0,\yd) \big)}
    \sim\\&\hspace{4cm}\sim
    2 \tr{\xi(Y) \star [R^{(1)}(0,\yd), \omega(y,0)]} + (\omega^2\text{ and exact terms})\,.
\end{align*}
Thus the problem amounts to the construction of a two-form $\Psi(Y)$ from dynamical and background fields, such that
\begin{equation}\label{Psi-eq}
    \tr{\xi(Y) \star D_\Omega\Psi(Y)} \sim \tr{\xi(Y) \star [R^{(1)}(0,\yd), \omega(y,0)]}
        + (\omega^2\text{ and exact terms})\,,
\end{equation}
This equation can be solved iff the first term on its RHS is zero. Since functions of $y$ and $\yd$ commute  with respect to the star product, $[R^{(1)}(0,\yd), \omega(y,0)]$ only contains the commutator in $A$. If $\xi(Y)$ is proportional to $Id \in A$ this term indeed vanishes due to the cyclic property of trace over $A$. Thus $\cJ_\xi$ can be rewritten as a function of $\omega$ once $\xi$ is in the center of $A$.

Let us compare bilinear corrections to the Fronsdal equations resulting from the action \eqref{S_def} against those from the nonlinear system \eqref{Vasiliev}. Since their form is essentially different: the former are represented by three-forms while the latter by the two-forms, we simply compare the number of space-time derivatives in the respective currents. The variation of the action \eqref{S_def} at the second order is
\begin{align}\label{dS}
\begin{split}
    \delta S_3 \sim 
        \sum_{m,n}\int \tr\big\{\delta\omega_{m,n}(Y) \star
        \big[
            -&[\sgn(m-n)-\sgn(m-n+2)] ih(p,\yd) R^{(2)}_{m+1,n-1}(Y) -\\
            -&[\sgn(m-n)-\sgn(m-n-2)] ih(y,\pd) R^{(2)}_{m-1,n+1}(Y) +\\
            +&[\sgn(m-n)-1] [R^{(1)}(y,0),\omega(Y)]_{\star\ m,n}+\\
            +&[\sgn(m-n)+1] [R^{(1)}(0,\yd),\omega(Y)]_{\star\ m,n}
        \big]\big\}\,.
\end{split}
\end{align}
Using the counting analogous  to that of section \ref{Vertex_content} we find that in this case currents are on-shell non-zero iff the triangle inequalities \eqref{tri_ineq} hold true and $\#_{der} \leqslant [s_1] + [s_2] + [s_3] - 2\min\{s_1, s_2, s_3\}$. This result matches that of Table \ref{spins_1} and agrees with the light-cone results of \cite{Metsaev:2005ar,Metsaev:2018xip} discussed at the end of the previous section. Thus, the action of \cite{Fradkin:1986qy} is shown to reproduce the vertex with the minimal number of derivatives in  Metsaev's classification \cite{Metsaev:2005ar,Metsaev:2018xip}.

\subsection{Gauge fixing}\label{Gauge_fixing}%%%%%%%%%%%%%%%%%%%%%%%%%%%%%%%%%%%%%%%%%%%%%%%%

Unfolded equations \eqref{omega_eq}, \eqref{C_eq} imply the gauge symmetry \cite{Vasiliev:1988sa}
\begin{align}
    \delta_\epsilon \omega &=\label{omega_gauge}
        D_\Omega \epsilon + \lp \epsilon \cdot \pdv{}{\omega}\rp \Upsilon_\omega\,,\\
    \delta_\epsilon C &=\label{C_gauge}
        \lp \epsilon \cdot \pdv{}{\omega}\rp \Upsilon_C\,,
\end{align}
where $\cdot$ denotes contraction of all spinor indices; $\Upsilon_\omega$ and $\Upsilon_C$ are the RHSs of the unfolded equations \eqref{omega_eq} and \eqref{C_eq}, respectively. At the lowest order it takes the form
\begin{align}
    \delta_\epsilon \omega &=\label{omega_gauge_1}
        D_\Omega \epsilon\,,\\
    \delta_\epsilon C &=\label{C_gauge_1}
        0\,.
\end{align}

For the further analysis it is useful to gauge fix the first-order fields $\omega$. The aim of this section is to prove the following 
\begin{prop}\label{Gauge_Prop}
Consider unfolded system $D_\Omega \omega = H(p,p) F(Y) + \Hd(\pd,\pd) \Bar{F}(Y)$ with $F$ and $\Bar{F}$ being any gauge invariant functions of $Y$, $K$, $x$, that obey the Bianchi identity \eqref{Bianchi_id}. This system admits such a gauge that
\begin{equation}
    \omega_{m,n}(Y;K) = -h(p,\pd)\phi_{m+1,n+1}(Y;K)\,\label{omega-phi_gauged}\qquad \forall m,\, n = 0,\,1,\dots
\end{equation}
\end{prop}

Let us  stress that the gauge \eqref{omega-phi_gauged} is available both for 
 bosons and for fermions.
 Though the proof is insensitive to the choice of $F$ and $\Bar{F}$, 
  to make it easier to apply in the sequel we will assume that
 \begin{equation}\label{fbf}
 H(p,p) F(Y) + \Hd(\pd,\pd) \Bar{F}(Y) = \Upsilon(\Omega,\Omega,C)\,.
 \end{equation}
with $\Upsilon(\Omega,\Omega,C)$ \eqref{WWC_res}.

The proof becomes straightforward using the decomposition of the lowest order gauge transformations \eqref{omega_gauge_1} 
 and unfolded equations \eqref{omega_n_eq} into one- and two- frame forms. We start with some auxiliary formulae.

The following simple corollary of the Schouten identity 
\begin{equation*}
    y_\ga(pp_1) + (p_1)_\ga(yp) + (p_1y)p_\ga = 0\,,
\end{equation*}
expressing the fact that antisymmetrization over 
any three two-component indices gives zero, will be useful:
\begin{equation}\label{Sch_1}
    F_{\ga|\gg(n)}y^{\gg(n)} =
        \frac{1}{n+1} \lp y_\ga (p_1p) - p_\ga (p_1y) \rp F_{\gb|\gg(n)}y_1^\gb y^{\gg(n)}\,
\end{equation}
(recall that  $p$ and $p_1$ are defined in \eqref{p_def}). Identity \eqref{Sch_1} has the following consequence:
\begin{align}
\begin{split}
    F_{\ga(2)|\gg(n)}y^{\gg(n)} = \frac12\lb
            \frac{1}{n(n+1)} y_{\ga(2)}(p_1p)^2 -
            \frac{2}{n(n+2)} y_\ga p_\ga (p_1y)(p_1p) +\rd\\\ld+
            \frac{1}{(n+1)(n+2)} p_{\ga(2)} (p_1y)^2
        \rb F_{\gb(2)|\gg(n)}y_1^{\gb(2)}y^{\gg(n)}\,.
\end{split}\label{Sch_2}
\end{align}

By virtue of \eqref{Sch_1}, a  one-form $\omega$ has the following four components within the decomposition into the frame one-forms:
\begin{align}
\begin{split}
    \omega_{m,n}(Y) =
        \frac{1}{(m+1)(n+1)} \big[
            &h(y,\yd) (p_1p)(\pd_1\pd) -
            h(y,\pd) (p_1p)(\pd_1\yd) -\\-
            &h(p,\yd) (p_1y)(\pd_1\pd) +
            h(p,\pd) (p_1y)(\pd_1\yd)
        \big] \omega_{m,n}(Y_1|Y)
        \equiv\\\equiv
        -\frac{1}{(m+1)(n+1)} \big[
            &h(y,\yd) \omega_{m,n}(p,\pd|Y) -
            h(y,\pd) \omega_{m,n}(p,\yd|Y) -\\-
            &h(p,\yd) \omega_{m,n}(y,\pd|Y) +
            h(p,\pd) \omega_{m,n}(y,\yd|Y)
        \big] \,.
\end{split}\label{omega_decomp}
\end{align}
Here $\omega_{m,n}(Y_1|Y)$ is bilinear in $y_1$ and $\bar y_1$ with the labels $m$ and $n$ referring to
the degrees in $y$ and $\bar y$, respectively.
Analogously, 
\begin{align}
\begin{split}
    (D_\Omega\epsilon)_{m,n}(Y) =
        -\frac{1}{(m+1)(n+1)} \big[
            &h(y,\yd) D(p,\pd) \epsilon_{m,n}(Y) -
            h(y,\pd) D(p,\yd) \epsilon_{m,n}(Y) -\\-
            &h(p,\yd) D(y,\pd) \epsilon_{m,n}(Y) +
            h(p,\pd) D(y,\yd) \epsilon_{m,n}(Y)
        \big] +\\+
        &ih(y,\pd) \epsilon_{m-1,n+1}(Y) +
        ih(p,\yd) \epsilon_{m+1,n-1}(Y)\,,
\end{split}\label{D_epsilon_decomp}
\end{align}
where the components $ D(y_1,\yd_1)$ of $D_L$ are defined as follows:
\begin{equation}\label{D_expansion}
    D_L f(Y) \equiv h^{\ga\gad}D_{\ga\gad} f(Y) \equiv -h(p_1,\pd_1) D(y_1,\yd_1) f(Y)\,.
\end{equation}

As a result, by virtue of \eqref{Sch_2}, unfolded equations $D_\Omega \omega = \Upsilon(\Omega,\Omega,C)$ can be rewritten in the form
\begin{align}
\begin{split}
    -i(m+1)(n+1)(D_\Omega\omega)_{m,n}(Y) =\\=
        \frac{1}{m} H(y,y) \lb 
            D(p,\yd) \omega_{m,n}(p,\pd|Y) - D(p,\pd) \omega_{m,n}(p,\yd|Y)
                + i(m+1)(n+1) \omega_{m-1,n+1}(p,\pd|Y)
        \rb -\\-
        \frac{1}{m} H(y,p) \lb
            D(y,\yd) \omega_{m,n}(p,\pd|Y) - D(y,\pd) \omega_{m,n}(p,\yd|Y)
                + i(m+1)(n+1) \omega_{m-1,n+1}(y,\pd|Y)
        \rb -\\-
        \frac{1}{m+2} H(y,p) \lb 
            D(p,\yd) \omega_{m,n}(y,\pd|Y) - D(p,\pd) \omega_{m,n}(y,\yd|Y)
                - i(m+1)(n+1) \omega_{m+1,n-1}(p,\yd|Y)
        \rb +\\+
        \frac{1}{m+2} H(p,p) \lb
            D(y,\yd) \omega_{m,n}(y,\pd|Y) - D(y,\pd) \omega_{m,n}(y,\yd|Y)
                - i(m+1)(n+1) \omega_{m+1,n-1}(y,\yd|Y)
        \rb
    +\\+
        \frac{1}{n} \Hd(\yd,\yd) \lb
            D(y,\pd) \omega_{m,n}(p,\pd|Y) - D(p,\pd) \omega_{m,n}(y,\pd|Y)
                + i(m+1)(n+1) \omega_{m+1,n-1}(p,\pd|Y)
        \rb -\\-
        \frac{1}{n} \Hd(\yd,\pd) \lb
            D(y,\yd) \omega_{m,n}(p,\pd|Y) - D(p,\yd) \omega_{m,n}(y,\pd|Y)
                + i(m+1)(n+1) \omega_{m+1,n-1}(p,\yd|Y)
        \rb -\\-
        \frac{1}{n+2} \Hd(\yd,\pd) \lb
            D(y,\pd) \omega_{m,n}(p,\yd|Y) - D(p,\pd) \omega_{m,n}(y,\yd|Y)
                - i(m+1)(n+1) \omega_{m-1,n+1}(y,\pd|Y)
        \rb +\\+
        \frac{1}{n+2} \Hd(\pd,\pd) \lb
            D(y,\yd) \omega_{m,n}(p,\yd|Y) - D(p,\yd) \omega_{m,n}(y,\yd|Y)
                - i(m+1)(n+1) \omega_{m-1,n+1}(y,\yd|Y)
        \rb =\\=
    -\frac{(2s-1)}{2} \lb
            \delta_{n,0}\bar{\eta} H(p,p)C_{2s,0}(y)\bar{k} + \delta_{m,0}\eta \Hd(\pd,\pd)C_{0,2s}(\yd)k
        \rb\,.
\end{split}\label{D_omega_decomp}
\end{align}
Note that, although $m$ and $n$ are in denominators of some terms, this formula is well-defined even at $m=0$ and $n=0$ because potentially problematic terms vanish as they contain more derivatives in $y$ or $\yd$ than the power degree in these variables.

Below it is shown  that to reach the gauge \eqref{omega-phi_gauged} one has to perform the gauge transformation \eqref{omega_gauge} with some parameter $\epsilon$, that  verifies  the equations
\begin{equation}\label{eps_eqn_1}
    D(p,\pd)D(y,\yd) \epsilon_{2s-2,0}(y,0) = -D(p,\pd) \omega_{2s-2,0}(y,\yd|y,0)\,,
\end{equation}
\begin{equation}\label{eps_eqn_2}
    \epsilon_{m,n}(Y) =-\frac{i}{(m+2)n}\lb D(p,\yd)\epsilon_{m+1,n-1}(Y) + \omega_{m+1,n-1}(p,\yd|Y) \rb\,,\qquad n>0\,,
\end{equation}
which, locally,  have a solution since by virtue of the Schouten identity equation \eqref{eps_eqn_1} has the Klein-Gordon form 
\begin{equation*}
    [\Box - 2(s-1)]\epsilon_{2s-2,0}(y,0)
        = -\frac{1}{2s} D(p,\pd) \omega_{2s-2,0}(y,\yd|y,0)\,
\end{equation*}
with $-2\Box = (p_1p_2)(\pd_1\pd_2)D(Y_1)D(Y_2)$, 
while \eqref{eps_eqn_2} determines $\epsilon_{m,n}(Y)$ via $\epsilon_{m+1,n-1}(Y)$ algebraically. Note that the conditions \eqref{eps_eqn_1}, \eqref{eps_eqn_2}
do not determine $\epsilon$ unambiguously. The residual gauge symmetry parameters obey
\begin{equation}\label{eps^res_eqn_1}
    D(p,\pd)D(y,\yd) \epsilon^{res}_{2s-2,0}(y,0) = 0\quad \Longleftrightarrow\quad
        [\Box - 2(s-1)]\epsilon^{res}_{2s-2,0}(y,0) = 0\,,
\end{equation}
\begin{equation}\label{eps^res_eqn_2}
    \epsilon^{res}_{m,n}(Y) =-\frac{i}{(m+2)n} D(p,\yd)\epsilon^{res}_{m+1,n-1}(Y)\,,\qquad n>0\,.
\end{equation}

Since equations \eqref{eps_eqn_1} and \eqref{eps_eqn_2} do not respect the reality conditions (i.e, the symmetry with respect to $y \leftrightarrow \yd$), in the rest of this section we consider complexified fields. In the end, they can be made real because in this gauge the resulting expressions for $\phi_{s+m,s-m}(Y)$ and $\phi_{s-m,s+m}(Y)$ (see \eqref{comp-to-der}) are $y \leftrightarrow \yd$ symmetric. Indeed, let $\omega(Y)$ and $\overline{\omega(Y)}$ be complex conjugated fields in the gauge \eqref{omega-phi_gauged} obtained from the same \emph{real} field $\omega_*(Y)$ but in different ways: $\omega(Y) = \omega_*(Y) + D_\Omega \epsilon(Y)$, while $\overline{\omega(Y)} = \omega_*(Y) + D_\Omega \overline{\epsilon(Y)}$, where the gauge parameters $\epsilon(Y)$ and $\overline{\epsilon(Y)}$ obey equations \eqref{eps_eqn_1}, \eqref{eps_eqn_2} and their complex conjugated counterparts, respectively. This allows us to introduce a real field $\omega_*^{gauged}(Y) = 
\frac{1}{2}(\omega(Y) + \overline{\omega(Y)})$ which is gauge equivalent to  $\omega_*(Y)$. 
The corresponding gauge parameter $\frac{1}{2}(\epsilon(Y) + \overline{\epsilon(Y)})$ can be made real as well.

Now we turn to the proof of Proposition \ref{Gauge_Prop}. Comparison of the $\omega$ decomposition \eqref{omega_decomp} with the gauge transformation decomposition \eqref{D_epsilon_decomp} implies that the gauge transformation with the parameter satisfying \eqref{eps_eqn_1} sets
\begin{equation}\label{just_another_formula_1}
    D(p,\pd) \omega_{2s-2,0}(y,\yd|y,0) = 0
\end{equation}
for the gauge fixed field. The gauge transformations obeying \eqref{eps_eqn_2} make it possible to set
\begin{equation}\label{just_another_formula_2}
    \omega_{m,n}(p,\yd|Y)=0\qquad \text{for all allowed $m$, $n$.}
\end{equation}
As a result, the proof of Proposition \ref{Gauge_Prop} amounts to showing that \eqref{just_another_formula_1}, \eqref{just_another_formula_2} along with the unfolded equations \eqref{D_omega_decomp} (i.e., on-shell) lead to $\omega$ of the form \eqref{omega-phi_gauged}. This will be shown by induction.

As the induction base case we prove that the gauge condition \eqref{omega-phi_gauged} can be reached for $n=0,1$. Let us consider unfolded equations \eqref{D_omega_decomp} at $n=0$, assuming \eqref{just_another_formula_1} and \eqref{just_another_formula_2}:
\begin{align}
\begin{split}
    (D_\Omega\omega)_{2s-2,0}(Y) =&\\=
        -\frac{1}{2s-2} &H(y,y) \omega_{2s-3,1}(p,\pd|Y) +\\+
        \frac{1}{2s-2} &H(y,p) \omega_{2s-3,1}(y,\pd|Y) -\\-
        \frac{i}{2s(2s-1)} &H(p,p) D(y,\pd) \omega_{2s-2,0}(y,\yd|Y)
    -\\-
        \frac{1}{2} &\Hd(\pd,\pd) \lb
            \frac{i}{(2s-1)} D(p,\yd) \omega_{2s-2,0}(y,\yd|Y)
                - \omega_{2s-3,1}(y,\yd|Y)
        \rb
    =\\=
        -\frac{i}{2} \bar{\eta} &H(p,p)C_{2s,0}(y)\bar{k}\,.
\end{split}\label{D_omega_decomp:n=0}
\end{align}
($\Hd(\yd,\yd)$- and $\Hd(\yd,\pd)$-components of \eqref{D_omega_decomp} vanish at $n=0$ because in this case they contain $\pd$ acting on a $\yd$-independent expression.) $H(y,y)$- and $H(y,p)$-components of \eqref{D_omega_decomp:n=0} without gauge fixing would express $\omega_{2s-3,1}(p,\pd|Y)$ and $\omega_{2s-3,1}(y,\pd|Y)$ via $D(p,\pd) \omega_{2s-2,0}(y,\yd|Y)$, $D(p,\pd) \omega_{2s-2,0}(p,\yd|Y)$ and $D(y,\pd) \omega_{2s-2,0}(p,\yd|Y)$ (see eq. \eqref{D_omega_decomp}). Now, by \eqref{just_another_formula_1}, \eqref{just_another_formula_2} from these components it follows that $\omega_{2s-3,1}(p,\pd|Y)$ and $\omega_{2s-3,1}(y,\pd|Y)$ are zero. Since also $\omega_{2s-3,1}(p,\yd|Y) = 0$, only $\omega_{2s-3,1}(y,\yd|Y)$ survives in $\omega_{2s-3,1}(Y)$ \eqref{omega_decomp}, hence the gauge \eqref{omega-phi_gauged} is proven to hold for $n=0,1$.

The induction step: if \eqref{just_another_formula_2} is true and for some  $m>0$ and $n>0$ $\omega_{m,n}$ and $\omega_{m+1,n-1}$ obey \eqref{omega-phi_gauged}, then $\omega_{m-1,n+1}$ obeys the same condition as well. Indeed, under these conditions, unfolded equations \eqref{D_omega_decomp} take the form
\begin{align}
\begin{split}\label{a_very_useful_formula}
    (D_\Omega\omega)_{m,n}(Y) =\\=
        -\frac{1}{m} H(y,y)& 
            \omega_{m-1,n+1}(p,\pd|Y) +\\+
        \frac{1}{m} H(y,p)& \lb
            \frac{im}{(m+2)(m+1)(n+1)} D(p,\pd) \omega_{m,n}(y,\yd|Y) + \omega_{m-1,n+1}(y,\pd|Y)
        \rb -\\-
        \frac{1}{m+2} H(p,p)& \lb
            \frac{i}{(m+1)(n+1)} D(y,\pd) \omega_{m,n}(y,\yd|Y)
                - \omega_{m+1,n-1}(y,\yd|Y)
        \rb
    +\\+
        \frac{1}{n+2} \Hd(\yd,\pd)& \lb
            \frac{i}{(m+1)(n+1)} D(p,\pd) \omega_{m,n}(y,\yd|Y)
                - \omega_{m-1,n+1}(y,\pd|Y)
        \rb -\\-
        \frac{1}{n+2} \Hd(\pd,\pd)& \lb
            \frac{i}{(m+1)(n+1)} D(p,\yd) \omega_{m,n}(y,\yd|Y)
                - \omega_{m-1,n+1}(y,\yd|Y)
        \rb = 0\,.
\end{split}
\end{align}
Its  $H(y,y)$ component implies that $\omega_{m-1,n+1}(p,\pd|Y) = 0$. $H(y,p)$- and $\Hd(\yd,\pd)$- components form a linear system on $\omega_{m-1,n+1}(y,\pd|Y)$ and $D(p,\pd) \omega_{m,n}(y,\yd|Y)$ with the unique solution $\omega_{m-1,n+1}(y,\pd|Y) = 0$ and $D(p,\pd) \omega_{m,n}(y,\yd|Y) = 0$. Hence, the gauge condition \eqref{omega-phi_gauged} holds true for $\omega_{m-1,n+1}(Y)$ as well. Thus, it is proven that equations \eqref{eps_eqn_1} and \eqref{eps_eqn_2} along with the unfolded HS equations  imply the gauge \eqref{omega-phi_gauged}.

Note that, apart from cancelling three out of four components of the one-form $\omega$, this gauge has another nice feature. As a consequence of \eqref{just_another_formula_1}, \eqref{a_very_useful_formula}, it fixes $\phi$ in \eqref{omega-phi_gauged} to be transversal:
\begin{equation}\label{transversality}
    D(p,\pd) \phi(Y) = 0\,.
\end{equation}
Hence, it represents the frame-like counterpart of the metric-like TT gauge (see, e.g., \cite{Joung:2011ww}). Note that $\omega$ of the form \eqref{omega-phi_gauged}, \eqref{transversality} was considered in \cite{Misuna:2017bjb}. However, unlike the present paper, where we reach such a form of $\omega$ by the gauge transformation, in \cite{Misuna:2017bjb} it was formulated as a reduced representation `keeping track of' the traceless components of $\omega$ \eqref{omega-phi_gauged}.

It should be noticed that the suggested gauge fixing procedure does not directly apply in the flat limit because in \eqref{eps_eqn_2} we implicitly divide over the cosmological constant $\Lambda$ (set to $-1$ in this paper). However, once the Fronsdal field can be restricted by the TT-constraints, in the flat limit  the gauge \eqref{omega-phi_gauged} can be reached 
too by moving from $\omega_{s-1,s-1}(Y)$ (or $\omega_{s-1/2,s-3/2}(Y)$ and $\omega_{s-3/2,s-1/2}(Y)$ for fermions) to $\omega_{2s-2,0}(y,0)$ and $\omega_{0,2s-2}(0,\yd)$. More precisely, the gauge parameter for the Fronsdal fields corresponds to $\epsilon_{s-1,s-1}(Y)$ (or $\epsilon_{s-1/2,s-3/2}(Y)$ and $\epsilon_{s-3/2,s-1/2}(Y)$). After fixing it so that the Fronsdal field is transverse-traceless, it remains to eliminate $h(y,\pd)$- and $h(p,\yd)$- component of $\omega(Y)$. This induces algebraic constraints on the rest components of $\epsilon(Y)$ analogous to \eqref{eps_eqn_2}. As a result, the frame-like TT gauge 
\eqref{omega-phi_gauged} is applicable in the flat limit as well.

\subsection{Nontriviality of the gauge non-invariant currents}\label{Nontriviality}%%%%%%%%%%%%%%

In this section we show that $\omega^2$ and $\omega C$ currents cannot be removed by a local field redefinition $\omega \to \omega + \Delta\omega$ with a second-order $\Delta\omega$. This will be proven by contradiction. Namely, supposing that there exists a proper change of variables, we will reduce the problem to the zero-form sector then showing that a general local field redefinition does not solve the problem.

The free unfolded HS equations in the sector of one-forms $\omega$ have the form \eqref{omega_eq_free}, \eqref{WWC_res}. This fact, called First on-shell theorem, known since \cite{Vasiliev:1986td,Vasiliev:1988sa}, plays important role in the theory. Its complete proof was presented recently in \cite{Bychkov:2021zvd} in terms of $\sigma_-$-cohomology. For the reader's convenience an alternative  proof is presented in Appendix \ref{FoMSTheorem}.

From the proof of the First on-shell theorem  it follows that the unfolded equation on $\omega$ with trivial current in $\Upsilon(\omega,\omega) + \Upsilon(\Omega,\omega,C)$ can be reduced  by a redefinition of $\omega$ to the free equations with possibly deformed RHS of the form $\Upsilon(\Omega,\Omega,C + \Delta C)$, where $\Delta C$ is a local function bilinear in the dynamical fields and their derivatives (in principle, components of both $\omega$ and $C$ can contribute to $\Delta C$). For spin $s_1 \geqslant 1$, equation on $C$ results from the Bianchi identity, and hence  must have the form $D_\Omega(C + \Delta C) = 0$. Since the original equation on $C$ reads as $D_\Omega C = \Upsilon(\omega,C) + O(C^2)$, the change of variables $C \to C + \Delta C$ must eliminate its RHS.

Now we are in a position to show that the equation $D_\Omega \Delta C = -\Upsilon(\omega,C) + O(C^2)$ admits no solution with any local $\Delta C$. Though this can be done directly by checking a general local Ansatz bilinear in the dynamical fields,  we follow a simpler way avoiding involved calculations.

From the gauge transformation law \eqref{C_gauge}, the expression $\Upsilon(\omega,C) = -[\omega, C]_\star$, and the fact
that the free field strength equal in our case  to $C + \Delta C$ is invariant under the linearized gauge transformation 
$\delta_\epsilon$, it follows that the following has to be true
\begin{equation}\label{DeltaC_gauge}
    \delta_\epsilon \Delta C = [\epsilon, C]_\star\,.
\end{equation}

Now we  show that  $\Delta C$ obeying equation \eqref{DeltaC_gauge} does not exist. This is a kind of obvious since the RHS of \eqref{DeltaC_gauge} is free of space-time derivatives of $\epsilon$ while the linearized gauge transformation law of $\omega$ contains $D_\Omega \epsilon$ while $C$ is gauge invariant.
More in detail, \eqref{DeltaC_gauge} along with the  gauge transformation law \eqref{omega_gauge_1}, \eqref{C_gauge_1} imply that the gauge non-invariant part of $\Delta C$ is bilinear in $\omega$ and $C$. In the formula \eqref{D_epsilon_decomp}, that represents the decomposition of the $\omega$ gauge transformation \eqref{omega_gauge_1} into the frame one-forms, only $h(y,\pd)$- and $h(p,\yd)$-components include $\epsilon$ free from Lorentz derivatives. Hence, $\epsilon$ in \eqref{DeltaC_gauge} can only be generated via these components of $\delta_\epsilon \omega$, namely
\begin{subequations}\label{eps_making}
\begin{align}
    \delta_\epsilon \omega_{m+1,n-1}(p,\yd|Y) =\label{eps_making_1}
        D(p,\yd)\epsilon_{m+1,n-1}(Y)-in(m+2)\epsilon_{m,n}(Y)\,,\\
    \delta_\epsilon \omega_{m-1,n+1}(y,\pd|Y) =\label{eps_making_2}
        D(y,\pd)\epsilon_{m-1,n+1}(Y)-im(n+2)\epsilon_{m,n}(Y)\,.
\end{align}
\end{subequations}
Though, in principle, one can also consider an Ansatz with Lorentz derivatives of $\omega$, but commuting  derivatives and using Schouten identity, one always gets \eqref{eps_making}. Using  \eqref{eps_making_1} or \eqref{eps_making_2} to generate $\epsilon_{m,n}(Y)$ we have to account an extra term $D(p,\yd)\epsilon_{m+1,n-1}(Y)$ or $D(y,\pd)\epsilon_{m-1,n+1}(Y)$. Then, \eqref{eps_making} has to be used again  to generate this term, and so on indefinitely. As a result,  the Ansatz must contain an infinite number of derivatives of $\omega(Y_1|Y)$ to  obey \eqref{DeltaC_gauge}. Hence, no local $\Delta C$ exists that eliminates $\Upsilon(\omega,C)$.

Thus, it is shown that $\Upsilon(\Omega,\omega,C)$ and $\Upsilon(\omega,\omega)$ cannot be trivial. Note that analogous question for the $C^2$ currents has been worked out in \cite{Vasiliev:2017cae} where an explicit solution for the unfolded equations with the RHS quadratic in $C$ was found in the absence of $\Upsilon(\omega,C)$, i.e., at $s_1\geqslant s_2+s_3$. That the $C^2$ currents were shown in \cite{Vasiliev:2017cae} be removable by a nonlocal field redefinition implied their nontriviality as well.

%%%%%%%%%%%%%%%%%%%%%%%%%%%%%%%%
\section{Projection to Fronsdal currents}\label{Projection}
%%%%%%%%%%%%%%%%%%%%%%%%%%%%%%%%

\subsection{Strategy}\label{Strategy}%%%%%%%%%%%%%%%%%%%%%%%%%%%%%%%%%%%%%%%%%%%%%%%%%%%%%%

The goal of this section is to derive  Fronsdal equations with the current deformation resulting
from the HS unfolded system, that fixes  relative coefficients in front of the different currents.

\subsubsection{Bosonic case}\label{Bosonic_case}%%%%%%%%%%%%%%%%%%%%%%%%%%%%%%%%%%%%%%%%%

The decomposition of $\Upsilon_{m,n}$ into the frame two-forms resulting from \eqref{Sch_2} yields
\begin{align}
\begin{split}\label{Y_decomp}
    \Upsilon_{m,n}(Y) = \frac12\bigg[
            \frac{1}{m(m+1)} H(y,y)(p_1p)^2 -
            \frac{2}{m(m+2)} H(y,p) (p_1y)(p_1p) +\\+
            \frac{1}{(m+1)(m+2)} H(p,p) (p_1y)^2 +\\+
            \frac{1}{n(n+1)} \Hd(\yd,\yd)(\pd_1\pd)^2 -
            \frac{2}{n(n+2)} \Hd(\yd,\pd) (\pd_1\yd)(\pd_1\pd) +\\+
            \frac{1}{(n+1)(n+2)} \Hd(\pd,\pd) (\pd_1\yd)^2
        \bigg]\cdot\\\cdot \lp \Upsilon_{m,n}(y_1,y_1|Y) + \Upsilon_{m,n}(\yd_1,\yd_1|Y) \rp\,.
\end{split}
\end{align}

For integer $s_1$, the first step towards derivation of the Fronsdal currents consists of the elimination of torsion $\Upsilon_{s_1-1,s_1-1}$. Making the field redefinition 
$$\omega_{s_1-1\pm1,s_1-1\mp1} \to \omega_{s_1-1\pm1,s_1-1\mp1} + \Delta\omega_{s_1-1\pm1,s_1-1\mp1}$$
to reach zero torsion we impose
\begin{align}
\begin{split}\label{Delta_Omega}
    \Upsilon_{s_1-1,s_1-1}(Y) =
        i &h(p,\yd)\Delta\omega_{s_1,s_1-2}(Y) + i h(y,\pd)\Delta\omega_{s_1-2,s_1}(Y)
    =\\=
        \bigg[ \frac {1}{s_1-1} &\Hd(\yd,\yd) (p_1p)(\pd_1\pd) - \frac{1}{s_1+1} H(p,p) (p_1y)(\pd_1\yd)  +\\
            + \frac{1}{s_1+1} &H(p,y) (p_1p)(\pd_1\yd) - \frac{1}{s_1-1} \Hd(\yd,\pd) (p_1p)(\pd_1\yd)\bigg]
            \Delta\omega_{s_1,s_1-2}(Y_1|Y) +\\+
        \bigg[ \frac {1}{s_1-1} &H(y,y) (p_1p)(\pd_1\pd) - \frac{1}{s_1+1} \Hd(\pd,\pd) (p_1y)(\pd_1\yd) -\\
            - \frac{1}{s_1-1} &H(y,p) (p_1y)(\pd_1\pd) + \frac{1}{s_1+1} \Hd(\pd,\yd) (p_1y)(\pd_1\pd) \bigg]
            \Delta\omega_{s_1-2,s_1}(Y_1|Y)\,.
\end{split}
\end{align}
Comparison of this expression with  $\Upsilon_{s_1-1,s_1-1}$ \eqref{Y_decomp} yields 
\begin{align*}
    \Delta\omega_{s_1,s_1-2}(Y) = \frac{1}{2s_1(s_1^2-1)} \bigg[
            &h(y,\yd)(\pd_1\pd)^2 +\\+
            \frac12&h(y,\pd)\lb (s_1+1)(p_1y)(p_1p) + (s_1-1)(\pd_1\yd)(\pd_1\pd) \rb -\\-
            &h(p,\pd)(p_1y)^2
        \bigg]\lp \Upsilon_{s_1-1,s_1-1}(y_1,y_1|Y) + \Upsilon_{s_1-1,s_1-1}(\yd_1,\yd_1|Y) \rp\,,\\
    \Delta\omega_{s_1-2,s_1}(Y) = \frac{1}{2s_1(s_1^2-1)} \bigg[
            &h(y,\yd)(p_1p)^2 +\\+
            \frac12&h(y,\pd)\lb (s_1+1)(\pd_1\yd)(\pd_1\pd) + (s_1-1)(p_1y)(p_1p) \rb -\\-
            &h(p,\pd)(\pd_1\yd)^2
        \bigg]\lp \Upsilon_{s_1-1,s_1-1}(y_1,y_1|Y) + \Upsilon_{s_1-1,s_1-1}(\yd_1,\yd_1|Y) \rp\,.
\end{align*}

Let
\begin{equation}\label{U}
    U := \Upsilon - D_\Omega \Delta\omega
\end{equation}
be the redefined torsion-free vertex.
To find its form we start with the Lorentz derivative of $\Delta\omega$: 
\begin{align*}
    D_L\Delta\omega_{s_1,s_1-2}(Y) =
        \frac{-i}{2s_1^2(s_1^2-1)} \bigg[
            &H(y,y)(p_2p)(\yd\pd_2)(\pd_1\pd)^2 -\\-
            \frac12&H(y,y)(p_2p)(\pd\pd_2)\lb (s_1+1)(p_1y)(p_1p) + (s_1-1)(\pd_1\yd)(\pd_1\pd) \rb +\\+
            \frac12&\Hd(\pd,\pd)(\pd_2\yd)(yp_2)\lb (s_1+1)(p_1y)(p_1p) + (s_1-1)(\pd_1\yd)(\pd_1\pd) \rb +\\+
            &\Hd(\pd,\pd)(\pd_2\yd)(pp_2)(p_1y)^2 + ...
        \bigg] \cdot\\\qquad\cdot
    &D_L(Y_2)\lp \Upsilon_{s_1-1,s_1-1}(y_1,y_1|Y) + \Upsilon_{s_1-1,s_1-1}(\yd_1,\yd_1|Y) \rp\,.
\end{align*}
Here ellipsis denotes the $H(y,p)$- and $\Hd(\yd,\pd)$-terms  irrelevant for our purpose. Indeed, the LHSs of the Fronsdal equations belong to the $H(y,y)$- and $\Hd(\pd,\pd)$-components of the $m=s_1\,,\ n=s_1-2$ sector of the unfolded system and its complex conjugate (for  detail see Appendix \ref{FoMSTheorem}). As a result, the current deformation of the Fronsdal equations
is determined by the following four equations:
\begin{align}
\begin{split}
    iD(p,\yd) \omega_{s_1,s_1-2}(p,\pd|Y) - iD(p,\pd) \omega_{s_1,s_1-2}(p,\yd|Y)
            - (s_1^2-1) \omega_{s_1-1,s_1-1}(p,\pd|Y) =\\=
                -(s_1-1) U_{s_1,s_1-2}(p,p|Y)\,,\\
    iD(y,\yd) \omega_{s_1,s_1-2}(p,\yd|Y) - iD(p,\yd) \omega_{s_1,s_1-2}(y,\yd|Y)
            + (s_1^2-1) \omega_{s_1-1,s_1-1}(y,\yd|Y) =\\=
                -(s_1+1) U_{s_1,s_1-2}(\yd,\yd|Y)\,,\\
    iD(y,\pd) \omega_{s_1-2,s_1}(p,\pd|Y) - iD(p,\pd) \omega_{s_1-2,s_1}(y,\pd|Y)
            - (s_1^2-1) \omega_{s_1-1,s_1-1}(\pd,\pd|Y) =\\=
                -(s_1-1) U_{s_1-2,s_1}(\pd,\pd|Y)\,,\\
    iD(y,\yd) \omega_{s_1-2,s_1}(y,\pd|Y) - iD(y,\pd) \omega_{s_1-2,s_1}(y,\yd|Y)
            + (s_1^2-1) \omega_{s_1-1,s_1-1}(y,\yd|Y) =\\=
                -(s_1+1) U_{s_1-2,s_1}(y,y|Y)\,.
\end{split}\label{Fr_eq_unf}
\end{align}
It is not hard to see that the first and 
 third equations  as well as the second and  fourth are equivalent as a consequence of the Bianchi identities for the 
zero-torsion condition.

To obtain Fronsdal equations in their conventional form \cite{Fronsdal:1978rb} $\omega_{s_1-1\pm1,s_1-1\mp1}$ has to be expressed via the Fronsdal fields $\phi_{s_1,s_1}(Y)\equiv\phi$ and $\phi'_{s_1-2,s_1-2}(Y)\equiv\phi'$ \eqref{omega-phi}, that can be easily done using zero-torsion condition (or \eqref{D_omega_decomp} for $m=n=s_1-1$). The final result, simplified with the help of the
Schouten identity, is
\begin{align*}
    2s_1&\Box \phi - D(y,\yd)D(p,\pd) \phi + D(y,\yd)D(y,\yd) \phi'
        - 2s_1(s_1^2-2s_1-2) \phi =
            2s_1J_{s_1,s_1}(Y)\,;\\
    2&\Box \phi' - D(y,\yd)D(p,\pd) \phi' + D(p,\pd)D(p,\pd) \phi
        - 2s_1^3 \phi' =
            2J_{s_1-2,s_1-2}(Y)\,,
\end{align*}
where $-2\Box \equiv (p_1p_2)(\pd_1\pd_2)D(Y_1)D(Y_2)$ and the Fronsdal currents are related to the unfolded ones as 
follows
\begin{align*}
    J_{s_1,s_1}(Y) &= 
        \frac{1}{2s_1} \lp U_{s_1,s_1-2}(\yd,\yd|Y) + U_{s_1-2,s_1}(y,y|Y) \rp\,,\\
    J_{s_1-2,s_1-2}(Y) &= 
        \frac12 \lp U_{s_1,s_1-2}(p,p|Y) + U_{s_1-2,s_1}(\pd,\pd|Y) \rp\,.
\end{align*}

The final expressions resulting from the substitution \eqref{U} read as (recall that $\Upsilon_{m,n}(\bullet|Y)$ are defined in \eqref{Y_decomp})
\begin{align}
\begin{split}
    &J_{s_1,s_1}(Y) =
        \frac{1}{2s_1}\Upsilon_{s_1,s_1-2}(\yd,\yd|Y) +\\&+
        \frac{i}{2s_1^2(s_1+1)}
            \big[
                D(p,\yd)\Upsilon_{s_1-1,s_1-1}(y,y|Y)
                + s_1 D(y,\yd)\Upsilon_{s_1-1,s_1-1}(y,p|Y)
            \big]
        + c.c.\,,
\end{split}\label{J_s}
\end{align}

\begin{align}
\begin{split}
    &J_{s_1-2,s_1-2}(Y) =
        \frac12 \Upsilon_{s_1,s_1-2}(p,p|Y) +\\&+
        \frac{i}{2s_1(s_1-1)}
            \big[
                D(p,\yd)\Upsilon_{s_1-1,s_1-1}(\pd,\pd|Y)
                - s_1 D(p,\pd)\Upsilon_{s_1-1,s_1-1}(\yd,\pd|Y)
            \big]
        +c.c.
\end{split}\label{J_s-2}
\end{align}
Obviously,  at $s_1\leqslant 1$ these formulae make no sense. This is because, in this case, Fronsdal equations belong to the equation on $C$ \eqref{C_eq}, so the discussed construction is not applicable. We do not consider derivation of currents for $s_1 = 0,\,1$ in detail because  its idea is analogous. The final results for the Klein-Gordon and Yang-Mills equations
read, respectively, as
\begin{equation}\label{KG_eq}
    \Box \phi + 2 \phi = J_{0,0}\,,\qquad 
    J_{0,0} = -\frac12 [i\Upsilon^C_{1,1}(p,\pd|Y) - D(p,\pd)\Upsilon^C_{0,0}(y,\yd|0)]\,,
\end{equation}
\begin{equation}\label{YM_eq}
    \Box \phi + \frac12 D(y,\yd)D(p,\pd) \phi + 3 \phi = J_{1,1}\,,\qquad 
    J_{1,1} = -\frac{\eta}{4}\Upsilon^C_{0,2}(y,\pd|0,\yd)
        + \frac i4 D(y,\pd)\Upsilon_{0,0}(\yd,\yd|0) + c.c.
\end{equation}
Here $\Upsilon^C$ stands for the RHS of the equation on $C$, namely $\Upsilon^C=\Upsilon(\omega,C) + \Upsilon(\Omega,C,C)$.

\subsubsection{Fermionic case}\label{Fermionic_case}%%%%%%%%%%%%%%%%%%%%%%%%%%%%%%%%%%%%%%%%%

For half-integer $s_1$ the problem is far simpler due to the absence of torsion. The analogue of equations \eqref{Fr_eq_unf} reads as 
\begin{align}\label{Fang-Fr_eq}
\begin{split}
    \pushleft{D(p,\yd) \omega_{s_1-1/2,s_1-3/2}(p,\pd|Y)
        - D(p,\pd) \omega_{s_1-1/2,s_1-3/2}(p,\yd|Y)
        +}\\\hspace{2cm}
            + i(s_1^2-1/4) \omega_{s_1-3/2,s_1-1/2}(p,\pd|Y) =
                -\frac{1}{s_1+1/2} \Upsilon_{s_1-1/2,s_1-3/2}(p,p|Y)\,,\\
    \pushleft{D(y,\yd) \omega_{s_1-1/2,s_1-3/2}(p,\yd|Y)
        - D(p,\yd) \omega_{s_1-1/2,s_1-3/2}(y,\yd|Y)
        -}\\\hspace{2cm}
            - i(s_1^2-1/4) \omega_{s_1-3/2,s_1-1/2}(y,\yd|Y) =
                -\frac{1}{s_1-1/2} \Upsilon_{s_1-1/2,s_1-3/2}(\yd,\yd|Y)\,,\\
    \pushleft{D(y,\pd) \omega_{s_1-1/2,s_1-3/2}(p,\yd|Y)
        + \frac{(s_1-1/2)^2}{2s_1} D(p,\pd) \omega_{s_1-1/2,s_1-3/2}(y,\yd|Y)
        -}\\\pushleft{- \frac{(s_1+1/2)^2}{2s_1} D(y,\yd) \omega_{s_1-1/2,s_1-3/2}(p,\pd|Y)
        - i(s_1^2-1/4)\omega_{s_1-3/2,s_1-1/2}(y,\pd|Y) =}
            \\=
                -i\frac{(s_1^2-1/4)}{2s_1} \lb
                    (s_1+1/2) \Upsilon_{s_1-1/2,s_1-3/2}(y,p|Y)
                    + (s_1-1/2) \Upsilon_{s_1-1/2,s_1-3/2}(\yd,\pd|Y)
                \rb\,.
\end{split}
\end{align}
(The complex conjugated  equations are analogous.)
By virtue of the relations
\begin{equation*}
    A_{\ga\gad} = A_{\underline{n}} \sigma^{\underline{n}}_{\ga\gad}\,,\qquad
    \gamma^{\underline{n}}{}_\ga{}^\gb = 0\,,\quad
    \gamma^{\underline{n}}{}_\ga{}^\gbd = i\sigma^{\underline{n}}{}_\ga{}^\gbd\,,\quad
    \gamma^{\underline{n}}{}_\gad{}^\gbd = 0\,,\quad
    \gamma^{\underline{n}}{}_\gad{}^\gb = i\sigma^{\underline{n}}{}^\gb{}_\gad\,,
\end{equation*}
where  $\underline{n}$ is a space-time vector index and $\gamma^{\underline{n}}$ are Dirac matrices,
to reproduce the $\gamma^{\underline{n}}D_{\underline{n}}$-term in the Fang-Fronsdal equation \cite{Fang:1978wz} on the RHS of \eqref{Fang-Fr_eq}, the second of these equations should be multiplied by $(-1)$. This yields the following result
\begin{align}
\begin{split}
    J_{s_1-1/2,s_1+1/2}(Y) =
        \frac{1}{s_1-1/2} \Upsilon_{s_1-1/2,s_1-3/2}(\yd,\yd|Y)\,,
\end{split}\label{J_s-1/2}
\end{align}
\begin{align}
\begin{split}
    J_{s_1-5/2,s_1-3/2}(Y) =
        -\frac{1}{s_1+1/2} \Upsilon_{s_1-1/2,s_1-3/2}(p,p|Y)\,,
\end{split}\label{J_s-5/2}
\end{align}
\begin{align}
\begin{split}
    J_{s_1-1/2,s_1-3/2}(Y) =
        -i\frac{(s_1^2-1/4)}{2s_1} \big[
                    &(s_1+1/2) \Upsilon_{s_1-1/2,s_1-3/2}(y,p|Y)
                    +\\+ &(s_1-1/2) \Upsilon_{s_1-1/2,s_1-3/2}(\yd,\pd|Y)
                \big]\,.
\end{split}\label{J_s-3/2}
\end{align}

The singularity at $s_1=1/2$ has the same origin as that at $s_1=0,1$ in section \ref{Bosonic_case}. 
Dirac equation resulting from the equation on $C$ has the form (we again set $\Upsilon^C=\Upsilon(\omega,C) + \Upsilon(\Omega,C,C)$)
\begin{align}\label{Dirac_eq}
\begin{split}
    D(y,\pd) C_{0,1} = J_{1,0}\,,\qquad
    J_{1,0} = \Upsilon^C_{0,1}(y,\pd|0,\yd)\,;\\
    D(p,\yd) C_{1,0} = J_{0,1}\,,\qquad
    J_{0,1} = \Upsilon^C_{1,0}(p,\yd|y,0)\,.
\end{split}
\end{align}

\subsection{Calculation results}\label{Calculation_results}%%%%%%%%%%%%%%%%%%%%%%%%%%%%%%%%%%%%%%%

Here we present explicit formulae for all bilinear currents resulting from the nonlinear $AdS_4$ HS theory. To pass from the master-field language to the Fronsdal field derivatives we use equations of motion and the TT-gauge  \eqref{omega-phi_gauged}. Namely, \eqref{D_omega_decomp}, along with the decomposition of the equation on $C$ into one-forms, leads to the following relations at the free level:
\begin{subequations}
\begin{align}
    &\phi_{s+m+\{s\},s-m-\{s\}}(Y;K) =
        i^m (s-\{s\})\frac{(s-m-1-\{s\})!}{(s+m+\{s\})!}
            D^m(y,\pd) \phi_{s+\{s\},s-\{s\}}(Y;K)\,,\\
    &\phi_{s-m-\{s\},s+m+\{s\}}(Y;K) =
        i^m (s-\{s\})\frac{(s-m-1-\{s\})!}{(s+m+\{s\})!}
            D^m(p,\yd) \phi_{s-\{s\},s+\{s\}}(Y;K)\,;\\
    &C_{2s+n,n}(Y;K) = 2s \frac{i^{s-1-n-\{s\}}}{\bar{\eta}(2s-1)(2s+n)!n!}
        D^n(y,\yd) D^{s-\{s\}}(y,\pd) \phi_{s+\{s\},s-\{s\}}(Y;K)\bar{k}\,,\\
    &C_{n,2s+n}(Y;K) = 2s \frac{i^{s-1-n-\{s\}}}{\eta(2s-1)(2s+n)!n!}
        D^n(y,\yd) D^{s-\{s\}}(p,\yd) \phi_{s-\{s\},s+\{s\}}(Y;K)k\,,
\end{align}\label{comp-to-der}
\end{subequations}
where $\{s\}$ is  the fractional part of $s$.

Recall that all fields in the original system can take values in any associative algebra $A$ \cite{Vasiliev:1988sa}. To make this structure manifest we introduce color indices $a,b,c\ldots$ carried by the fields valued in $A$ and define the structure constants of $A$ as usual,
$$(fg)^a = c^a_{bc}f^bg^c \quad\forall f,g\in A\,.$$ 
Taking into account \eqref{DynW}
it is convenient to describe the dependence on the Klein operators analogously setting 
$$\phi^a(Y;K) = \phi^{a,0}(Y) + \phi^{a,1}(Y)k\Bar{k}\,.$$ Defining a collective index 
$\cA$ in place of $(a,0)$ and $(a,1)$, we introduce new `structure constants' $c^\cA_{\cB\cC}(s_1,s_2,s_3)$ and $f^\cA_{\cB\cC}(s_1,s_2,s_3)$:
\begin{align}
\begin{split}\label{StructConst}
    c^{a,0}_{\cB\cC}(s_1,s_2,s_3) :=&
        (\delta^{b,0}_{\cB}\delta^{c,0}_{\cC} + (-1)^{2s_3}\delta^{b,1}_{\cB}\delta^{c,1}_{\cC})
            c^a_{bc}
        + (-1)^{s_2+s_3-s_1+4s_2s_3}
            (\delta^{b,0}_{\cB}\delta^{c,0}_{\cC} + (-1)^{2s_2}\delta^{b,1}_{\cB}\delta^{c,1}_{\cC})
                c^a_{cb}\,,\\
    c^{a,1}_{\cB\cC}(s_1,s_2,s_3) :=&
        (\delta^{b,0}_{\cB}\delta^{c,1}_{\cC} + (-1)^{2s_3}\delta^{b,1}_{\cB}\delta^{c,0}_{\cC})
            c^a_{bc}
        + (-1)^{s_2+s_3-s_1+4s_2s_3}
            (\delta^{b,1}_{\cB}\delta^{c,0}_{\cC} +
            (-1)^{2s_2}\delta^{b,0}_{\cB}\delta^{c,1}_{\cC})
                c^a_{cb}\,;\\
    f^{a,0}_{\cB\cC}(s_1,s_2,s_3) :=&
        (\delta^{b,0}_{\cB}\delta^{c,1}_{\cC} + (-1)^{2s_3}\delta^{b,1}_{\cB}\delta^{c,0}_{\cC})
            c^a_{bc}
        + (-1)^{s_1+s_2+s_3+4s_2s_3}
            (\delta^{b,1}_{\cB}\delta^{c,0}_{\cC} +
            (-1)^{2s_2}\delta^{b,0}_{\cB}\delta^{c,1}_{\cC})
                c^a_{cb}\,,\\
    f^{a,1}_{\cB\cC}(s_1,s_2,s_3) :=&
        (\delta^{b,0}_{\cB}\delta^{c,0}_{\cC} + (-1)^{2s_3}\delta^{b,1}_{\cB}\delta^{c,1}_{\cC})
            c^a_{bc}
        + (-1)^{s_1+s_2+s_3+4s_2s_3}
            (\delta^{b,0}_{\cB}\delta^{c,0}_{\cC} + (-1)^{2s_2}\delta^{b,1}_{\cB}\delta^{c,1}_{\cC})
                c^a_{cb}\,,
\end{split}
\end{align}
that appear below in the currents \eqref{J_s-2_res}, \eqref{J_s_res}.

In the purely bosonic system, the exterior Klein operators can be factored out \cite{Vasiliev:1999ba} since in that case $k\bar{k}$ is central. So, setting $\phi(Y;K) = \phi(Y)(1+k\bar{k})$, one arrives at $C(Y;K) = C(Y)(k + \bar{k})$. In practice, this simply means that $\phi(Y;K)$ can be replaced by $\phi(Y)$ while $C(Y;K)k$ and $C(Y;K)\bar{k}$ can be replaced by $C(Y)$. In that case $c^\cA_{\cB\cC}(s_1,s_2,s_3)$ and $f^\cA_{\cB\cC}(s_1,s_2,s_3)$ reduce to $c^a_{bc} + (-1)^{s_1+s_2+s_3} c^a_{cb}$.

As a result, inserting expressions \eqref{ww_exp}, \eqref{wWC_exp}, and \eqref{WWCC_exp} into formulae \eqref{J_s} and \eqref{J_s-2} and using \eqref{comp-to-der} to express master-fields via Fronsdal field derivatives, one obtains the following expressions for the Fronsdal currents in the TT-gauge (we set $\theta(0) = 1$ and extract $|\eta|^2$ from the fields):
\begin{align}\label{J_s-2_res}
\begin{split}
    J^\cA_{s_1-2,s_1-2}(Y) =&
        |\eta|^2\sum_{s_2,s_3,m,n} \lp\prod_{i,j,k} \theta(s_i+s_j-s_k-1)\rp
            c^\cA_{\cB\cC}(s_1,s_2,s_3)
            \mathbf{A}(s_1,s_2,s_3|m,n|\phi^\cB_{s_2}, \phi^\cC_{s_3})\,;
\end{split}
\end{align}
\begin{align}\label{J_s_res}
\begin{split}
    J^\cA_{s_1,s_1}(Y) =
        |\eta|^2\sum_{s_2,s_3,m,n} &\lp\prod_{i,j,k} \theta(s_i+s_j-s_k-1)\rp
            c^\cA_{\cB\cC}(s_1,s_2,s_3)
            \mathbf{B}(s_1,s_2,s_3|m,n|\phi^\cB_{s_2}, \phi^\cC_{s_3})
    +\\+
        |\eta|^2\sum_{s_2,s_3,n} &\theta(s_2+s_3-s_1-1)c^\cA_{\cB\cC}(s_1,s_2,s_3)
        \mathbf{C}(s_1,s_2,s_3|n|\phi^\cB_{s_2}, \phi^\cC_{s_3})
    +\\+
        |\eta|^2\sum_{s_2,s_3,n} &\theta(s_1-s_2-s_3)c^\cA_{\cB\cC}(s_1,s_2,s_3)
        \mathbf{D}(s_1,s_2,s_3|n|\phi^\cB_{s_2}, \phi^\cC_{s_3})
    +\\+
        |\eta|^2\cos{2\vartheta}\sum_{s_2,s_3,n} &\theta(s_2-s_1-s_3-1)f^{\cA}_{\cB\cC}(s_1,s_2,s_3)\frac12 
        \mathbf{E}(s_1,s_2,s_3|n|\phi^\cB_{s_2}, \phi^\cC_{s_3})
    +\\+
        |\eta|^2\cos{2\vartheta}\sum_{s_2,s_3,n} &\theta(s_3-s_1-s_2-1)f^{\cA}_{\cB\cC}(s_1,s_3,s_2)\frac12 
        \mathbf{E}(s_1,s_3,s_2|n|\phi^\cB_{s_3}, \phi^\cC_{s_2})
    +\\+
        |\eta|^2\cos{2\vartheta}\sum_{s_2,s_3,n} &\theta(s_1+s_2-s_3)\theta(s_1+s_3-s_2)
            f^{\cA}_{\cB\cC}(s_1,s_2,s_3)
            \mathbf{F}(s_1,s_2,s_3|n|\phi^\cB_{s_2}, \phi^\cC_{s_3})
    +\\+
        |\eta|^2\sin{2\vartheta}\sum_{s_2,s_3,n} &\theta(s_2-s_1-s_3-1)f^{\cA}_{\cB\cC}(s_1,s_2,s_3)\frac12 
        \Tilde{\mathbf{E}}(s_1,s_2,s_3|n|\phi^\cB_{s_2}, \phi^\cC_{s_3})
    +\\+
        |\eta|^2\sin{2\vartheta}\sum_{s_2,s_3,n} &\theta(s_3-s_1-s_2-1)f^{\cA}_{\cB\cC}(s_1,s_3,s_2)\frac12 
        \Tilde{\mathbf{E}}(s_1,s_3,s_2|n|\phi^\cB_{s_3}, \phi^\cC_{s_2})
    +\\+
        |\eta|^2\sin{2\vartheta}\sum_{s_2,s_3,n} &\theta(s_1+s_2-s_3)\theta(s_1+s_3-s_2)
            f^{\cA}_{\cB\cC}(s_1,s_2,s_3)
            \Tilde{\mathbf{F}}(s_1,s_2,s_3|n|\phi^\cB_{s_2}, \phi^\cC_{s_3})\,.
\end{split}
\end{align}
To simplify formulae,  Lorentz derivatives and other details are hidden in the coefficient functions $\mathbf{A},\dots\mathbf{F}$ presented in Appendix \ref{BosCur}. Note that $J_{s_1-2,s_1-2}$ is generated solely by $\Upsilon(\omega,\omega)$ while 
different terms in $J_{s_1,s_1}$ \eqref{J_s_res} correspond to $\Upsilon(\omega,\omega)$ ($\mathbf{B}$-terms), $\Upsilon(\Omega,\omega,C)$ ($\mathbf{C}$-terms and $\mathbf{E}$-terms) and $\Upsilon(\Omega,\Omega,C,C)$ ($\mathbf{D}$-terms and $\mathbf{F}$-terms). This implies that in the TT-gauge all currents in the HS theory are conformal except for 
 $\Upsilon(\omega,\omega)$ which conclusion is an agreement with the known fact that massless fields of spins $s\leq 1$,
 that are free of the $\Upsilon(\omega,\omega)$ currents,  are conformally invariant while (super)gravity is not. In turn, this  suggests that conformal symmetry should be extendable  to the $\Upsilon(\omega,\omega)$ sector as well though in a somewhat nontrivial way (for more detail see  \cite{Vasiliev:2007yc}).

Tilded coefficient functions result from the original ones by replacing $+c.c.$  by $-c.c.$ These  correspond to $P$-odd currents while the rest are $P$-even. So, in the Fronsdal equations of the form $$\Box \phi + \ldots = g(\eta)J + c.c. = 2\Re (g) \Re (J) - 2\Im (g) \Im (J)$$ the last term contains an odd number of Levi-Civita symbols thus being $P$-odd.\footnote{This can be seen as follows. Restoration of the tensor indices in spinor expressions amounts to calculation of traces of Pauli matrices combinations $\tr{\sigma_\Um\bar{\sigma}_\Un\sigma_\Uk\Bar{\sigma}_\Ul\dots}$, in which Levi-Civita symbol is realised by the imaginary unit $i$. In these terms, symbol $c.c.$  used so far becomes genuine complex conjugation.} This agrees with the fact that the parity automorphism changes a sign of the phase of $\eta$ in the HS equations \eqref{Vas4}. Correspondingly, the $\vartheta$-independent and $\cos{2\vartheta}$-dependent terms in \eqref{J_s-2_res}, \eqref{J_s_res} describe $P$-even currents while those proportional to $\sin{2\vartheta}$ are $P$-odd. It is worth mentioning that this structure is analogous to that of boundary correlators found in \cite{Maldacena:2012sf}.
Also note that the $P$-odd vertex is absent at $\vartheta=0$ and $\vartheta=\pi/2$ which correspond to so-called $A$ and $B$ HS models \cite{Sezgin:2003pt}. The highest derivative $P$-even vertex vanishes at $\vartheta=\frac{\pi}{4}$, which is `maximally parity-breaking' case in the terminology of \cite{Misuna:2017bjb}.

In formulae \eqref{J_s-2_res} and \eqref{J_s_res} we keep track only of the structure constants $c^\cA_{\cB\cC}$ and $f^\cA_{\cB\cC}$, $\theta$-functions restricting spins of the fields, and the factors of $|\eta|^2$, $\cos{2\vartheta}=\Re(\eta/\Bar{\eta})$ or $\sin{2\vartheta}=\Im(\eta/\Bar{\eta})$. In terms associated with $\Upsilon(\Omega, \omega, C)$, $\cos{2\vartheta}$ or $\sin{2\vartheta}$ appears when $\eta$ in \eqref{wWC_exp} is divided by $\bar{\eta}$ coming from $C$ via \eqref{comp-to-der}. Then, the sum  or difference of  the two conjugated parts gives $\cos{2\vartheta}$ or $\sin{2\vartheta}$, respectively. Analogously for $\Upsilon(\Omega,\Omega,C,C)$: if both $C$ in \eqref{WWCC_exp} come with $1/\eta$ or with $1/\bar{\eta}$ then due to the prefactor $\eta\bar{\eta}$ the result is proportional to $\cos{2\vartheta}$ or $\sin{2\vartheta}$. Comparing the effect of $\theta$-functions from here with Table \ref{spins_1} one can see that the terms with the factor of $\cos{2\vartheta}$ or $\sin{2\vartheta}$ contain $\leqslant [s_1]+[s_2]+[s_3]$ derivatives while the $\vartheta$-independent ones contain $\leqslant [s_1]+[s_2]+[s_3]-2\min\{s_1,s_2,s_3\}$ derivatives. Hence, remembering that, in accordance with the Metsaev's classification \cite{Metsaev:2005ar}, there are two independent types of HS currents in the $P$-even sector, we see how the original nonlinear equations fix their couplings. (Recall that in the flat (Minkowski) limit only the terms with the maximal number of derivatives survive.)

One can reinterpret these results from the light-cone cubic Lagrangian perspective known \cite{Bengtsson:1986kh, Metsaev:2018xip} to contain two pairs of complex conjugated vertices (with two pairs of complex conjugated coupling constants) that admit covariant form. These vertices have just the same number of derivatives as the aforementioned $P$-even ones. Since light-cone fields are defined up to a $U(1)$-factor, one can adjust this factor in such a way  that the coupling constants of the lower-derivative vertices become real. Then, the sum of two conjugated lower-derivative vertices will produce a single $P$-even vertex. The rest higher-derivative vertices, however, will have the complex couplings producing the combination of $P$-even and $P$-odd vertices. This analysis is in agreement with the fact that the phase $\vartheta$ in the HS equations can be interpreted as the parameter of
electric-magnetic duality transformation on the zero-forms $C$,
uncovering the relation of the complex coupling $\eta$ in the $AdS_4$ HS theory with the structure of the cubic interaction in the $4d$ light-cone approach.

The expressions for the currents with half-integer $s_1$, $J_{s_1-1/2,s_1+1/2}(Y)$ are analogous to $J_{s_1,s_1}(Y)$, while $J_{s_1-5/2,s_1-3/2}(Y)$ and $J_{s_1-1/2,s_1-3/2}(Y)$ are analogous to $J_{s_1-2,s_1-2}(Y)$. The difference is only in the coefficient functions presented in Appendix \ref{FerCur}. As a result, the discussion of the structure constants and the number of derivatives remains valid for this case as well.

\subsection{Examples}\label{Examples}%%%%%%%%%%%%%%%%%%%%%%%%%%%%%%%%%%%%%%%%%%%%%%%%%%%%%%%%%%%%%%

\subsubsection{Yang-Mills theory}\label{Yang-Mills}

Let $s_1 = s_2 = s_3 = 1$ and exterior Klein operators are turned off in the way described in section \ref{Calculation_results}. Then, the current in Yang-Mills equations reads as
\begin{align}
\begin{split}
    J^a_{1,1}(Y) =
        -(c^a_{bc} - c^a_{cb})\bigg\{
            |\eta|^2\frac i8&\big[
                D(y,\pd) (p_1p_2)(\pd_1\yd)(\pd_2\yd) +
                2(p_1y)(\pd_1\pd_2)(\pd_2\yd) D_2(p_2,\yd_2) + c.c.
            \big] +\\+
            \frac{1}{96} |\eta|^2\cos{2\vartheta}&
            \big[
                i(p_1p_2) D_1(p,\yd)(p_1y)(p_2y)
                    D_1(y_1,\pd_1) D_2(y_2,\pd_2)
                +\\&+ (p_1p_2)^2(\yd\pd_1)
                    D_1(y_1,\yd_1) D_1(y_1,\pd_1) D_2(y_2,\pd_2) + c.c.
            \big] +\\+
            \frac{i}{96} |\eta|^2\sin{2\vartheta}&
            \big[
                i(p_1p_2) D_1(p,\yd)(p_1y)(p_2y)
                    D_1(y_1,\pd_1) D_2(y_2,\pd_2)
                +\\&+ (p_1p_2)^2(\yd\pd_1)
                    D_1(y_1,\yd_1) D_1(y_1,\pd_1) D_2(y_2,\pd_2) - c.c.
            \big]
        \bigg\}\phi^b_{1,1}(Y_1)\phi^c_{1,1}(Y_2)\,.
\end{split}
\end{align}
It is not hard to see that in terms of usual Yang-Mills connection this yields
\begin{align}
\begin{split}
    J^{a\ \Um} =
    &|\eta|^2(2 \eta^{\Um\Ul}\eta^{\Uk\Un} - \eta^{\Um\Uk}\eta^{\Un\Ul})
            (c^a_{bc} - c^a_{cb}) \phi^b_\Un D_\Uk \phi^c_\Ul +\\+
    \frac13 &|\eta|^2\cos{2\vartheta} \big[
            \eta^{\Uk\Ul} \eta^{\Up\Uq} - \eta^{\Up\Ul} \eta^{\Uq\Uk}
        \big] (c^a_{bc} - c^a_{cb}) (D^\Um D_\Uk \phi^b_\Up) (D_\Ul \phi^c_\Uq) -\\-
    \frac13 &|\eta|^2\sin{2\vartheta}
        \epsilon^{\Uk\Ul\Up\Uq}
            (c^a_{bc} - c^a_{cb}) (D^\Um D_\Uk \phi^b_\Up) (D_\Ul \phi^c_\Uq)\,.
\end{split}
\end{align}
Here the first term is associated with the minimal Yang-Mills coupling while the 
second and the third  result from the $P$-even and $P$-odd $F^3$ couplings, respectively.
Note that in this case only the antisymmetric part of the structure coefficients contributes.

That the two types of vertices have different number of derivatives implies that there
has to be a dimensionful coupling constant in front of the highest derivative one. Indeed,
this is $\Lambda^{-1}$  which is implicit in our formulae since in this paper it is set $
\Lambda=-1$. The relative coefficients of these vertices are uniquely determined by the 
HS symmetry of the whole theory.

\subsubsection{Gravity}\label{Gravity}

Let $s_1 = s_2 = s_3 = 2$ and the color group as well as the dependence on the Klein operators be trivial.
Then formulae \eqref{J_s_res}, \eqref{J_s_res} yield
\begin{align}
\begin{split}
    J_{0,0}(Y) =
        \frac{i}{4}|\eta|^2\big[
            (p_1p_2)^3(\pd_1\pd_2) D_1(y_1,\pd_1)D_2(y_2,\pd_2) -
            4 (p_1p_2)^2(\pd_1\pd_2)^2 + c.c.
        \big]\phi_{2,2}(Y_1)\phi_{2,2}(Y_2)\,.
\end{split}
\end{align}
\begin{align}
\begin{split}
    J_{2,2}(Y) =
        \Bigg\{
            \frac{i}{18}|\eta|^2 \big[
                 18 \mathbf{K}(1,1;1,1|2,2)
                - 9 \mathbf{K}(1,1;0,2|2,2)D_1(p_1,\yd_1)D_2(p_2,\yd_2) -\\
                + 2\mathbf{K}(2,0;0,2|2,2)D_2^2(p_2,\yd_2)
                + c.c.
            \big]
            +\\+
            |\eta|^2\cos 2\vartheta \frac{1}{(15\cdot 6!)^2}\big[
                10i \mathbf{K}(0,0;0,4|2,2)
                    D_1^2(p_1,\yd_1) D_2^2(y_2,\yd_2)D_2^2(p_2,\yd_2) -\\-
                12i \mathbf{K}(1,1;0,4|2,2)
                    D_1(y_1,\yd_1)D_1^2(p_1,\yd_1) D_2(y_2,\yd_2)D_2^2(p_2,\yd_2) -\\-
                15 D(y,\pd)\mathbf{K}(0,1;0,3|1,3)
                    D_1^2(p_1,\yd_1) D_2(y_2,\yd_2)D_2^2(p_2,\yd_2)
                + c.c.
            \big]
            -\\-
            |\eta|^2\sin 2\vartheta \frac{i}{(15\cdot 6!)^2}\big[
                10i \mathbf{K}(0,0;0,4|2,2)
                    D_1^2(p_1,\yd_1) D_2^2(y_2,\yd_2)D_2^2(p_2,\yd_2) -\\-
                12i \mathbf{K}(1,1;0,4|2,2)
                    D_1(y_1,\yd_1)D_1^2(p_1,\yd_1) D_2(y_2,\yd_2)D_2^2(p_2,\yd_2) -\\-
                15 D(y,\pd)\mathbf{K}(0,1;0,3|1,3)
                    D_1^2(p_1,\yd_1) D_2(y_2,\yd_2)D_2^2(p_2,\yd_2)
                - c.c.
            \big]
        \Bigg\}\phi_{2,2}(Y_1)\phi_{2,2}(Y_2)\,,
\end{split}
\end{align}
where
\begin{equation*}
    \mathbf{K}(a,b;c,d|g,h) :=
        \frac{(p_1y)^a(\pd_1\yd)^b (p_2y)^{g-a}(\pd_2\yd)^{h-b} (p_1p_2)^c(\pd_1\pd_2)^d}
            {a!b!c!d!(g-a)!(h-b)!}\,.
\end{equation*}

Here the $\vartheta$-independent terms contain two derivatives of the dynamical Fronsdal field (i.e., metric). These describe the so-called gravitational pseudo stress tensor which is known to be gauge (diffeomorphism) non-invariant \cite{Landau1987-cd}. The $\vartheta$-dependent terms on the other hand contain six derivatives of the dynamical Fronsdal fields. Such terms, resulting from the Lagrangian vertices cubic in the Weyl tensor, are absent in the standard Einstein GR, which is a low-energy limit of a to be constructed fundamental theory to which the HS theory may have  direct relation. Since higher derivative terms naturally appear in string theory it would be  interesting to explore parallels between these theories in more detail. The immediate difference is
that the  dimensionful coupling constant in front of the highest derivative vertex (that 
contains six derivatives in the spin two${}^3$ case) is proportional to 
 $\Lambda^{-2}$ in the HS theory while in string theory it is expressed via string tension.
Again, the relative coefficients of these vertices are  determined by the 
symmetry of the whole HS theory.

%%%%%%%%%%%%%%%%%%%%%%%%%%%%%%%%
\section{Conclusion}\label{Conclusion}
%%%%%%%%%%%%%%%%%%%%%%%%%%%%%%%%

In this paper, the properties of the bilinear currents in $AdS_4$ HS theory with the complex coupling constant $\eta=|\eta|\exp{i\vartheta}$ have been studied with the emphasis on the gauge non-invariant sector formed by vertices $\Upsilon(\omega,\omega)$, $\Upsilon(\Omega,\omega,C)$, and $\Upsilon(\omega,C)$ containing the HS gauge potentials $\omega$ since the gauge invariant $C^2$ sector has been  extensively studied in the literature (see e.g. \cite{Vasiliev:2016xui, Gelfond:2017wrh}, \cite{Misuna:2017bjb}, \cite{Didenko:2018fgx,Gelfond:2018vmi,Didenko:2019xzz,Gelfond:2019tac,Didenko:2020bxd}). We have shown that the vertices in question generate nontrivial currents in the Fronsdal equations and, by  straightforward calculations, found their explicit spinor form (see \eqref{J_s-2_res}, \eqref{J_s_res} and Appendix \ref{CoefFunc}). The resulting  expressions decompose into three types of currents with two independent coupling constants 
\begin{equation*}
    J = g_1 J^{\text{even}}_{\text{min}} + g_2 J^{\text{even}}_{\text{max}}
        + g_3 J^{\text{odd}}_{\text{max}}\,,\qquad
    g_1^2 = g_2^2 +g_3^2\,.
\end{equation*}
These currents are labelled by parity and the number of derivatives which, in agreement with the results of light-cone 
approach of \cite{Bengtsson:1986kh, Metsaev:2005ar, Metsaev:2018xip}, is bounded from above by $[s_1]+[s_2]+[s_3]-2\min\{s_1,s_2,s_3\}$ for $J^{\text{even}}_{\text{min}}$ and by $[s_1]+[s_2]+[s_3]$ for $J^{\text{even}}_{\text{max}}$ and $J^{\text{odd}}_{\text{max}}$. The couplings are expressed in terms of two real parameters, namely $g_1 = |\eta|^2$, $g_2 = |\eta|^2\cos 2\vartheta$ and $g_3 = |\eta|^2\sin 2\vartheta$. The identification of the two types of independent parity-invariant currents in the Metsaev classification with the couplings of the nonlinear theory is a new result. It would be interesting to compare our results with those of \cite{Sleight:2016dba} where the cubic HS couplings were found for the case of $A$-model with $\vartheta=0$.

So far, explicit expressions for the HS currents of arbitrary spins and phase of $\eta$ were not available in the literature. These are important from the perspective of holographic duality between $AdS_4$ HS theory with non-zero $\vartheta$ in the bulk and parity-non-invariant vector sigma-model with Chern-Simons interaction  on the boundary \cite{Aharony:2011jz, Giombi:2011kc}. The obtained  expressions for vertices  make it possible to check the  HS-Chern-Simons holographic duality conjecture directly. Note that the structure of the higher-derivative currents we obtained is analogous to that of three-point function of \cite{Maldacena:2012sf}.
(See also \cite{Giombi:2012ms} for the correlators with $s_3=0$).

To simplify expressions for the currents we fixed the gauge at the free level in the way that generalizes the TT-gauge in the metric-like formalism \cite{Joung:2011ww} to the unfolded formulation  of free  HS equations for all components of $\omega$, which is another new result of the paper. It is shown that in this gauge  only one out of the four components of the one-forms $\omega$ in the four-dimensional space, namely, the rank-$s$ traceless two-row Young diagrams in tensor notations, survive.

An interesting output of our analysis is that all currents in the HS theory turn out to be conformal 
(i.e., traceless) in the TT-gauge except for those bilinear in HS gauge
potentials, $\Upsilon(\omega,\omega)$. This conclusion agrees with the known fact that massless fields of spins $s\leq 1$,
 that are free of the $\Upsilon(\omega,\omega)$ currents,  are conformally invariant while (super)gravity is not. In turn, this suggests that conformal symmetry should be extendable  to the $\Upsilon(\omega,\omega)$ sector as well though in a somewhat nontrivial way as  discussed in \cite{Vasiliev:2007yc}.

\acknowledgments{
We acknowledge the collaboration with Theoder Razorenov at the initial stage of this work.
We are grateful to V. Didenko, O. Gelfond, A. Korybut, R. Metsaev and T. Razorenov for
careful reading the draft and useful comments and especially to the referee for pointing
out some terminological inaccuracies in the original version. Also we thank N. Misuna and
W. Wachowski for helpful comments. MV wishes to thank for hospitality Ofer Aharony,
Theoretical High Energy Physics Group of Weizmann Institute of Science where the substantial
part of this work has been done. This work was supported by Theoretical Physics and
Mathematics Advancement Foundation “BASIS” Grant No 24-1-1-9-5.
}

\begin{appendices}
%%%%%%%%%%%%%%%%%%%%%%%%%%%%%%%%
\section{Details of vertices derivation}\label{VertDer}
%%%%%%%%%%%%%%%%%%%%%%%%%%%%%%%%

At the first order, by virtue of the conventional homotopy with zero shift parameter, formulae \eqref{B_n}, \eqref{S_n}, \eqref{W_n} take form
\begin{align}
    B_1 &=\label{C_1}
        C_1\,,\\
    S_1 &=\label{S_1}
        -\frac 12 \hmt_0 \lp\eta C_1\star\gamma + \bar{\eta} C_1\star\bar{\gamma}\rp + \dd_Z \epsilon_1\,,\\
    W_1 &=\label{W_1}
        \omega_1
        + \frac i4 \hmt_0 D_\Omega \hmt_0 \lp \eta C_1\star\gamma + \bar{\eta} C_1\star\bar{\gamma} \rp
        - \frac i2 \hmt_0 D_\Omega \dd_Z \epsilon_1\,.
\end{align}
To obtain $\Upsilon(\Omega,\Omega,C)$ one  plugs these expressions into equations \eqref{omega_n_eq} and \eqref{C_n_eq} setting $Z=0$ on the RHS using the remark after \eqref{C_n_eq},
\begin{align}
    D_\Omega \omega_1 &=\label{omega_1_eq}
        -\frac i4 D_\Omega \hmt_0 D_\Omega \hmt_0
            \lp \eta C_1\star\gamma + \bar{\eta} C_1\star\bar{\gamma} \rp \bigg|_{Z=0}\,,\\
    D_\Omega C_1 &=\label{C_1_eq}
        0\,.
\end{align}
After simple calculations the RHS of \eqref{omega_1_eq} is transformed to the form of \eqref{WWC_res}.

Since we are interested in $\Upsilon(\Omega,\omega,C)$, in what follows we only keep track of the terms linear both in $C_1$ and in $\omega_1$, discarding the gauge $\epsilon$-terms. Thus,
\begin{align}
    B_2 &\simeq \label{B_2}
        0\,,\\
    S_2 &\simeq \label{S_2}
        0\,,\\
    W_2 &\simeq \label{W_2}
        \frac i4 \hmt_0 [\omega_1,
            \hmt_0 \lp\eta C_1\star\gamma + \bar{\eta} C_1\star\bar{\gamma}\rp]_\star\,.
\end{align}
The resulting contribution to the RHS of \eqref{omega_n_eq} is
\begin{equation}
    \Upsilon(\Omega,\omega,C) =\label{wWC_alm-res}
        - \eta \frac i4 [\omega_1,
            \hmt_0 [\Omega, \hmt_0 (C_1\star\gamma)]_\star]_{\star}\bigg|_{z=0}
        - \eta \frac i4 [\Omega,
            \hmt_0 [\omega_1, \hmt_0 (C_1\star\gamma)]_\star]_{\star}\bigg|_{z=0}
        + c.c.
\end{equation}
To bring it to the form of \eqref{wWC_res} we use the following lemma.

\begin{lemma}\label{lemma1} Let $f = f(Y; K)$, $g = g(\theta, Z, Y; K)$ and $u$ be an auxiliary 
spinor variable that anticommutes with the exterior Klein operator $k$, $\{u, k\} = 0$. Then the following is true:
\begin{equation*}
    [f,(u z)g]_\star=
        (u z)[f,g]_\star + [i(u \partial_y) f,g]_\star\,,
\end{equation*}
where $\partial_y$ acts only on $f$.
\end{lemma}

Its proof is straightforward. 

In each $\hmt_0$, one can rewrite the spinor contraction as
\begin{equation*}
    -z^\ga\pdv{}{\theta^\ga} =
    \lp z \pdv{}{\theta} \rp \equiv
    (z \partial_\theta) = (\partial_u \partial_v)(uz)(v\partial_\theta)
\end{equation*}
with $u$ and $v$ being auxiliary spinor variables anticommuting with $k$. 
 Lemma \ref{lemma1} applied to each $z$ contraction in the resulting expression yields
\begin{align*}
    \Upsilon(\Omega,\omega,C) =
        &\eta \frac i2 (\partial_u \partial_v)
            \lb
                (u \partial_y)\omega_1,
                \int_0^1 d\tau
                    \lb
                        (v\partial_y)\Omega,
                        \int_0^1 dt t(C_1\star\varkappa)_{z\to tz}k
                    \rb_{\star\ z\to \tau z}
            \rb_{\star\ z=0}
        +\\+
        &\eta \frac i2 (\partial_u \partial_v)
            \lb
                (u \partial_y)\Omega,
                \int_0^1 d\tau
                    \lb
                        (v \partial_y)\omega_1,
                        \int_0^1 dt t(C_1\star\varkappa)_{z\to tz}k
                    \rb_{\star\ z\to \tau z}
            \rb_{\star\ z=0}
        + c.c.
\end{align*}
The expression for $\Omega$ \eqref{W_0} yields that the first term reads as
\begin{align}\label{wWC_1}
\begin{split}
    -\eta h^{\ga\gad}\frac i2 (\partial_u\partial_v)
        &\lb
            (u \partial_y)\omega_1,
            v_\ga (\bar{\partial}_y)_\gad \int_0^1 dt(1-t)(C_1\star\varkappa)_{z\to tz}k
        \rb_{\star\ z=0}
    -\\-
    \eta \varpi^{\ga\ga}\frac i2 (\partial_u\partial_v)
        &\lb
            (u \partial_y)\omega_1,
            v_\ga(\partial_y)_\ga \int_0^1 dt (1-t)(C_1\star\varkappa)_{z\to tz}k
        \rb_{\star\ z=0}\,.
\end{split}
\end{align}
The second one is
\begin{align}\label{wWC_2}
\begin{split}
    -\eta h^{\ga\gad} \frac i2
        (\partial_u)_\ga(\bar{\partial}_y)_\gad
        &\lb
            (u\partial_y)\omega_1,
            \int_0^1 dt t(C_1\star\varkappa)_{z\to tz}k
        \rb_{\star\ z=0}
    -\\-
    \eta \varpi^{\ga\ga} \frac i2
        (\partial_u)_\ga(\partial_y)_\ga
        &\lb
            (u\partial_y)\omega_1,
            \int_0^1 dt t(C_1\star\varkappa)_{z\to tz}k
        \rb_{\star\ z=0}\,.
\end{split}
\end{align}
The final step  is to sum up \eqref{wWC_1} and \eqref{wWC_2} noticing that $C_1\star\varkappa = C_1(-z, \yd; K)e^{izy}$. Then the terms containing Lorentz connection  cancel out and one is left with  \eqref{wWC_res}.

%%%%%%%%%%%%%%%%%%%%%%%%%%%%%%%%
\section{On the consistency conditions}\label{ConsCond}
%%%%%%%%%%%%%%%%%%%%%%%%%%%%%%%%

Here we illustrate how $D_\Omega\Upsilon(\Omega,\Omega,C)$, $D_\Omega\Upsilon(\Omega,\omega,C)$ and $D_\Omega\Upsilon(\omega,\omega)$ were obtained. The simplest case is that of $\Upsilon(\omega,\omega)$:
\begin{equation*}
    D_\Omega\Upsilon(\omega,\omega) = -[\Upsilon(\Omega,\Omega,C), \omega]_\star\,,
\end{equation*}
which by using \eqref{expform} gives 
\begin{equation*}
    (D_\Omega \Upsilon(\omega,\omega))_{\omega C} = -\frac{i\eta}{4}
        (\pd_3\pd_2)^2
        e^{ - i(p_1y) - i (\pd_1\pd_2) + i (\pd_1\yd) + i (\pd_2\yd)}
    k H(Y_3)\omega(Y_1)C(Y_2)\bigg|_{Y_i = 0} + c.c.
\end{equation*}

Next consider $\Upsilon(\Omega,\Omega,C)$. ($D_\Omega\Upsilon(\Omega,\omega,C)$ is derived analogously.) With the aid of \eqref{expform}, formula \eqref{WWC_res} can be rewritten as
\begin{equation*}
    \Upsilon(\Omega,\Omega,C) = i\eta (p_1p_2)(\pd_1\pd_3)(\pd_2\pd_3)\exp{i(\pd_3\yd)}
        \Omega(Y_1)\Omega(Y_2)C(Y_3)\bigg|_{Y_i=0} + c.c.
\end{equation*}
Then, using the flatness condition \eqref{AdS_eqn} for $\Omega$ and equation of motion \eqref{C_eq} for $C$, one obtains
\begin{align}
\begin{split}\label{DY(WWC)}
    D_\Omega\Upsilon(\Omega,\Omega,C) =
        -i\eta (p_1p_2)(\pd_1\pd_3)(\pd_2\pd_3)\exp{i(\pd_3\yd)}
            (\Omega(Y_1)\star\Omega(Y_1))\cdot \Omega(Y_2)\cdot C(Y_3)k\bigg|_{Y_i=0} +\\
        +i\eta (p_1p_2)(\pd_1\pd_3)(\pd_2\pd_3)\exp{i(\pd_3\yd)}
            \Omega(Y_1)\cdot(\Omega(Y_2)\star\Omega(Y_2))\cdot C(Y_3)k\bigg|_{Y_i=0} -\\
        -i\eta (p_1p_2)(\pd_1\pd_3)(\pd_2\pd_3)\exp{i(\pd_3\yd)}
            \Omega(Y_1)\cdot\Omega(Y_2)\cdot[\Omega(Y_3)+\omega(Y_3),C(Y_3)]_\star k\bigg|_{Y_i=0} +\\
        +i\eta [\Omega(Y), (p_1p_2)(\pd_1\pd_3)(\pd_2\pd_3)\exp{i(\pd_3\yd)}
            \Omega(Y_1)\Omega(Y_2)C(Y_3)k]_\star\bigg|_{Y_i=0} + c.c.
\end{split}
\end{align}
Since in this illustrative analysis we only consider the expressions bilinear in $\omega$ and $C$ with some fixed ordering of $\omega$ and $C$ (other orderings can be considered analogously), the only contributing term to the $\omega C$ ordering from 
the \eqref{DY(WWC)} is
\begin{align*}
    (D_\Omega\Upsilon(\Omega,\Omega,C))_{\omega C} = 
        -i\eta (p_1p_2)(\pd_1\pd_3)(\pd_2\pd_3)\exp{i(\pd_3\yd)}
            \Omega(Y_1)\cdot \Omega(Y_2)\cdot (\omega(Y_3)\star C(Y_3))k\bigg|_{Y_i=0} + c.c.=\\=
        \frac{i\eta}{4}
        \big(
            (\pd_3\pd_1) + (\pd_3\pd_2)
        \big)^2
        e^{ -i (p_1p_2) - i (\pd_1\pd_2) + i (\pd_1\yd) + i (\pd_2\yd)}
    k H(Y_3)\omega(Y_1)C(Y_2)\bigg|_{Y_i = 0} + c.c.
\end{align*}
In the last equality, \eqref{expform} was used again.

%%%%%%%%%%%%%%%%%%%%%%%%%%%%%%%%
\section{First on-mass shell theorem}\label{FoMSTheorem}
%%%%%%%%%%%%%%%%%%%%%%%%%%%%%%%%

To begin with, we recall the decomposition of $\Upsilon_{m,n}$ into the frame two-forms:
\begin{align}
\begin{split}\label{Y_decomp_appendix}
    \Upsilon_{m,n}(Y) = \frac12\bigg[
            \frac{1}{m(m+1)} H(y,y)(p_1p)^2 -
            \frac{2}{m(m+2)} H(y,p) (p_1y)(p_1p) +\\+
            \frac{1}{(m+1)(m+2)} H(p,p) (p_1y)^2 +\\+
            \frac{1}{n(n+1)} \Hd(\yd,\yd)(\pd_1\pd)^2 -
            \frac{2}{n(n+2)} \Hd(\yd,\pd) (\pd_1\yd)(\pd_1\pd) +\\+
            \frac{1}{(n+1)(n+2)} \Hd(\pd,\pd) (\pd_1\yd)^2
        \bigg]\cdot\\\cdot \lp \Upsilon_{m,n}(y_1,y_1|Y) + \Upsilon_{m,n}(\yd_1,\yd_1|Y) \rp\,
\end{split}
\end{align}
with the coefficients resulting from \eqref{Sch_2}.

As a first step, suppose that there exists such a $\Delta \omega$ that removes currents from the Fronsdal equations contained in the sectors of the unfolded equations \eqref{D_omega_decomp} with  $m=s_1-1\pm1,\,n=s_1-1\mp1$ that are free from $\omega_{s_1-1\pm2,s_1-1\mp2}(Y)$. Specifically, from \eqref{D_omega_decomp} it follows that Fronsdal equations are contained in $H(y,y)$- and $\Hd(\pd,\pd)$-components for $m=s_1$, and in $\Hd(\yd,\yd)$- and $H(p,p)$-components for $m=s_1-2$. So, $\Delta \omega$ should eliminate the RHS of these sectors of the unfolded equations. Denoting the RHS of the unfolded equations by $\Upsilon$, it is assumed that the HS zero-torsion condition $\Upsilon_{s-1,s-1} = 0$ holds true, that can always be achieved by a redefinition of the Lorentz-like HS fields with $|m-n|=2$ (see section \ref{Strategy}).

Let us consider the $m>n$ case in what follows. Keeping in mind \eqref{D_omega_decomp}, one can adjust $\Delta \omega_{s_1+1,s_1-3}(Y)$ to eliminate $H(y,p)$-, $\Hd(\yd,\yd)$- and $H(p,p)$-components of $\Upsilon_{s_1,s_1-2}(Y)$. 
The rest $\Hd(\yd,\pd)$-component is zero too by virtue of the Bianchi identity for $m=s_1-1$. 
In this case the Bianchi identity takes the form $h(p,\yd)\Hd(\yd,\pd)\Upsilon(\pd,\yd|Y)_{s_1,s_1-2} + c.c. = 0$.

Next consider the Bianchi identity for $m=s_1$. Since it is shown that $\Upsilon_{s_1,s_1-2}(Y)=\Upsilon_{s_1-1,s_1-1}(Y) = 0$, it implies $h(p,\yd)\Upsilon_{s_1+1,s_1-3}=0$. The decomposition of this condition with the help of the three-forms $\mathcal{H}_{\ga \gad} := h_{\ga\gbd}H^\gbd{}_\gad$ (notice that $\mathcal{H}_{\ga \gad} =- h_{\gb\gad}H^\gb{}_\ga$) yields
\begin{align}\label{h-Upsilon}
\begin{split}
     h(p,\yd) \Upsilon_{m,n}(Y) &=
        \frac{2}{3m} \mathcal{H}(y,\yd) \Upsilon_{m.n}(p,p|Y)
        -\frac{2}{3(m+2)} \mathcal{H}(p,\yd) \Upsilon_{m.n}(y,p|Y) -\\&
        -\frac{2}{3(n+2)} \mathcal{H}(p,\yd) \Upsilon_{m.n}(\yd,\pd|Y) +
        \frac{2}{3(n+2)} \mathcal{H}(p,\pd) \Upsilon_{m.n}(\yd,\yd|Y)
        = 0\,,   
\end{split}
\end{align}
where the coefficients are determined with the help of \eqref{Y_decomp_appendix}. From \eqref{h-Upsilon} it is clear that $\Upsilon_{s_1+1,s_1-3}(p,p|Y) = \Upsilon_{s_1+1,s_1-3}(\yd,\yd|Y) = 0$. In the $H$-language this means that $H(y,y)$- and $\Hd(\pd,\pd)$-components of $\Upsilon_{s_1+1,s_1-3}(Y)$ are zero. Thus, the situation is analogous to that in the beginning of the previous paragraph.

Analogously, one can reach $\Upsilon_{m,n}(Y) = 0$ for $m,n>0$ by adjusting $\Delta \omega$. Moreover, $\Upsilon_{2s_1-2,0}(\yd,\pd|y) = \Upsilon_{2s_1-2,0}(\pd,\pd|y) = 0$. The rest parts of $\Upsilon$ are fixed by the Bianchi identity. For instance, for $n=0$, the analogue of \eqref{h-Upsilon} takes the form
\begin{align}\label{h-Upsilon:n=0}
\begin{split}
     h(p,\yd) \Upsilon_{2s_1-2,0}(Y) &=
        \frac{1}{3(s_1-1)} \mathcal{H}(y,\yd) \Upsilon_{2s_1-2,0}(p,p|Y)
        -\frac{1}{3s_1} \mathcal{H}(p,\yd) \Upsilon_{2s_1-2,0}(y,p|Y) +\\&+
        \frac{1}{3} \mathcal{H}(p,\pd) \Upsilon_{2s_1-2,0}(\yd,\yd|Y)
        = 0\,.   
\end{split}
\end{align}
As a result, the only nonzero component of $\Upsilon_{2s_1-2,0}(y)$ is $\Upsilon_{2s_1-2,0}(y,y|y)$. This is just
 $\Upsilon(\Omega,\Omega,C)$ \eqref{WWC_res}.

%%%%%%%%%%%%%%%%%%%%%%%%%%%%%%%%
\section{Coefficient functions in the expressions of currents}\label{CoefFunc}
%%%%%%%%%%%%%%%%%%%%%%%%%%%%%%%%

Here we collect expressions for the coefficients in the currents \eqref{J_s-2_res}, \eqref{J_s_res} (App. \ref{BosCur}) and their half-integer spin analogues  (App. \ref{FerCur}). These are derived by straightforward computations main steps of which are sketched above \eqref{J_s-2_res}. We present expressions only for the HS currents, i.e, for the case $s_1 > 1$, $s_2 \geqslant 1$, $s_3 \geqslant 1$. Currents involving lower spin fields can be obtained in the same way and have the structure analogous to \eqref{J_s-2_res} and \eqref{J_s_res}.

Parameters $m,n$ in $\mathbf{A}(s_1,s_2,s_3|m,n|\phi^\cB_{s_2}, \phi^\cC_{s_3})$ and $\mathbf{B}(s_1,s_2,s_3|m,n|\phi^\cB_{s_2}, \phi^\cC_{s_3})$ are assumed to be integer and half-integer. Those in other functions are non-negative integers. We do not specify the allowed domain of the parameters and spins explicitly. Instead we allow them to take any values such that the power degrees and arguments of factorials are non-negative integers.

To simplify the formulae, we introduce $\Tilde{s}_i := s_1 + s_2 + s_3 - 2s_i$ with $i=1,2,3$; $D_{s_i}$ which denotes the Lorentz covariant derivative acting only on the spin $s_i$ field; and 
\begin{equation}\label{K_def}
    \mathbf{K}(a,b;c,d|g,h) :=
        \frac{(p_1y)^a(\pd_1\yd)^b (p_2y)^{g-a}(\pd_2\yd)^{h-b} (p_1p_2)^c(\pd_1\pd_2)^d}
            {a!b!c!d!(g-a)!(h-b)!}\,.
\end{equation}

\subsection{Bosonic currents}\label{BosCur}%%%%%%%%%%%%%%%%%%%%%%%%%%%%%%%%%%%%%%%%%%%%%%%%%%%%%%%%
The coefficient function in $J_{s_1-2,s_1-2}(Y)$ is
\begin{align}
\begin{split}
    \mathbf{A}&(s_1,s_2,s_3|m,n|\phi^\cB_{s_2}, \phi^\cC_{s_3}) =
        (1-\delta_{s_2,s_3}/2)(-1)^{s_1-s_3+1}i^{-m+n+|m|+|n|}
        \frac{(s_2-|m|-1)!(s_3-|n|-1)!}{16 s_1(s_1+1)(s_2+|m|)!(s_3+|n|)!} \cdot\\\cdot&
        \big[
        (s_1-1)(-2-m-n-\tilde{s}_1)(-2+m+n-\tilde{s}_1)(m+n+\tilde{s}_1) \cdot\\\cdot&
            \mathbf{K}((-2+m-n+\tilde{s}_3)/2,(-2-m+n+\tilde{s}_3)/2;
                (2+m+n+\tilde{s}_1)/2,(2-m-n+\tilde{s}_1)/2|s_1-2,s_1-2)+\\&+
        (-1+m+n-\tilde{s}_1)(1+m+n+\tilde{s}_1)(3+m+n+\tilde{s}_1) D(y,\pd) \cdot\\\cdot&
            \mathbf{K}((-3+m-n+\tilde{s}_3)/2,(-1-m+n+\tilde{s}_3)/2;
                (3+m+n+\tilde{s}_1)/2,(1-m-n+\tilde{s}_1)/2|s_1-3,s_1-1) -\\&-
        s_1(-1+m+n-\tilde{s}_1)(1+m+n+\tilde{s}_1)(m-n+s_2-s_3) D(p,\pd) \cdot\\\cdot&
            \mathbf{K}((-1+m-n+\tilde{s}_3)/2,(-1-m+n+\tilde{s}_3)/2;
                (1+m+n+\tilde{s}_1)/2,(1-m-n+\tilde{s}_1)/2|s_1-1,s_1-1)
        \big] \cdot\\\cdot&
        \bigg[
            s_2s_3
            \lb \theta(m)D_{s_2}^m(y_1,\pd_1) + \theta(-1-m)D_{s_2}^{-m}(p_1,\yd_1) \rb 
                \cdot\\\cdot&
            \lb \theta(n)D_{s_3}^n(y_2,\pd_2) + \theta(-1-n)D_{s_3}^{-n}(p_2,\yd_2) \rb
                \phi^\cB_{s_2,s_2}(Y_1) \phi^\cC_{s_3,s_3}(Y_2)
            -\\-&
            i(s_2-1/2)(s_3-1/2) \cdot\\\cdot&
            \lb \theta(m-1/2)D_{s_2}^{m-1/2}(y_1,\pd_1) \phi^\cB_{s_2+1/2,s_2-1/2}(Y_1)
                + \theta(-1/2-m)D_{s_2}^{1/2-m}(p_1,\yd_1) \phi^\cB_{s_2-1/2,s_2+1/2}(Y_1) \rb
                 \cdot\\\cdot&
            \lb \theta(n-1/2)D_{s_3}^{n-1/2}(y_2,\pd_2) \phi^\cC_{s_3+1/2,s_3-1/2}(Y_2)
                + \theta(-1/2-n)D_{s_3}^{1/2-n}(p_2,\yd_2) \phi^\cC_{s_3-1/2,s_3+1/2}(Y_2) \rb
        \bigg] + c.c.\,;
\end{split}
\end{align}
The coefficient functions in $J_{s_1,s_1}(Y)$ are
\begin{align}
\begin{split}
    \mathbf{B}&(s_1,s_2,s_3|m,n|\phi^\cB_{s_2}, \phi^\cC_{s_3}) =
        (1-\delta_{s_2,s_3}/2)(-1)^{s_1-s_3+1}i^{-m+n+|m|+|n|}
        \frac{(s_2-|m|-1)!(s_3-|n|-1)!}{16 s_1^2(s_1+1)(s_2+|m|)!(s_3+|n|)!} \cdot\\\cdot&
        \big[
        is_1(s_1+1)(m+n-\tilde{s}_1)(m-n+\tilde{s}_3)(-m+n+\tilde{s}_2) \cdot\\\cdot&
            \mathbf{K}((m-n+\tilde{s}_3)/2,(-m+n+\tilde{s}_3)/2;
                (m+n+\tilde{s}_1)/2,(-m-n+\tilde{s}_1)/2|s_1,s_1)+\\&+
        i(-1+m+n-\tilde{s}_1)(-1+m-n-\tilde{s}_2)(1+m-n+\tilde{s}_3) D(p,\yd) \cdot\\\cdot&
            \mathbf{K}((1+m-n+\tilde{s}_3)/2,(-1-m+n+\tilde{s}_3)/2;
                (-1+m+n+\tilde{s}_1)/2,(1-m-n+\tilde{s}_1)/2|s_1+1,s_1-1) +\\&+
        s_1(s_3-s_2+n-m)(-1-m-n-\tilde{s}_1)(-1+m+n-\tilde{s}_1) D(y,\yd) \cdot\\\cdot&
            \mathbf{K}((-1+m-n+\tilde{s}_3)/2,(-1-m+n+\tilde{s}_3)/2;
                (1+m+n+\tilde{s}_1)/2,(1-m-n+\tilde{s}_1)/2|s_1-1,s_1-1)
        \big] \cdot\\\cdot&
        \bigg[
            s_2s_3
            \lb \theta(m)D_{s_2}^m(y_1,\pd_1) + \theta(-1-m)D_{s_2}^{-m}(p_1,\yd_1) \rb 
                \cdot\\\cdot&
            \lb \theta(n)D_{s_3}^n(y_2,\pd_2) + \theta(-1-n)D_{s_3}^{-n}(p_2,\yd_2) \rb
                \phi^\cB_{s_2,s_2}(Y_1) \phi^\cC_{s_3,s_3}(Y_2)
            -\\-&
            i(s_2-1/2)(s_3-1/2) \cdot\\\cdot&
            \lb \theta(m-1/2)D_{s_2}^{m-1/2}(y_1,\pd_1) \phi^\cB_{s_2+1/2,s_2-1/2}(Y_1)
                + \theta(-1/2-m)D_{s_2}^{1/2-m}(p_1,\yd_1) \phi^\cB_{s_2-1/2,s_2+1/2}(Y_1) \rb
                 \cdot\\\cdot&
            \lb \theta(n-1/2)D_{s_3}^{n-1/2}(y_2,\pd_2) \phi^\cC_{s_3+1/2,s_3-1/2}(Y_2)
                + \theta(-1/2-n)D_{s_3}^{1/2-n}(p_2,\yd_2) \phi^\cC_{s_3-1/2,s_3+1/2}(Y_2) \rb
        \bigg] + c.c.\,;
\end{split}
\end{align}
\begin{align}
\begin{split}
    \mathbf{C}&(s_1,s_2,s_3|n|\phi^\cB_{s_2}, \phi^\cC_{s_3}) =
       i^{3 s_1 - 2 s_2 + s_3 + |n + s_1 - s_2| - n - 1}\cdot\\\cdot&
        \frac{s_3 (-s_1+2s_2-n)(s_1-1)!(s_2-|s_1-s_2+n|-1)!}
            {2s_1(2s_3-1) (2s_3+n)!(s_1+n-1)!(s_2+|s_1-s_2+n|)!}\cdot\\\cdot&
        \Big[
            is_1
                \mathbf{K}(2(s_2-s_3-n),s_1;-s_1+2s_2-n,n|s_1,s_1) D_{s_3}(y_2,\yd_2)+\\+&
            (n + 2s_3) D(y,\pd)
                \mathbf{K}(2(s_2-s_3-n),s_1+1;-s_1+2s_2-n,n-1|s_1-1,s_1+1)
        \Big]\cdot\\\cdot&
        \bigg[
        s_2 \lb \theta(s_1-s_2+n)D_{s_2}^{|n + s_1 - s_2|}(p_1,\yd_1)
            + \theta(-s_1+s_2-n-1)D_{s_2}^{|n + s_1 - s_2|}(y_1,\pd_1) \rb\cdot\\\cdot&
        D_{s_3}^{n - 1}(y_2,\yd_2) D_{s_3}^{s_3}(p_2,\yd_2)
        \phi^\cB_{s_2,s_2}(Y_1) \phi^\cC_{s_3,s_3}(Y_2)
    +\\+&
        (s_2-1/2) \Big[ \theta(s_1-s_2+n-1/2)D_{s_2}^{|n + s_1 - s_2|-1/2}(p_1,\yd_1)
            \phi^\cB_{s_2-1/2,s_2+1/2}(Y_1)
            +\\+& \theta(-s_1+s_2-n-1/2)D_{s_2}^{|n + s_1 - s_2|-1/2}(y_1,\pd_1)
                \phi^\cB_{s_2+1/2,s_2-1/2}(Y_1) \Big]\cdot\\\cdot&
        D_{s_3}^{n - 1}(y_2,\yd_2) D_{s_3}^{s_3-1/2}(p_2,\yd_2) \phi^\cC_{s_3-1/2,s_3+1/2}(Y_2)
         \bigg]
        + (s_2 \leftrightarrow s_3) + c.c.\,;
\end{split}
\end{align}
\begin{align}
\begin{split}
    \mathbf{D}&(s_1,s_2,s_3|m|\phi^\cB_{s_2}, \phi^\cC_{s_3}) =
        i^{3 s_1 + 2 m - 1}
        \frac{s_2 s_3 (s_1-1)! (s_1 - s_2 - s_3)! (s_1 + s_2 + s_3)! (s_3 - s_2 - 1)!}
            {2s_1 (2 s_2 - 1) (2 s_3 - 1) (2 s_1 + 1)! (s_1 - s_2 + s_3 - 1)!}\cdot\\\cdot&
        \frac{1}
            {m! (m - 2 s_2)! (s_1 - m)!(s_1 + 2 s_3 - m)!}
        \Big[
            s_1(s_3-s_2)
                \mathbf{K}(m,-s_2-s_3+m;0,-s_2+s_3|s_1,s_1)D_{s_3}(y_2,\yd_2) +\\+&
            i (2s_3+s_1-m)(s_1-m)D(y,\pd)
                \mathbf{K}(m,-s_2-s_3+m+1;0,-s_2+s_3-1|s_1-1,s_1+1)
        \Big]\cdot\\\cdot&
         D_{s_2}^{m-2s_2}(y_1,\yd_1) D_{s_3}^{s_1-m-1}(y_2,\yd_2)
         \Big[
            D_{s_2}^{s_2}(p_1,\yd_1) D_{s_3}^{s_3}(p_2,\yd_2)
                \phi^\cB_{s_2,s_2}(Y_1) \phi^\cC_{s_3,s_3}(Y_2) +\\+&
            iD_{s_2}^{s_2-1/2}(p_1,\yd_1) D_{s_3}^{s_3-1/2}(p_2,\yd_2)
                \phi^\cB_{s_2-1/2,s_2+1/2}(Y_1) \phi^\cC_{s_3-1/2,s_3+1/2}(Y_2)
        \Big]
        + (s_2 \leftrightarrow s_3) + c.c.\,;
\end{split}
\end{align}
\begin{align}
\begin{split}
    \mathbf{E}&(s_1,s_2,s_3|n|\phi^\cB_{s_2}, \phi^\cC_{s_3}) =
        i^{3 s_1 - 2 s_2 + s_3 + |n + s_1 - s_2 + 2 s_3| - n - 1}\cdot\\\cdot&
        \frac{s_3(-s_1+2s_2-2s_3-n)(s_1-1)!(s_2-|s_1-s_2+2s_3+n|-1)!}
            {2s_1(2s_3-1) n!(s_1+2s_3+n-1)!(s_2+|s_1-s_2+2s_3+n|)!}\cdot\\\cdot&
        \Big[
            is_1
                \mathbf{K}(2(s_2-s_3-n),s_1;-s_1+2s_2-2s_3-n,n+2s_3|s_1,s_1) D_{s_3}(y_2,\yd_2)+\\+&
            n D(y,\pd)
                \mathbf{K}(2(s_2-s_3-n),s_1+1;-s_1+2s_2-2s_3-n,2s_3+n-1|s_1-1,s_1+1)
        \Big]\cdot\\\cdot&
        \bigg[
            s_2 \lb
                \theta(s_1-s_2+2s_3+n)D_{s_2}^{|n + s_1 - s_2 + 2s_3|}(p_1,\yd_1)
                + \theta(-s_1+s_2-2s_3-n-1)D_{s_2}^{|n + s_1 - s_2 + 2s_3|}(y_1,\pd_1)
            \rb\cdot\\\cdot&
            D_{s_3}^{n-1}(y_2,\yd_2) D_{s_3}^{s_3}(y_2,\pd_2)
                \phi^\cB_{s_2,s_2}(Y_1) \phi^\cC_{s_3,s_3}(Y_2) 
            +\\+&
            (s_2-1/2) \big[ \theta(s_1-s_2+2s_3+n-1/2)D_{s_2}^{|n + s_1 - s_2 + 2s_3|-1/2}(p_1,\yd_1)
            \phi^\cB_{s_2-1/2,s_2+1/2}(Y_1) 
            +\\+& \theta(-s_1+s_2-2s_3-n-1/2)D_{s_2}^{|n + s_1 - s_2 + 2s_3|-1/2}(y_1,\pd_1)
                \phi^\cB_{s_2+1/2,s_2-1/2}(Y_1) \big]\cdot\\\cdot&
        D_{s_3}^{n-1}(y_2,\yd_2) D_{s_3}^{s_3-1/2}(y_2,\pd_2)
            \phi^\cC_{s_3+1/2,s_3-1/2}(Y_2)
        \bigg]
    + (s_2 \leftrightarrow s_3) + c.c.\,;
\end{split}
\end{align}
\begin{align}
\begin{split}
    \mathbf{F}&(s_1,s_2,s_3|m|\phi^\cB_{s_2}, \phi^\cC_{s_3}) =
        (1-\delta_{s_2,s_3}/2) i^{3 s_1 + 2 m - 1} \cdot\\\cdot&
        \frac{s_2 s_3 (s_1 - 1)! (s_1 + s_2 - s_3)! (s_1 - s_2 + s_3)! (s_2 + s_3 - 1)!}
            {2s_1 (2 s_2 - 1) (2 s_3 - 1) (2 s_1 + 1)! (s_1 + s_2 + s_3 - 1)!
                m! (m + 2 s_2)! (s_1 - m)!(s_1 + 2 s_3 - m)!}
        \cdot\\\cdot&
        \Big[
            s_1(s_2+s_3)
                \mathbf{K}(m,s_2-s_3+m;0,s_2+s_3|s_1,s_1)D_{s_3}(y_2,\yd_2) +\\+&
            i (2s_3+s_1-m)(s_1-m)D(y,\pd)
                \mathbf{K}(m,s_2-s_3+m+1;0,s_2+s_3-1|s_1-1,s_1+1)
        \Big]\cdot\\\cdot&
        D_{s_2}^m(y_1,\yd_1) D_{s_3}^{s_1-m-1}(y_2,\yd_2)
        \Bigg[
            D_{s_2}^{s_2}(p_1,\yd_1) D_{s_3}^{s_3}(p_2,\yd_2)
                \phi^\cB_{s_2,s_2}(Y_1) \phi^\cC_{s_3,s_3}(Y_2)
            +\\+&i D_{s_2}^{s_2-1/2}(p_1,\yd_1) D_{s_3}^{s_3-1/2}(p_2,\yd_2)
                \phi^\cB_{s_2-1/2,s_2+1/2}(Y_1) \phi^\cC_{s_3-1/2,s_3+1/2}(Y_2)
        \Bigg] + c.c.
\end{split}
\end{align}

\subsection{Fermionic currents}\label{FerCur}%%%%%%%%%%%%%%%%%%%%%%%%%%%%%%%%%%%%%%%%%%%%%%%%%%%
The coefficient function in $J_{s_1-5/2,s_1-3/2}(Y)$ is
\begin{align}
\begin{split}
    \mathbf{A}&(s_1,s_2,s_3|m,n|\phi^\cB_{s_2}, \phi^\cC_{s_3}) =
        (1-\delta_{s_2,s_3}/2)(-1)^{s_1-s_3}i^{-m+n+|m|+|n|}
        \frac{(s_2-|m|-1)!(s_3-|n|-1)!}{16 s_1(s_1+1)(s_2+|m|)!(s_3+|n|)!} \cdot\\\cdot&
        (3/2-m-n+\tilde{s}_1)(5/2+m+n+\tilde{s}_1)(m+n+\tilde{s}_1+1/2) \cdot\\\cdot&
            \mathbf{K}((-5/2+m-n+\tilde{s}_3)/2,(-3/2-m+n+\tilde{s}_3)/2;\\&\hspace{2cm}
                (5/2+m+n+\tilde{s}_1)/2,(3/2-m-n+\tilde{s}_1)/2|s_1-5/2,s_1-3/2)\cdot\\\cdot&
        \bigg[
            s_2(s_3-1/2)
            \lb \theta(m)D_{s_2}^m(y_1,\pd_1) + \theta(-1-m)D_{s_2}^{-m}(p_1,\yd_1) \rb 
                \phi^\cB_{s_2,s_2}(Y_1)
                \cdot\\\cdot&
            \lb \theta(n-1/2)D_{s_3}^{n-1/2}(y_2,\pd_2) \phi^\cC_{s_3+1/2,s_3-1/2}(Y_2)
                + \theta(-n-1/2)D_{s_3}^{-n+1/2}(p_2,\yd_2)\phi^\cC_{s_3-1/2,s_3+1/2}(Y_2) \rb
            +\\+&
            (s_2-1/2)s_3 \cdot\\\cdot&
            \lb \theta(m-1/2)D_{s_2}^{m-1/2}(y_1,\pd_1) \phi^\cB_{s_2+1/2,s_2-1/2}(Y_1)
                + \theta(-m-1/2)D_{s_2}^{1/2-m}(p_1,\yd_1) \phi^\cB_{s_2-1/2,s_2+1/2}(Y_1) \rb
                 \cdot\\\cdot&
            \lb \theta(n)D_{s_3}^{n}(y_2,\pd_2) + \theta(-n-1)D_{s_3}^{-n}(p_2,\yd_2) \rb 
                \phi^\cC_{s_3,s_3}(Y_2)
        \bigg] + c.c.\,;
\end{split}
\end{align}
The coefficient function in $J_{s_1-1/2,s_1-3/2}(Y)$ is
\begin{align}
\begin{split}
    \mathbf{A}&(s_1,s_2,s_3|m,n|\phi^\cB_{s_2}, \phi^\cC_{s_3}) =
        (1-\delta_{s_2,s_3}/2)(-1)^{s_1-s_3}i^{-m+n+|m|+|n|}
        \frac{(s_1^2-1/4)(s_2-|m|-1)!(s_3-|n|-1)!}{32 s_1(s_2+|m|)!(s_3+|n|)!} \cdot\\\cdot&
        (3/2-m-n+\tilde{s}_1)(1/2+m+n+\tilde{s}_1)(-m+n+2s_1(s_3-s_2))) \cdot\\\cdot&
            \mathbf{K}((-1/2+m-n+\tilde{s}_3)/2,(-3/2-m+n+\tilde{s}_3)/2;\\&\hspace{2cm}
                (1/2+m+n+\tilde{s}_1)/2,(3/2-m-n+\tilde{s}_1)/2|s_1-1/2,s_1-3/2)\cdot\\\cdot&
        \bigg[
            s_2(s_3-1/2)
            \lb \theta(m)D_{s_2}^m(y_1,\pd_1) + \theta(-1-m)D_{s_2}^{-m}(p_1,\yd_1) \rb 
                \phi^\cB_{s_2,s_2}(Y_1)
                \cdot\\\cdot&
            \lb \theta(n-1/2)D_{s_3}^{n-1/2}(y_2,\pd_2) \phi^\cC_{s_3+1/2,s_3-1/2}(Y_2)
                + \theta(-n-1/2)D_{s_3}^{-n+1/2}(p_2,\yd_2)\phi^\cC_{s_3-1/2,s_3+1/2}(Y_2) \rb
            +\\+&
            (s_2-1/2)s_3 \cdot\\\cdot&
            \lb \theta(m-1/2)D_{s_2}^{m-1/2}(y_1,\pd_1) \phi^\cB_{s_2+1/2,s_2-1/2}(Y_1)
                + \theta(-m-1/2)D_{s_2}^{1/2-m}(p_1,\yd_1) \phi^\cB_{s_2-1/2,s_2+1/2}(Y_1) \rb
                 \cdot\\\cdot&
            \lb \theta(n)D_{s_3}^{n}(y_2,\pd_2) + \theta(-n-1)D_{s_3}^{-n}(p_2,\yd_2) \rb 
                \phi^\cC_{s_3,s_3}(Y_2)
        \bigg] + c.c.\,;
\end{split}
\end{align}
The coefficient functions in $J_{s_1+1/2,s_1-1/2}(Y)$ are
\begin{align}
\begin{split}
    \mathbf{B}&(s_1,s_2,s_3|m,n|\phi^\cB_{s_2}, \phi^\cC_{s_3}) =
        (1-\delta_{s_2,s_3}/2)(-1)^{s_1-s_3}i^{1-m+n+|m|+|n|}
        \frac{(s_2-|m|-1)!(s_3-|n|-1)!}{16 (2s_1-1)(s_2+|m|)!(s_3+|n|)!} \cdot\\\cdot&
        (1-2m-2n+2\tilde{s}_1)(1-2m+2n+2\tilde{s}_2)(1+2m-2n+2\tilde{s}_3) \cdot\\\cdot&
            \mathbf{K}((m-n+\tilde{s}_3+1/2)/2,(-m+n+\tilde{s}_3-1/2)/2;\\&\hspace{2cm}
                (m+n+\tilde{s}_1-1/2)/2,(-m-n+\tilde{s}_1+1/2)/2|s_1+1/2,s_1-1/2) \cdot\\\cdot&
        \bigg[
            s_2(s_3-1/2)
            \lb \theta(m)D_{s_2}^m(y_1,\pd_1) + \theta(-1-m)D_{s_2}^{-m}(p_1,\yd_1) \rb
                \phi^\cB_{s_2,s_2}(Y_1)
                \cdot\\\cdot&
            \lb \theta(n-1/2)D_{s_3}^{n-1/2}(y_2,\pd_2) \phi^\cC_{s_3+1/2,s_3-1/2}(Y_2)
                + \theta(-1/2-n)D_{s_3}^{1/2-n}(p_2,\yd_2) \phi^\cC_{s_3-1/2,s_3+1/2}(Y_2) \rb
            +\\+&
            (s_2-1/2)s_3 \cdot\\\cdot&
            \lb \theta(m-1/2)D_{s_2}^{m-1/2}(y_1,\pd_1) \phi^\cB_{s_2+1/2,s_2-1/2}(Y_1)
                + \theta(-1/2-m)D_{s_2}^{1/2-m}(p_1,\yd_1) \phi^\cB_{s_2-1/2,s_2+1/2}(Y_1) \rb
                 \cdot\\\cdot&
            \lb \theta(n)D_{s_3}^n(y_2,\pd_2) + \theta(-1-n)D_{s_3}^{-n}(p_2,\yd_2) \rb
                 \phi^\cC_{s_3,s_3}(Y_2)
        \bigg] + c.c.\,;
\end{split}
\end{align}
\begin{align}
\begin{split}
    \mathbf{C}&(s_1,s_2,s_3|n|\phi^\cB_{s_2}, \phi^\cC_{s_3}) =
       i^{1 - n + 3 s_1 - 2 s_2 + s_3 + |n + s_1 - s_2 + 1/2|}\cdot\\\cdot&
        \frac{s_3 (-s_1+2s_2+n+1/2)(s_1+1/2)!(s_2-|s_1-s_2+n+1/2|-1)!}
            {(2s_1-1)(2s_3-1) (2s_3+n)!(s_1+n-1/2)!(s_2+|s_1-s_2+n+1/2|)!}
        \cdot\\\cdot&
        \mathbf{K}(2(s_2-s_3-n-1/2),s_1+1/2;-s_1+2s_2-n-1/2,n|s_1-1/2,s_1+1/2)
        \cdot\\\cdot&
        \bigg[
        s_2 \lb \theta(s_1-s_2+n-1/2)D_{s_2}^{|n + s_1 - s_2+1/2|}(p_1,\yd_1)
            + \theta(-s_1+s_2-n-3/2)D_{s_2}^{|n + s_1 - s_2+1/2|}(y_1,\pd_1) \rb\cdot\\\cdot&
        D_{s_3}^{n}(y_2,\yd_2) D_{s_3}^{s_3-1/2}(p_2,\yd_2)
        \phi^\cB_{s_2,s_2}(Y_1) \phi^\cC_{s_3,s_3}(Y_2)
    +\\+&
        (s_2-1/2) \Big[ \theta(s_1-s_2+n)D_{s_2}^{|n + s_1 - s_2+1/2|-1/2}(p_1,\yd_1)
            \phi^\cB_{s_2-1/2,s_2+1/2}(Y_1)
            +\\+& \theta(-s_1+s_2-n-1)D_{s_2}^{|n + s_1 - s_2+1/2|-1/2}(y_1,\pd_1)
                \phi^\cB_{s_2+1/2,s_2-1/2}(Y_1) \Big]\cdot\\\cdot&
        D_{s_3}^{n}(y_2,\yd_2) D_{s_3}^{s_3}(p_2,\yd_2) \phi^\cC_{s_3,s_3}(Y_2)
         \bigg]
        + (s_2 \leftrightarrow s_3) + c.c.\,;
\end{split}
\end{align}
\begin{align}
\begin{split}
    \mathbf{D}&(s_1,s_2,s_3|m|\phi^\cB_{s_2}, \phi^\cC_{s_3}) =
        i^{- s_1 + 2 m + 1/2} \frac{s_2 s_3}{(2s_1 - 1) (2 s_2 - 1) (2 s_3 - 1)} \cdot\\\cdot&
        \frac{(s_1+1/2)! (s_1 - s_2 - s_3)! (s_1 + s_2 + s_3)! (s_3 - s_2 - 1/2)!}
            {(2 s_1 + 1)! (s_1 - s_2 + s_3 - 1)!
                m! (m - 2 s_2)! (s_1 - m - 1/2)!(s_1 + 2 s_3 - m - 1/2)!}\cdot\\\cdot&
        \mathbf{K}(m,-s_2-s_3+m+1/2;0,-s_2+s_3-1/2|s_1-1/2,s_1+1/2)D_{s_3}(y_2,\yd_2)
        \cdot\\\cdot&
         D_{s_2}^{m-2s_2}(y_1,\yd_1) D_{s_3}^{s_1-m-1/2}(y_2,\yd_2)
         \Big[
            D_{s_2}^{s_2}(p_1,\yd_1) D_{s_3}^{s_3-1/2}(p_2,\yd_2)
                \phi^\cB_{s_2,s_2}(Y_1) \phi^\cC_{s_3-1/2,s_3+1/2}(Y_2) -\\-&
            D_{s_2}^{s_2-1/2}(p_1,\yd_1) D_{s_3}^{s_3}(p_2,\yd_2)
                \phi^\cB_{s_2-1/2,s_2+1/2}(Y_1) \phi^\cC_{s_3-1/2,s_3+1/2}(Y_2)
        \Big]
        + (s_2 \leftrightarrow s_3) + c.c.\,;
\end{split}
\end{align}
\begin{align}
\begin{split}
    \mathbf{E}&(s_1,s_2,s_3|n|\phi^\cB_{s_2}, \phi^\cC_{s_3}) =
        i^{1 - n + 3 s_1 - 2 s_2 + s_3 + |n + s_1 - s_2 + 2 s_3+1/2|}\cdot\\\cdot&
        \frac{s_3(-s_1+2s_2-2s_3-n-1/2)(s_1-1)!(s_2-|s_1-s_2+2s_3+n+1/2|-1)!}
            {(2s_1-1)(2s_3-1) n!(s_1+2s_3+n-1/2)!(s_2+|s_1-s_2+2s_3+n+1/2|)!}\cdot\\\cdot&
        \mathbf{K}(2(s_2-s_3-n-1/2),s_1+1/2;-s_1+2s_2-2s_3-n-1/2,n+2s_3|s_1-1/2,s_1+1/2)
        \cdot\\\cdot&
        \bigg[
            s_2 \big[
                \theta(s_1-s_2+2s_3+n-1/2)D_{s_2}^{|n + s_1 - s_2 + 2s_3+1/2|}(p_1,\yd_1)
                +\\+& \theta(-s_1+s_2-2s_3-n-3/2)D_{s_2}^{|n + s_1 - s_2 + 2s_3+1/2|}(y_1,\pd_1)
            \big]\cdot\\\cdot&
            D_{s_3}^{n}(y_2,\yd_2) D_{s_3}^{s_3-1/2}(p_2,\yd_2)
                \phi^\cB_{s_2,s_2}(Y_1) \phi^\cC_{s_3-1/2,s_3+1/2}(Y_2) 
            +\\+&
            (s_2-1/2) \big[ \theta(s_1-s_2+2s_3+n)D_{s_2}^{|n + s_1 - s_2 + 2s_3+1/2|-1/2}(p_1,\yd_1)
            \phi^\cB_{s_2-1/2,s_2+1/2}(Y_1) 
            +\\+& \theta(-s_1+s_2-2s_3-n-1)D_{s_2}^{|n + s_1 - s_2 + 2s_3+1/2|-1/2}(y_1,\pd_1)
                \phi^\cB_{s_2+1/2,s_2-1/2}(Y_1) \big]\cdot\\\cdot&
        D_{s_3}^{n-1}(y_2,\yd_2) D_{s_3}^{s_3}(p_2,\yd_2)
            \phi^\cC_{s_3,s_3}(Y_2)
        \bigg]
    + (s_2 \leftrightarrow s_3) + c.c.\,;
\end{split}
\end{align}
\begin{align}
\begin{split}
    \mathbf{F}&(s_1,s_2,s_3|m|\phi^\cB_{s_2}, \phi^\cC_{s_3}) =
        (1-\delta_{s_2,s_3}/2) i^{-s_1 + 2 m + 3/2}
        \frac{s_2 s_3}{(2s_1 - 1) (2 s_2 - 1) (2 s_3 - 1)}\cdot\\\cdot&
        \frac{(s_1 + 1/2)! (s_1 + s_2 - s_3)! (s_1 - s_2 + s_3)! (s_2 + s_3 - 1/2)!}
            {(2 s_1 + 1)! (s_1 + s_2 + s_3 - 1)!
                m! (m + 2 s_2)! (s_1 - m - 1/2)!(s_1 + 2 s_3 - m - 1/2)!}
        \cdot\\\cdot&
        \mathbf{K}(m,s_2-s_3+m+1/2;0,s_2+s_3-1/2|s_1-1/2,s_1+1/2)D_{s_3}(y_2,\yd_2)
        \cdot\\\cdot&
        D_{s_2}^m(y_1,\yd_1) D_{s_3}^{s_1-m-1/2}(y_2,\yd_2)
        \Bigg[
            D_{s_2}^{s_2}(p_1,\yd_1) D_{s_3}^{s_3-1/2}(p_2,\yd_2)
                \phi^\cB_{s_2,s_2}(Y_1) \phi^\cC_{s_3-1/2,s_3+1/2}(Y_2)
            -\\-& D_{s_2}^{s_2-1/2}(p_1,\yd_1) D_{s_3}^{s_3}(p_2,\yd_2)
                \phi^\cB_{s_2-1/2,s_2+1/2}(Y_1) \phi^\cC_{s_3,s_3}(Y_2)
        \Bigg] + c.c.
\end{split}
\end{align}
\end{appendices}

\bibliographystyle{JHEP}

\begin{thebibliography}{10}

\bibitem{Vasiliev:1990en}
M.~A. Vasiliev, \emph{{Consistent equation for interacting gauge fields of all
  spins in (3+1)-dimensions}},
  \href{http://dx.doi.org/10.1016/0370-2693(90)91400-6}{\emph{Phys. Lett. B}
  {\bf 243} (1990) 378--382}.

\bibitem{Vasiliev:1992av}
M.~A. Vasiliev, \emph{{More on equations of motion for interacting massless
  fields of all spins in (3+1)-dimensions}},
  \href{http://dx.doi.org/10.1016/0370-2693(92)91457-K}{\emph{Phys. Lett. B}
  {\bf 285} (1992) 225--234}.

\bibitem{Vasiliev:1999ba}
M.~A. Vasiliev, \emph{{Higher spin gauge theories: Star product and AdS
  space}},  \href{http://arxiv.org/abs/hep-th/9910096}{{\tt hep-th/9910096}}.

\bibitem{Gelfond:2018vmi}
O.~A. Gelfond and M.~A. Vasiliev, \emph{{Homotopy Operators and Locality
  Theorems in Higher-Spin Equations}},
  \href{http://dx.doi.org/10.1016/j.physletb.2018.09.038}{\emph{Phys. Lett. B}
  {\bf 786} (2018) 180--188}, [\href{http://arxiv.org/abs/1805.11941}{{\tt
  1805.11941}}].

\bibitem{Didenko:2018fgx}
V.~E. Didenko, O.~A. Gelfond, A.~V. Korybut and M.~A. Vasiliev, \emph{{Homotopy
  Properties and Lower-Order Vertices in Higher-Spin Equations}},
  \href{http://dx.doi.org/10.1088/1751-8121/aae5e1}{\emph{J. Phys. A} {\bf 51}
  (2018) 465202}, [\href{http://arxiv.org/abs/1807.00001}{{\tt 1807.00001}}].

\bibitem{Didenko:2019xzz}
V.~E. Didenko, O.~A. Gelfond, A.~V. Korybut and M.~A. Vasiliev, \emph{{Limiting
  Shifted Homotopy in Higher-Spin Theory and Spin-Locality}},
  \href{http://dx.doi.org/10.1007/JHEP12(2019)086}{\emph{JHEP} {\bf 12} (2019)
  086}, [\href{http://arxiv.org/abs/1909.04876}{{\tt 1909.04876}}].

\bibitem{Gelfond:2019tac}
O.~A. Gelfond and M.~A. Vasiliev, \emph{{Spin-Locality of Higher-Spin Theories
  and Star-Product Functional Classes}},
  \href{http://dx.doi.org/10.1007/JHEP03(2020)002}{\emph{JHEP} {\bf 03} (2020)
  002}, [\href{http://arxiv.org/abs/1910.00487}{{\tt 1910.00487}}].

\bibitem{Didenko:2020bxd}
V.~E. Didenko, O.~A. Gelfond, A.~V. Korybut and M.~A. Vasiliev,
  \emph{{Spin-locality of $\eta^{2}$ and $ {\overline{\eta}}^2 $ quartic
  higher-spin vertices}},
  \href{http://dx.doi.org/10.1007/JHEP12(2020)184}{\emph{JHEP} {\bf 12} (2020)
  184}, [\href{http://arxiv.org/abs/2009.02811}{{\tt 2009.02811}}].

\bibitem{Didenko:2022qga}
V.~E. Didenko, \emph{{On holomorphic sector of higher-spin theory}},
  \href{http://dx.doi.org/10.1007/JHEP10(2022)191}{\emph{JHEP} {\bf 10} (2022)
  191}, [\href{http://arxiv.org/abs/2209.01966}{{\tt 2209.01966}}].

\bibitem{Didenko:2023vna}
V.~E. Didenko and A.~V. Korybut, \emph{{Interaction of symmetric higher-spin
  gauge fields}},
  \href{http://dx.doi.org/10.1103/PhysRevD.108.086031}{\emph{Phys. Rev. D} {\bf
  108} (2023) 086031}, [\href{http://arxiv.org/abs/2304.08850}{{\tt
  2304.08850}}].

\bibitem{Sleight:2017pcz}
C.~Sleight and M.~Taronna, \emph{{Higher-Spin Gauge Theories and Bulk
  Locality}},
  \href{http://dx.doi.org/10.1103/PhysRevLett.121.171604}{\emph{Phys. Rev.
  Lett.} {\bf 121} (2018) 171604}, [\href{http://arxiv.org/abs/1704.07859}{{\tt
  1704.07859}}].

\bibitem{Lysov:2022nsv}
V.~Lysov and Y.~Neiman, \emph{{Bulk locality and gauge invariance for
  boundary-bilocal cubic correlators in higher-spin gravity}},
  \href{http://dx.doi.org/10.1007/JHEP12(2022)142}{\emph{JHEP} {\bf 12} (2022)
  142}, [\href{http://arxiv.org/abs/2209.00854}{{\tt 2209.00854}}].

\bibitem{Neiman:2023orj}
Y.~Neiman, \emph{{Quartic locality of higher-spin gravity in de Sitter and
  Euclidean anti-de Sitter space}},
  \href{http://dx.doi.org/10.1016/j.physletb.2023.138048}{\emph{Phys. Lett. B}
  {\bf 843} (2023) 138048}, [\href{http://arxiv.org/abs/2302.00852}{{\tt
  2302.00852}}].

\bibitem{Vasiliev:1988sa}
M.~A. Vasiliev, \emph{{Consistent Equations for Interacting Massless Fields of
  All Spins in the First Order in Curvatures}},
  \href{http://dx.doi.org/10.1016/0003-4916(89)90261-3}{\emph{Annals Phys.}
  {\bf 190} (1989) 59--106}.

\bibitem{Boulanger:2015ova}
N.~Boulanger, P.~Kessel, E.~D. Skvortsov and M.~Taronna, \emph{{Higher spin
  interactions in four-dimensions: Vasiliev versus Fronsdal}},
  \href{http://dx.doi.org/10.1088/1751-8113/49/9/095402}{\emph{J. Phys. A} {\bf
  49} (2016) 095402}, [\href{http://arxiv.org/abs/1508.04139}{{\tt
  1508.04139}}].

\bibitem{Misuna:2017bjb}
N.~Misuna, \emph{{On current contribution to Fronsdal equations}},
  \href{http://dx.doi.org/10.1016/j.physletb.2018.01.019}{\emph{Phys. Lett. B}
  {\bf 778} (2018) 71--78}, [\href{http://arxiv.org/abs/1706.04605}{{\tt
  1706.04605}}].

\bibitem{Sleight:2016dba}
C.~Sleight and M.~Taronna, \emph{{Higher Spin Interactions from Conformal Field
  Theory: The Complete Cubic Couplings}},
  \href{http://dx.doi.org/10.1103/PhysRevLett.116.181602}{\emph{Phys. Rev.
  Lett.} {\bf 116} (2016) 181602}, [\href{http://arxiv.org/abs/1603.00022}{{\tt
  1603.00022}}].

\bibitem{Metsaev:2005ar}
R.~R. Metsaev, \emph{{Cubic interaction vertices of massive and massless higher
  spin fields}},
  \href{http://dx.doi.org/10.1016/j.nuclphysb.2006.10.002}{\emph{Nucl. Phys. B}
  {\bf 759} (2006) 147--201}, [\href{http://arxiv.org/abs/hep-th/0512342}{{\tt
  hep-th/0512342}}].

\bibitem{Fronsdal:1978rb}
C.~Fronsdal, \emph{{Massless Fields with Integer Spin}},
  \href{http://dx.doi.org/10.1103/PhysRevD.18.3624}{\emph{Phys. Rev. D} {\bf
  18} (1978) 3624}.

\bibitem{Bengtsson:1983pd}
A.~K.~H. Bengtsson, I.~Bengtsson and L.~Brink, \emph{{Cubic Interaction Terms
  for Arbitrary Spin}},
  \href{http://dx.doi.org/10.1016/0550-3213(83)90140-2}{\emph{Nucl. Phys. B}
  {\bf 227} (1983) 31--40}.

\bibitem{Bengtsson:1986kh}
A.~K.~H. Bengtsson, I.~Bengtsson and N.~Linden, \emph{{Interacting Higher Spin
  Gauge Fields on the Light Front}},
  \href{http://dx.doi.org/10.1088/0264-9381/4/5/028}{\emph{Class. Quant. Grav.}
  {\bf 4} (1987) 1333}.

\bibitem{Metsaev:2018xip}
R.~R. Metsaev, \emph{{Light-cone gauge cubic interaction vertices for massless
  fields in AdS(4)}},
  \href{http://dx.doi.org/10.1016/j.nuclphysb.2018.09.021}{\emph{Nucl. Phys. B}
  {\bf 936} (2018) 320--351}, [\href{http://arxiv.org/abs/1807.07542}{{\tt
  1807.07542}}].

\bibitem{Berends:1984wp}
F.~A. Berends, G.~J.~H. Burgers and H.~Van~Dam, \emph{{ON SPIN THREE
  SELFINTERACTIONS}}, \href{http://dx.doi.org/10.1007/BF01410362}{\emph{Z.
  Phys. C} {\bf 24} (1984) 247--254}.

\bibitem{Berends:1984rq}
F.~A. Berends, G.~J.~H. Burgers and H.~van Dam, \emph{{On the Theoretical
  Problems in Constructing Interactions Involving Higher Spin Massless
  Particles}},
  \href{http://dx.doi.org/10.1016/0550-3213(85)90074-4}{\emph{Nucl. Phys. B}
  {\bf 260} (1985) 295--322}.

\bibitem{Berends:1985xx}
F.~A. Berends, G.~J.~H. Burgers and H.~van Dam, \emph{{Explicit Construction of
  Conserved Currents for Massless Fields of Arbitrary Spin}},
  \href{http://dx.doi.org/10.1016/S0550-3213(86)80019-0}{\emph{Nucl. Phys. B}
  {\bf 271} (1986) 429--441}.

\bibitem{Metsaev:1991mt}
R.~R. Metsaev, \emph{{Poincare invariant dynamics of massless higher spins:
  Fourth order analysis on mass shell}},
  \href{http://dx.doi.org/10.1142/S0217732391000348}{\emph{Mod. Phys. Lett. A}
  {\bf 6} (1991) 359--367}.

\bibitem{Metsaev:1993ap}
R.~R. Metsaev, \emph{{Generating function for cubic interaction vertices of
  higher spin fields in any dimension}},
  \href{http://dx.doi.org/10.1142/S0217732393003706}{\emph{Mod. Phys. Lett. A}
  {\bf 8} (1993) 2413--2426}.

\bibitem{Metsaev:1999ui}
R.~R. Metsaev, \emph{{Light cone form of field dynamics in Anti-de Sitter
  space-time and AdS / CFT correspondence}},
  \href{http://dx.doi.org/10.1016/S0550-3213(99)00554-4}{\emph{Nucl. Phys. B}
  {\bf 563} (1999) 295--348}, [\href{http://arxiv.org/abs/hep-th/9906217}{{\tt
  hep-th/9906217}}].

\bibitem{Bekaert:2005jf}
X.~Bekaert, N.~Boulanger and S.~Cnockaert, \emph{{Spin three gauge theory
  revisited}},
  \href{http://dx.doi.org/10.1088/1126-6708/2006/01/052}{\emph{JHEP} {\bf 01}
  (2006) 052}, [\href{http://arxiv.org/abs/hep-th/0508048}{{\tt
  hep-th/0508048}}].

\bibitem{Bengtsson:2006pw}
A.~K.~H. Bengtsson, \emph{{Structure of higher spin gauge interactions}},
  \href{http://dx.doi.org/10.1063/1.2751277}{\emph{J. Math. Phys.} {\bf 48}
  (2007) 072302}, [\href{http://arxiv.org/abs/hep-th/0611067}{{\tt
  hep-th/0611067}}].

\bibitem{Metsaev:2007rn}
R.~R. Metsaev, \emph{{Cubic interaction vertices for fermionic and bosonic
  arbitrary spin fields}},
  \href{http://dx.doi.org/10.1016/j.nuclphysb.2012.01.022}{\emph{Nucl. Phys. B}
  {\bf 859} (2012) 13--69}, [\href{http://arxiv.org/abs/0712.3526}{{\tt
  0712.3526}}].

\bibitem{Manvelyan:2010jr}
R.~Manvelyan, K.~Mkrtchyan and W.~Ruhl, \emph{{General trilinear interaction
  for arbitrary even higher spin gauge fields}},
  \href{http://dx.doi.org/10.1016/j.nuclphysb.2010.04.019}{\emph{Nucl. Phys. B}
  {\bf 836} (2010) 204--221}, [\href{http://arxiv.org/abs/1003.2877}{{\tt
  1003.2877}}].

\bibitem{Sagnotti:2010at}
A.~Sagnotti and M.~Taronna, \emph{{String Lessons for Higher-Spin
  Interactions}},
  \href{http://dx.doi.org/10.1016/j.nuclphysb.2010.08.019}{\emph{Nucl. Phys. B}
  {\bf 842} (2011) 299--361}, [\href{http://arxiv.org/abs/1006.5242}{{\tt
  1006.5242}}].

\bibitem{Fotopoulos:2010ay}
A.~Fotopoulos and M.~Tsulaia, \emph{{On the Tensionless Limit of String theory,
  Off - Shell Higher Spin Interaction Vertices and BCFW Recursion Relations}},
  \href{http://dx.doi.org/10.1007/JHEP11(2010)086}{\emph{JHEP} {\bf 11} (2010)
  086}, [\href{http://arxiv.org/abs/1009.0727}{{\tt 1009.0727}}].

\bibitem{Vasiliev:2011knf}
M.~A. Vasiliev, \emph{{Cubic Vertices for Symmetric Higher-Spin Gauge Fields in
  $(A)dS_d$}},
  \href{http://dx.doi.org/10.1016/j.nuclphysb.2012.04.012}{\emph{Nucl. Phys. B}
  {\bf 862} (2012) 341--408}, [\href{http://arxiv.org/abs/1108.5921}{{\tt
  1108.5921}}].

\bibitem{Joung:2012rv}
E.~Joung, L.~Lopez and M.~Taronna, \emph{{On the cubic interactions of massive
  and partially-massless higher spins in (A)dS}},
  \href{http://dx.doi.org/10.1007/JHEP07(2012)041}{\emph{JHEP} {\bf 07} (2012)
  041}, [\href{http://arxiv.org/abs/1203.6578}{{\tt 1203.6578}}].

\bibitem{Bengtsson:2012jm}
A.~K.~H. Bengtsson, \emph{{Systematics of Higher-spin Light-front
  Interactions}},  \href{http://arxiv.org/abs/1205.6117}{{\tt 1205.6117}}.

\bibitem{Buchbinder:2012xa}
I.~L. Buchbinder, T.~V. Snegirev and Y.~M. Zinoviev, \emph{{On gravitational
  interactions for massive higher spins in $AdS_3$}},
  \href{http://dx.doi.org/10.1088/1751-8113/46/21/214015}{\emph{J. Phys. A}
  {\bf 46} (2013) 214015}, [\href{http://arxiv.org/abs/1208.0183}{{\tt
  1208.0183}}].

\bibitem{Francia:2016weg}
D.~Francia, G.~L. Monaco and K.~Mkrtchyan, \emph{{Cubic interactions of
  Maxwell-like higher spins}},
  \href{http://dx.doi.org/10.1007/JHEP04(2017)068}{\emph{JHEP} {\bf 04} (2017)
  068}, [\href{http://arxiv.org/abs/1611.00292}{{\tt 1611.00292}}].

\bibitem{Buchbinder:2017nuc}
I.~L. Buchbinder, S.~J. Gates and K.~Koutrolikos, \emph{{Higher Spin Superfield
  interactions with the Chiral Supermultiplet: Conserved Supercurrents and
  Cubic Vertices}},
  \href{http://dx.doi.org/10.3390/universe4010006}{\emph{Universe} {\bf 4}
  (2018) 6}, [\href{http://arxiv.org/abs/1708.06262}{{\tt 1708.06262}}].

\bibitem{Kuzenko:2022hdv}
S.~M. Kuzenko, M.~Ponds and E.~S.~N. Raptakis, \emph{{Conformal Interactions
  Between Matter and Higher-Spin (Super)Fields}},
  \href{http://dx.doi.org/10.1002/prop.202200157}{\emph{Fortsch. Phys.} {\bf
  71} (2023) 2200157}, [\href{http://arxiv.org/abs/2208.07783}{{\tt
  2208.07783}}].

\bibitem{Buchbinder:2023xlb}
I.~L. Buchbinder and A.~A. Reshetnyak, \emph{{Consistent Lagrangians for
  Irreducible Interacting Higher-Spin Fields with Holonomic Constraints}},
  \href{http://dx.doi.org/10.1134/S1063779623060084}{\emph{Phys. Part. Nucl.}
  {\bf 54} (2023) 1066--1071}, [\href{http://arxiv.org/abs/2304.10358}{{\tt
  2304.10358}}].

\bibitem{Buchbinder:2024pjm}
I.~Buchbinder, E.~Ivanov and N.~Zaigraev, \emph{{$\mathcal{N} = 2$
  superconformal higher-spin multiplets and their hypermultiplet couplings}},
  \href{http://arxiv.org/abs/2404.19016}{{\tt 2404.19016}}.

\bibitem{Giombi:2012ms}
S.~Giombi and X.~Yin, \emph{{The Higher Spin/Vector Model Duality}},
  \href{http://dx.doi.org/10.1088/1751-8113/46/21/214003}{\emph{J. Phys. A}
  {\bf 46} (2013) 214003}, [\href{http://arxiv.org/abs/1208.4036}{{\tt
  1208.4036}}].

\bibitem{Maldacena:2012sf}
J.~Maldacena and A.~Zhiboedov, \emph{{Constraining conformal field theories
  with a slightly broken higher spin symmetry}},
  \href{http://dx.doi.org/10.1088/0264-9381/30/10/104003}{\emph{Class. Quant.
  Grav.} {\bf 30} (2013) 104003}, [\href{http://arxiv.org/abs/1204.3882}{{\tt
  1204.3882}}].

\bibitem{Klebanov:2002ja}
I.~R. Klebanov and A.~M. Polyakov, \emph{{AdS dual of the critical O(N) vector
  model}}, \href{http://dx.doi.org/10.1016/S0370-2693(02)02980-5}{\emph{Phys.
  Lett. B} {\bf 550} (2002) 213--219},
  [\href{http://arxiv.org/abs/hep-th/0210114}{{\tt hep-th/0210114}}].

\bibitem{Sezgin:2002rt}
E.~Sezgin and P.~Sundell, \emph{{Massless higher spins and holography}},
  \href{http://dx.doi.org/10.1016/S0550-3213(02)00739-3}{\emph{Nucl. Phys. B}
  {\bf 644} (2002) 303--370}, [\href{http://arxiv.org/abs/hep-th/0205131}{{\tt
  hep-th/0205131}}].

\bibitem{Sezgin:2003pt}
E.~Sezgin and P.~Sundell, \emph{{Holography in 4D (super) higher spin theories
  and a test via cubic scalar couplings}},
  \href{http://dx.doi.org/10.1088/1126-6708/2005/07/044}{\emph{JHEP} {\bf 07}
  (2005) 044}, [\href{http://arxiv.org/abs/hep-th/0305040}{{\tt
  hep-th/0305040}}].

\bibitem{Aharony:2011jz}
O.~Aharony, G.~Gur-Ari and R.~Yacoby, \emph{{d=3 Bosonic Vector Models Coupled
  to Chern-Simons Gauge Theories}},
  \href{http://dx.doi.org/10.1007/JHEP03(2012)037}{\emph{JHEP} {\bf 03} (2012)
  037}, [\href{http://arxiv.org/abs/1110.4382}{{\tt 1110.4382}}].

\bibitem{Giombi:2011kc}
S.~Giombi, S.~Minwalla, S.~Prakash, S.~P. Trivedi, S.~R. Wadia and X.~Yin,
  \emph{{Chern-Simons Theory with Vector Fermion Matter}},
  \href{http://dx.doi.org/10.1140/epjc/s10052-012-2112-0}{\emph{Eur. Phys. J.
  C} {\bf 72} (2012) 2112}, [\href{http://arxiv.org/abs/1110.4386}{{\tt
  1110.4386}}].

\bibitem{Joung:2011ww}
E.~Joung and M.~Taronna, \emph{{Cubic interactions of massless higher spins in
  (A)dS: metric-like approach}},
  \href{http://dx.doi.org/10.1016/j.nuclphysb.2012.03.013}{\emph{Nucl. Phys. B}
  {\bf 861} (2012) 145--174}, [\href{http://arxiv.org/abs/1110.5918}{{\tt
  1110.5918}}].

\bibitem{Didenko:2014dwa}
V.~E. Didenko and E.~D. Skvortsov, \emph{{Elements of Vasiliev theory}},
  \href{http://arxiv.org/abs/1401.2975}{{\tt 1401.2975}}.

\bibitem{Sezgin:2000hr}
E.~Sezgin and P.~Sundell, \emph{{On curvature expansion of higher spin gauge
  theory}}, \href{http://dx.doi.org/10.1088/0264-9381/18/16/314}{\emph{Class.
  Quant. Grav.} {\bf 18} (2001) 3241--3250},
  [\href{http://arxiv.org/abs/hep-th/0012168}{{\tt hep-th/0012168}}].

\bibitem{Bekaert:2004qos}
X.~Bekaert, S.~Cnockaert, C.~Iazeolla and M.~A. Vasiliev, \emph{{Nonlinear
  higher spin theories in various dimensions}},  in \emph{{1st Solvay Workshop
  on Higher Spin Gauge Theories}}, pp.~132--197, 2004.
\newblock \href{http://arxiv.org/abs/hep-th/0503128}{{\tt hep-th/0503128}}.

\bibitem{Gutperle:2014aja}
M.~Gutperle and Y.~Li, \emph{{Higher Spin Lifshitz Theory and Integrable
  Systems}}, \href{http://dx.doi.org/10.1103/PhysRevD.91.046012}{\emph{Phys.
  Rev. D} {\bf 91} (2015) 046012}, [\href{http://arxiv.org/abs/1412.7085}{{\tt
  1412.7085}}].

\bibitem{Beccaria:2015iwa}
M.~Beccaria, M.~Gutperle, Y.~Li and G.~Macorini, \emph{{Higher spin Lifshitz
  theories and the Korteweg-de Vries hierarchy}},
  \href{http://dx.doi.org/10.1103/PhysRevD.92.085005}{\emph{Phys. Rev. D} {\bf
  92} (2015) 085005}, [\href{http://arxiv.org/abs/1504.06555}{{\tt
  1504.06555}}].

\bibitem{Vasiliev:1986td}
M.~A. Vasiliev, \emph{{Free Massless Fields of Arbitrary Spin in the De Sitter
  Space and Initial Data for a Higher Spin Superalgebra}}, {\emph{Fortsch.
  Phys.} {\bf 35} (1987) 741--770}.

\bibitem{Vasiliev:2016xui}
M.~A. Vasiliev, \emph{{Current Interactions and Holography from the 0-Form
  Sector of Nonlinear Higher-Spin Equations}},
  \href{http://dx.doi.org/10.1007/JHEP10(2017)111}{\emph{JHEP} {\bf 10} (2017)
  111}, [\href{http://arxiv.org/abs/1605.02662}{{\tt 1605.02662}}].

\bibitem{Gelfond:2017wrh}
O.~A. Gelfond and M.~A. Vasiliev, \emph{{Current Interactions from the One-Form
  Sector of Nonlinear Higher-Spin Equations}},
  \href{http://dx.doi.org/10.1016/j.nuclphysb.2018.04.017}{\emph{Nucl. Phys. B}
  {\bf 931} (2018) 383--417}, [\href{http://arxiv.org/abs/1706.03718}{{\tt
  1706.03718}}].

\bibitem{Vasiliev:1980as}
M.~A. Vasiliev, \emph{{'Gauge' form of description of massless fields with
  arbitrary spin}}, {\emph{Yad. Fiz.} {\bf 32} (1980) 855--861}.

\bibitem{Shaynkman:2000ts}
O.~V. Shaynkman and M.~A. Vasiliev, \emph{{Scalar field in any dimension from
  the higher spin gauge theory perspective}},
  \href{http://dx.doi.org/10.1007/BF02551402}{\emph{Theor. Math. Phys.} {\bf
  123} (2000) 683--700}, [\href{http://arxiv.org/abs/hep-th/0003123}{{\tt
  hep-th/0003123}}].

\bibitem{Gelfond:2003vh}
O.~A. Gelfond and M.~A. Vasiliev, \emph{{Higher rank conformal fields in the
  Sp(2M) symmetric generalized space-time}},
  \href{http://dx.doi.org/10.1007/s11232-005-0168-9}{\emph{Theor. Math. Phys.}
  {\bf 145} (2005) 1400--1424},
  [\href{http://arxiv.org/abs/hep-th/0304020}{{\tt hep-th/0304020}}].

\bibitem{Smirnov:2015waz}
P.~A. Smirnov and M.~A. Vasiliev, \emph{{Gauge Non-Invariant Higher-Spin
  Currents in $AdS_4$}},
  \href{http://dx.doi.org/10.3390/universe3040078}{\emph{Universe} {\bf 3}
  (2017) 78}, [\href{http://arxiv.org/abs/1512.07226}{{\tt 1512.07226}}].

\bibitem{Stasheff1963-ia}
J.~D. Stasheff, \emph{Homotopy associativity of {H-Spaces}. {I}}, {\emph{Trans.
  Am. Math. Soc.} {\bf 108} (Aug., 1963) 275}.

\bibitem{Stasheff1963-dp}
J.~D. Stasheff, \emph{Homotopy associativity of {H-Spaces}. {II}},
  {\emph{Trans. Am. Math. Soc.} {\bf 108} (Aug., 1963) 293}.

\bibitem{Fradkin:1986qy}
E.~S. Fradkin and M.~A. Vasiliev, \emph{{Cubic Interaction in Extended Theories
  of Massless Higher Spin Fields}},
  \href{http://dx.doi.org/10.1016/0550-3213(87)90469-X}{\emph{Nucl. Phys. B}
  {\bf 291} (1987) 141--171}.

\bibitem{Vasiliev:1986qx}
M.~A. Vasiliev, \emph{{Extended Higher Spin Superalgebras and Their
  Realizations in Terms of Quantum Operators}}, {\emph{Fortsch. Phys.} {\bf 36}
  (1988) 33--62}.

\bibitem{Bychkov:2021zvd}
A.~S. Bychkov, K.~A. Ushakov and M.~A. Vasiliev, \emph{{The
  \ensuremath{\sigma}\ensuremath{-} Cohomology Analysis for Symmetric
  Higher-Spin Fields}},
  \href{http://dx.doi.org/10.3390/sym13081498}{\emph{Symmetry} {\bf 13} (2021)
  1498}, [\href{http://arxiv.org/abs/2107.01736}{{\tt 2107.01736}}].

\bibitem{Vasiliev:2017cae}
M.~A. Vasiliev, \emph{{On the Local Frame in Nonlinear Higher-Spin Equations}},
  \href{http://dx.doi.org/10.1007/JHEP01(2018)062}{\emph{JHEP} {\bf 01} (2018)
  062}, [\href{http://arxiv.org/abs/1707.03735}{{\tt 1707.03735}}].

\bibitem{Fang:1978wz}
J.~Fang and C.~Fronsdal, \emph{{Massless Fields with Half Integral Spin}},
  \href{http://dx.doi.org/10.1103/PhysRevD.18.3630}{\emph{Phys. Rev. D} {\bf
  18} (1978) 3630}.

\bibitem{Vasiliev:2007yc}
M.~A. Vasiliev, \emph{{On Conformal, SL(4,R) and Sp(8,R) Symmetries of 4d
  Massless Fields}},
  \href{http://dx.doi.org/10.1016/j.nuclphysb.2007.10.017}{\emph{Nucl. Phys. B}
  {\bf 793} (2008) 469--526}, [\href{http://arxiv.org/abs/0707.1085}{{\tt
  0707.1085}}].

\bibitem{Landau1987-cd}
L.~D. Landau and E.~M. Lifshitz, \emph{The classical theory of fields}.
\newblock Butterworth-Heinemann, Oxford, England, 4~ed., Jan., 1987.

\end{thebibliography}

\providecommand{\href}[2]{#2}\begingroup\raggedright\endgroup

\end{document}